\documentclass{aastex7}
\usepackage[T1]{fontenc}
\usepackage{graphicx} 
\usepackage{natbib}
\usepackage{amsmath}
\usepackage{xcolor}

\newcommand{\kps}{km\,s$^{-1}$}

\begin{document}

\title{Quasi-periodic pulsations and three-dimensional magnetic reconnection\\ during 2022 March 31 flare observed by IRIS \& STIX}

\correspondingauthor{J. L\"{o}rin\v{c}\'{i}k} 
\email{lorincik@baeri.org}

\author[0000-0002-9690-8456]{Juraj L\"{o}rin\v{c}\'{i}k}\email{lorincik@baeri.org}
\affil{Bay Area Environmental Research Institute, NASA Ames Research Park, Moffett Field, CA 94035, USA}
\affil{Lockheed Martin Solar \& Astrophysics Laboratory, Org. A021S, Bldg. 203, 3251 Hanover St., Palo Alto, CA 94304, USA}
\author[0000-0001-5592-8023]{Hannah Collier}\email{hannah.collier@fhnw.ch}
\affil{University of Applied Sciences and Arts Northwestern Switzerland (FHNW), Bahnhofstrasse 6, 5210 Windisch, Switzerland}
\affil{ETH Zürich, Rämistrasse 101, 8092 Zürich, Switzerland}
\author[0000-0002-4980-7126]{Vanessa Polito}\email{polito@lmsal.com}
\affil{Lockheed Martin Solar \& Astrophysics Laboratory, Org. A021S, Bldg. 203, 3251 Hanover St., Palo Alto, CA 94304, USA}
\affil{Department of Physics, Oregon State University, 301 Weniger Hall, Corvallis, OR 97331}
\author[0000-0002-6835-2390]{Laura A. Hayes}\email{laura.hayes@dias.ie}
\affil{Astronomy $\&$ Astrophysics Section, School of Cosmic Physics, Dublin Institute for Advanced Studies, DIAS Dunsink Observatory, Dublin D15XR2R, Ireland}
\author[0000-0002-6368-939X]{William H. Ashfield IV}\email{william.ashfield@swri.org}
\affil{Southwest Research Institute, 1301 Walnut St, Suite 400, Boulder, CO 80302}
\author[0000-0002-6253-082X]{Nabil Freij}\email{nfreij@seti.org}
\affil{SETI Institute, 339 Bernardo Avenue, Mountain View, CA 94043, USA}
\affil{Lockheed Martin Solar \& Astrophysics Laboratory, Org. A021S, Bldg. 203, 3251 Hanover St., Palo Alto, CA 94304, USA}

\begin{abstract}

Apparent slipping motions of flare ribbon kernels and the formation of hard X-ray (HXR) footpoints are important signatures of magnetic reconnection in solar flares. Ultraviolet (UV) and HXR ribbon emission can show quasi-periodic pulsations (QPPs), but the link between HXR QPP sources and slipping reconnection remains poorly understood. In this work, we analyze high-cadence IRIS and STIX observations of the 2022 March 31 M9.6 flare. STIX detected non-thermal QPPs with periods of $\approx$35 s from two relatively stationary footpoints. The majority of the HXR QPPs were correlated with UV pulsations observed by the IRIS Slit Jaw Imager in ribbon regions encompassing these footpoints. In one such region observed under the IRIS slit, the \ion{Si}{4} 1402.77\,\AA~line exhibited pulsations in intensity, Doppler shift, and width, some of which were coincident with HXR QPPs. Apparent slipping motions of the UV kernels were also observed, but their locations, timing, and UV intensity variability showed lower correlation with the HXR QPPs. The ribbons along which the slipping kernels were found correspond well to footprints of quasi-separatrix layers (QSLs), regions of high magnetic connectivity gradients, identified using a NLFFF extrapolation. Our multi-instrument analysis suggests that strong deposition of energy by non-thermal electrons was concentrated in a specific loop system within a large-scale 3D reconnection structure. Slipping kernels were subjected to weaker, if any, energization by non-thermal electrons, offering new constraints on 3D reconnection and flare energy release.

\end{abstract}

\section{Introduction} \label{sec:introduction}

Solar flares are powerful explosions in the solar atmosphere. The energy release during flares is governed by magnetic reconnection, converting magnetic energy accumulated in the corona into particle acceleration, in-situ heating, magnetohydrodynamics waves (MHD) and bulk motions. Accelerated electrons and ions, waves, and conduction fronts then propagate along newly reconnected field lines and deposit their energy in the lower solar atmosphere \citep{Carmichael64, Sturrock66, Hirayama74, Kopp76}. There, this energy deposition promptly induces enhanced emission along flare ribbons, elongated bright structures typically observed in H$\alpha$ and ultraviolet (UV) emission \citep[see e.g., the review by][]{Fletcher11}. Ribbons often exhibit complicated morphology and dynamics governed by three-dimensional (3D) magnetic reconnection \citep[see e.g., the reviews by][]{Demoulin06, Janvier17, Dudik25}. 

In complex 3D configurations, magnetic reconnection occurs in quasi-separatrix layers \citep[QSLs;][]{Priest95, Demoulin96a, Titov02}, geometrical structures with high gradients of magnetic connectivity. Magnetic field lines reconnecting in QSLs exchange their connectivity in a sequence, driving apparent slipping motions of their footpoints along QSL footprints \citep[][]{Aulanier06, Janvier13}. Observationally, slipping reconnection manifests as apparent slipping motions of flare loops and their footpotins (`kernels') along polarity inversion lines (PILs). With characteristic sizes down to $\approx 100$\,km, kernels are fundamental constituents of flare ribbon structure. The observed slipping speeds $v_\mathrm{slip}$ range between tens \citep[e.g.][]{Dudik14, Dudik16, LiT16} to thousands \citep{Lorincik25a, ZhangY25} of kilometers per second. Those velocities were discovered using high-cadence observations from the Interface Region Imaging Spectrograph \citep[IRIS;][]{Depontieu14, Depontieu21}. 

Slipping reconnection can be quasi-periodic, exhibiting intermittent appearance and motions of slipping loops or kernels at timescales between $3-6$ minutes \citep{LiT15} and down to $\approx$8\,s \citep{Lorincik25b, ZhangY25}. Recent observations also provided compelling evidence that flare energy release associated with slipping reconnection can be bursty, or intermittent \citep[see also][]{Kashapova20, Huang24}. While slipping reconnection primarily reflects the spatial progression of energy release and deposition, its bursty nature leaves clear temporal signatures in flare emission. One of the most prominent temporal manifestations of bursty flare energy release is the detection of quasi-periodic signal variations collectively termed quasi-periodic pulsations \citep[QPPs; see e.g.,][and references therein]{Nakariakov09, VanDoorsselaere16, Hayes20}. QPPs are commonly observed in microwave radio, soft X-ray (SXR), hard X-ray (HXR), and UV wavelengths, with periods ranging between fractions of a second to tens of minutes. They are related to a variety of flare-driven processes, including MHD waves, bursty reconnection, or the combination of the two \citep[see e.g., reviews by][]{McLaughlin18, Zimovets21}. 

QPPs often occur concurrently in different parts of the electromagnetic spectrum, reflecting the coupling between the energy release and the response of the lower atmosphere. For example, HXR bursts can correlate with spikes in the UV emission from the chromosphere or the transition region \citep[TR; e.g.][]{Poland82, Tandberg83, Qiu12}. Combining multi-wavelength observations provide a comprehensive view of the flare energization (e.g., in HXR and radio wavelengths) and the corresponding response of the lower atmosphere (UV ribbon emission and spectra). This approach has proven to be particularly valuable in the identification of QPP drivers. Notably, \citet{Ashfield25} recently studied QPPs in chromospheric condensation \citep[e.g.,][]{Graham15, Warren16, Graham20} observed at high cadence by IRIS and the Swedish Solar Telescope. The condensation QPPs were well-correlated with HXR QPPs at timescales of about 32\,s, providing strong evidence that bursty reconnection acted as their physical driver \citep[see also][]{Lorincik22}.

HXR QPPs are of particular importance because they directly trace the acceleration of non-thermal electrons and their subsequent energy deposition in the lower atmosphere. The strongest non-thermal HXR emission is usually localized in compact HXR footpoints found along flare ribbons \citep[see e.g., the reviews][]{Krucker08, Benz17}. The dense chromosphere acts as a thick target, in which impacting particles decelerate and emit bremsstrahlung radiation. The HXR source reconstruction relies on indirect imaging techniques, more recently provided by the Spectrometer Telescope for Imaging X-rays \citep[STIX;][]{Krucker20} onboard Solar Orbiter \citep{Muller20}. For recent studies of HXR QPPs with STIX, see e.g., \citet[][]{Collier23, Collier24, French24, Szaforz25, Purkhart25}.

HXR footpoints can exhibit complex motions. When analyzed in tandem with ribbon dynamics, these motions offer valuable insights into the spatial and temporal progression of reconnection during flares \citep[see Section 3.4. in][and references therein]{Fletcher11}. In observations, HXR footpoints often occur in pairs and spread away \citep[e.g.,][]{Asai04, Krucker05, Inglis13} or move parallel \citep[see e.g.,][]{Fletcher02, Bogachev05, Inglis12} to the PIL. The former behavior is commonly associated with the ribbon separation, a consequence of the upward progression of the reconnection site in time. The parallel motions are consistent with the progression of reconnection along the ribbons as described by slipping reconnection in the framework of more complex 3D reconnection. Despite this conceptual link, the spatio-temporal correspondence between slipping reconnection and non-thermal HXR emission remains largely unexplored. Observational evidence of slipping reconnection traced by HXR footpoints is exceptionally rare, limiting our understanding of how energy released during slipping reconnection is distributed along ribbons.

For instance, slipping reconnection, associated thermodynamic evolution, and locations of HXR footpoints were analyzed by \citet{Jing17}. This study showed $\approx 10$\,MK plasma propagating along the ribbon concurrently with a $25-50$\,keV HXR footpoint over roughly 13 minutes. Recently, \citet{Purkhart25} analyzed the spatio-temporal evolution of ribbons, kernels, and HXR emission observed by STIX. They identified long-period ($3.4$ minute) QPPs in the $15-25$\,keV STIX energy range that correlated well with the appearance and subsequent slipping motions of kernels in one of the ribbons under study. Furthermore, HXR footpoints reconstructed during two QPPs roughly 4 minutes apart were co-spatial with compact UV kernels. Slipping reconnection is however an inherently dynamic process and can occur intermittently at much shorter timescales, down to only a few seconds \citep{Lorincik25b, ZhangY25}. How QPPs at such short periods manifest across multiple diagnostics remains largely unknown. Addressing this question requires high-cadence (e.g., < 10\,s) observations of slipping dynamics and simultaneously HXR measurements of non-thermal energy deposition. STIX has been involved in several successful flare observing campaigns \citep{Ryan25}, in some cases in coordination with IRIS, offering a unique opportunity to tackle this observational challenge.

In this study, we analyze observations of a large M-class flare from 2022 March 31 captured simultaneously by Solar Orbiter/STIX and IRIS. This dataset is well suited to study bursty flare energy release, as it combines multi-instrument coverage of QPPs and slipping reconnection at high cadence. STIX provides measurements of HXR flux and source reconstruction powered by indirect imaging, while IRIS surveys ribbon dynamics and spectra in chromospheric and TR lines. We mainly perform four complementary diagnostics that probe reconnection and energy transport: 1) HXR QPPs and imaging to study the spatio-temporal distribution of non-thermal electrons, 2) ribbon intensity pulsations to study the intermittency of UV emission, 3) slipping reconnection to study the progression of reconnection in QSLs, and finally 4) \ion{Si}{4} spectra to probe the response of the TR to the episodic energy deposition. By combining these diagnostics we provide a comprehensive view of quasi-periodic pulsations and slipping reconnection on short timescales. 

This article is structured as follows. In Section \ref{sec:data} we describe the data products used in our analysis. Section \ref{sec:overview} provides a brief overview of the event under study and the analysis of slipping kernels. Section \ref{sec:qpps} presents the analysis of QPPs detected by STIX and IRIS. In Section \ref{sec:qsls} we analyze the magnetic environment of the flare and its relation to the ribbons and HXR footpoints. Section \ref{sec:spectroscopy} delves into IRIS spectroscopy. Finally, Sections \ref{sec:disc} and \ref{sec:summary} present the discussion and summary of our results. 

\section{Data} \label{sec:data}

In this manuscript, we primarily focus on UV ribbon imaging observations from IRIS. IRIS is a UV spectrograph operating in two far-UV (FUV, 1331.6\,\AA--1358.4\,\AA~and 1380.6\,\AA--1406.8\,\AA) and one near-UV (NUV, 2782.6\,\AA--2833.9\,\AA) bands. These bands contain numerous lines formed across a broad range of temperatures between \mbox{log $T$ [K] = 3.7 and 7}. The default wavelength resolution of IRIS in the FUV and NUV bands is ~0.013\,\AA~and ~0.026\,\AA, respectively. IRIS provides imaging context observations via its Slit-Jaw Imager (SJI) that monitors the light reflected off the instrument CCD. The native pixel size of IRIS is 0.\arcsec 167 and its spatial resolution of $\approx 0.\arcsec 33$ is limited by the Nyquist criterion. The 2022 March 31 flare under study was observed in a high-cadence observing program with $\approx 0.8$\,s cadence for the spectroscopic sit-and-stare observations and 8\,s cadence for the SJI observations. The latter were taken in the 2796\,\AA~channel dominated by \ion{Mg}{2} emission formed in the chromosphere. In order to reduce the telemetry, the observing program employs spectral and spatial binning of 2, the latter increasing the pixel size to 0.\arcsec 33. Level-2 IRIS data were loaded and processed using the \texttt{SolarSoft} and \texttt{irispy-lmsal} libraries. 

Context imaging observations of the flare under study were taken by the Atmospheric Imaging Assembly \citep[AIA;][]{Lemen12} onboard the Solar Dynamics Observatory \citep[][]{Pesnell12}. AIA provides continuous full-disk images of the Sun at $\approx$1.\arcsec 5 spatial resolution in 3 UV and 7 extreme-UV (EUV) passbands sensitive to emission with characteristic temperatures of log($T$\,[K])\,=\,3.7 -- 7.2 \citep{Odwyer10}. AIA observations are carried out by default at a cadence of 12\,s or 24\,s, depending on the channel. AIA coordinate frames were used for reference to co-align the datasets used in this work. AIA and IRIS data were first corrected for differential rotation, with the reference time set to 18:30 UT during the late impulsive phase. SJI 2796\,\AA~snapshots taken close in time to this instant were then cross-correlated with the AIA 1600\,\AA~observations using the \texttt{auto\_align\_images} routine in \texttt{SolarSoft}. Shift of $-1.75\arcsec$ in both $X$ and $Y$ directions was applied to the SJI observations in order to reach a reasonable co-alignment between the IRIS and AIA data. In addition, we found that AIA 304\,\AA~channel data were slightly misaligned with other channels of AIA due to a brief failure of the limb detection algorithm (AIA team, private comm.). Manual comparison between locations of bright ribbon emission observed in AIA 1600\,\AA~and 304\,\AA~revealed that a shift of $X = -6\arcsec$ and $Y = -1\arcsec$ had to be applied to the 304\,\AA~channel observations to correct for this error.

\begin{figure}[h]
    \centering
    \includegraphics[width=8.95cm]{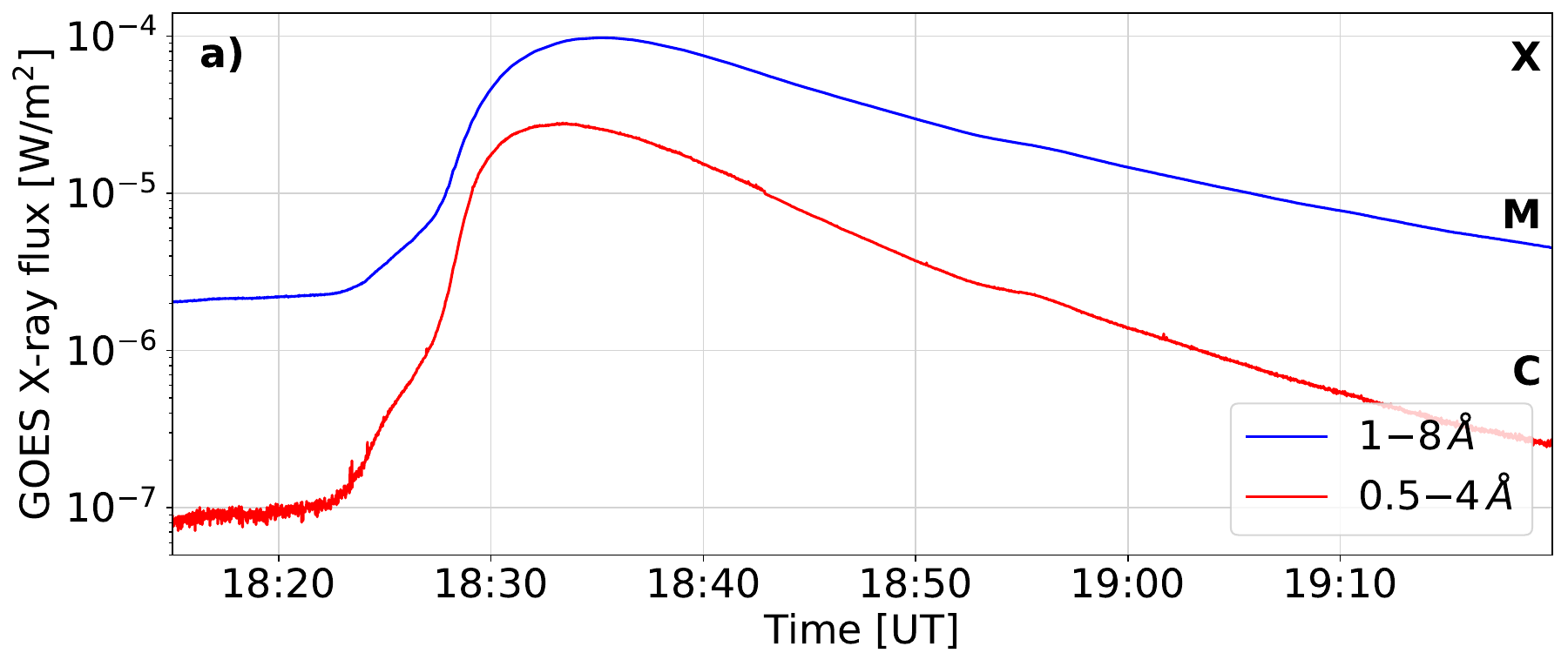}
    \includegraphics[width=8.95cm]{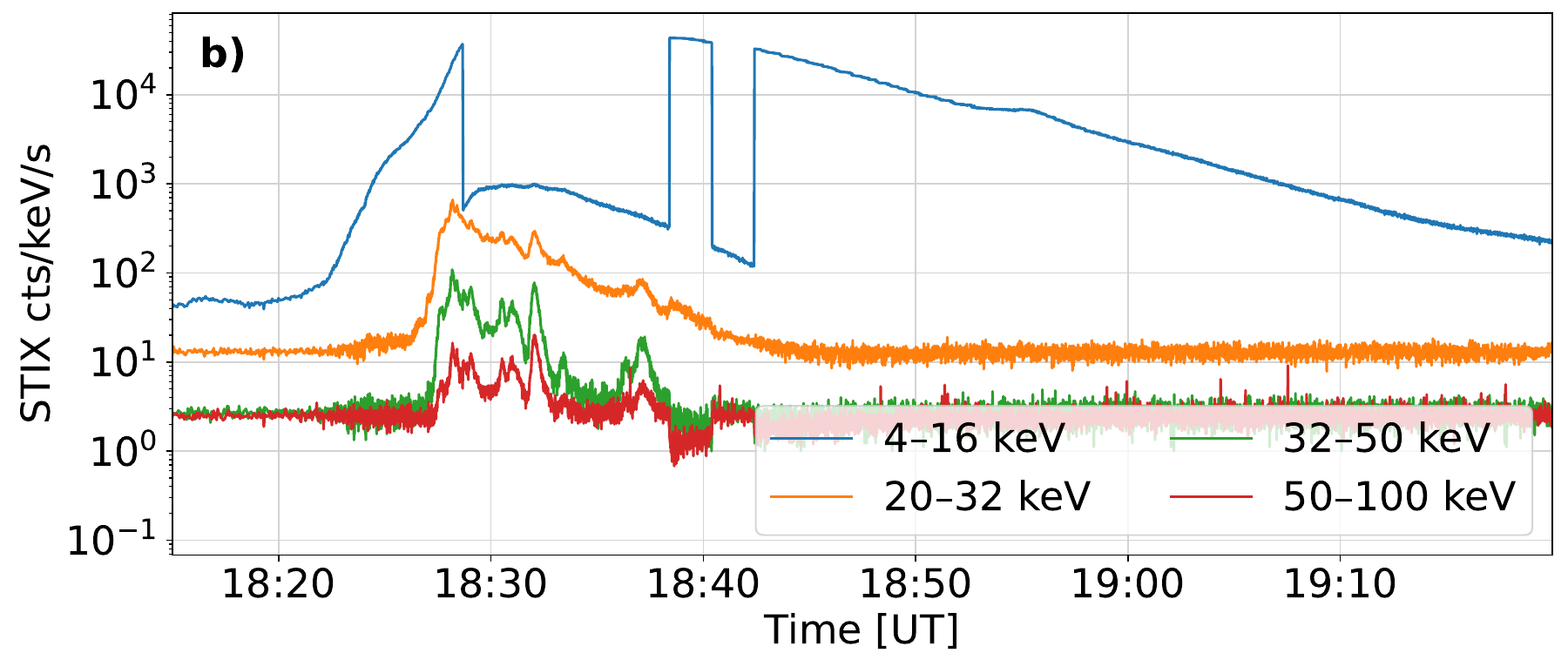} 
    \\
    \includegraphics[width=18cm, clip, viewport = 20 20 1270 1150]{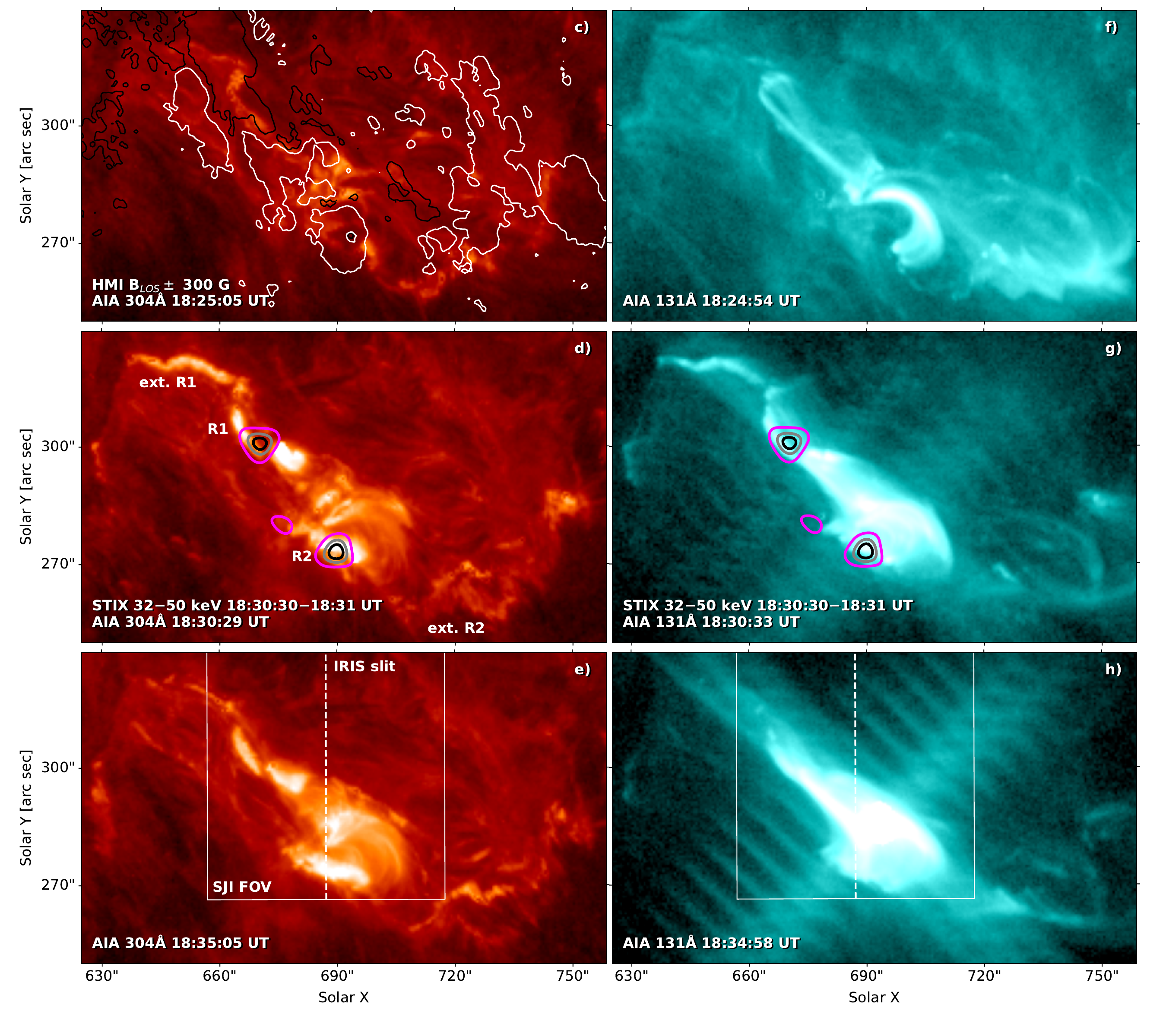}
    \caption{Context observations of the flare. Panels a and b present SXR and HXR lightcurves observed by GOES and STIX, respectively, in different wavelength and energy bands (see legend for details). Panels c -- e show the AIA 304\,\AA~channel observations of the ribbons, while panels f -- h detail hot emission of the flare in the AIA 131\,\AA~(the time progresses from top to bottom). The white and black contours in panel c correspond to HMI $B_{\textrm{LOS}} \pm 300$\,G. The colored contours in panels d, g represent STIX HXR footpoints at 25\% (black), 50\% (grey), and 75\% (magenta) of the maximum intensity in the integration period reprojected to the Earth's view. The white frame in panels e, h indicates the position of the SJI FOV, while the dashed line represents the IRIS slit. \\ Animated version of the 304\,\AA~{(panels c -- e)} and 131\,\AA~{(panels f -- h)} observations is available online. {The animations cover the time period between 18:15 -- 18:45 UT.}}
    \label{fig:overview}
\end{figure}

The magnetic environment of the flaring active region was investigated using observations from the Helioseismic and Magnetic Imager \citep[HMI;][]{Scherrer12} onboard the SDO. Here we used line-of-sight measurements of the photospheric magnetic field strength ($B_\textrm{LOS}$) and Space-weather HMI Active Region Patch (SHARP) magnetograms remapped to a Lambert Cylindrical Equal-Area (CEA) projection. AIA and HMI $B_\textrm{LOS}$ observations were loaded and visualized using \texttt{sunpy} \citep{Sunpy20}.

The HXR emission of this flare was investigated using data from Solar Orbiter/STIX \citep{Krucker20}. STIX detects photons of energies between $4-150$\,keV at an energy resolution of 1\,keV (at 6\,keV) at a cadence of up to 0.5\,s during flares. The dataset under study was acquired at a cadence of 0.5\,s. STIX employs a Fourier-based technique for source reconstruction that leads to a spatial resolution of $\sim7\arcsec$ when the finest subcollimator is used. The STIX data were processed and visualized using the \texttt{stixpy} package. Due to Solar Orbiter's unique orbit, STIX captured the flare from a different vantage point compared to AIA and IRIS. To enable qualitative data analysis, STIX observations were reprojected to Earth's view using \texttt{SunPy}. Reconstructed HXR footpoint sources (Section \ref{sec:qpp_sources}) were then manually co-aligned with the locations of bright ribbon emission as observed in the AIA 1600\,\AA~channel. All STIX data used in this work were corrected for the light travel time from the disk center of about 5.5 minutes. The difference between the light time from the the center disk and flare location was minimal. 

Finally, in this study we also used X-ray Sensor (XRS) on the Geostationary Operational Environmental Satellite (GOES) data with 1\,s cadence to analyze the soft X-ray (SXR) irradiance of the flare.

\section{2022 March 31 flare} \label{sec:overview}

\subsection{Event overview} 

The event under study occurred in the NOAA active region 12975 on 2022 March 31. Classified as an M9.6-class flare, it was one of the first major flares captured in the high-cadence observing programs of IRIS. Context observations of the flare are presented in Figure \ref{fig:overview}. The temporal evolution of the SXR fluxes observed in the $0.5 - 4$\,\AA~(red) and $1 - 8$\,\AA~(blue) GOES/XRS channels are plotted in panel a. The flare began after 18:22 UT. The impulsive phase was characterized by an initially lower gradient of the SXR flux, lasting until about 18:27:30 UT. The flux gradient then steepened before the flare reached its peak between 18:32 and 18:37 UT, after which the flux started to slowly decay. The $4-16$\,keV STIX lightcurve (blue, panel b) resembles those of the GOES SXR fluxes. Note that the STIX attenuator was inserted during the late impulsive and peak phases of the flare. This most notably effects the lower-energy bands of STIX. Signatures of the $20-32$\,keV emission (orange) were first detected at about 18:26 UT and decreased to pre-flare levels roughly 12 minutes later. Strong $32-50$\,keV (green) and $50-100$\,keV (red) signal was detected during the flare impulsive and peak phases, between 18:27 and 18:38 UT. The lightcurves show clear signatures of QPPs, further discussed in Section \ref{sec:stix_qpps}.

Panels c -- e and f -- h present AIA observations in the 304\,\AA~and 131\,\AA~channels, respectively (the time progresses from top to bottom). Two ribbons formed during the flare, one in the north-east (R1) and the other in the south-west (R2). R1 was spatially coincident with the negative-polarity concentrations of the photospheric field, indicated using the black contours in panel c. R2 corresponds to the positive flux shown using the white contours. The brightest ribbon portions were concentrated along the PIL in the central region of the flare, in the middle of the field-of-view (FOV) plotted in Figure \ref{fig:overview}d, e. The ribbons were highly sheared and asymmetric, with the brightest portions of R1 being at least twice as long as those of R2 during most of the flare. The brightest flare kernels were coincident with a pair of HXR footpoint sources (see also Section \ref{sec:qpp_sources}) plotted using the colored contours in panel d. The colors correspond to 25\% (black), 50\% (grey), and 75\% of the maximum intensity integrated between 18:30:30 -- 18:31 UT in the $32 - 50$\,keV energy bands (see Section \ref{sec:qpps} for details). Both ribbons developed faint extensions, labeled `ext. R1' and `ext. R2' in panel d, visible immediately after the flare onset (panel c). Ext. R1 exhibited a hooked $J$-shaped extension to the north-east, indicative of the extent of the flux rope footpoints rooted therein. The short R2 exhibited a separation motion directed towards the south. The separation speed of about $v_\perp \approx 20$\,\kps, estimated during the flare impulsive phase (18:28 -- 18:31 UT), decreased to $v_\perp \approx 4$\,\kps~during the peak and gradual phases (18:31 -- 18:40 UT). The FOV of IRIS, indicated in panel e, captured the central flare region exhibiting the brightest ribbon emission (panel e) and the spectrograph slit crossed the ribbon R2. 

Observations of the hot emission from AIA 131\,\AA~reveal a compact bundle of flare loops (Figure \ref{fig:overview}f) between R1 and R2 visible immediately after flare onset. Most of the flare loop arcade developed during the impulsive phase (panels g, h). Interestingly, while some faint flare loop emission stretched to ext. R1 in the north-east, a majority of their footpoints were spatially coincident with the HXR footpoints (panel g). This suggests more efficient chromospheric evaporation filling the hot loops from the ribbon regions corresponding to the HXR footpoints (Section \ref{sec:disc_deposition}). Since the \ion{Fe}{21} 1354.1\,\AA~IRIS line was not observed in this flare, we did not analyze it further. The flare was followed by a faint coronal mass ejection (CME) visible after $\approx$19:00 UT in coronagraphic observations from the The Large Angle Spectroscopic Coronagraph \citep[LASCO;][]{Brueckner95} onboard the SOHO mission.

\subsection{Flare kernel dynamics} \label{sec:slipreco}

\begin{figure}[h]
    \centering
    \includegraphics[width=18cm, clip, viewport = 00 10 1290 790]{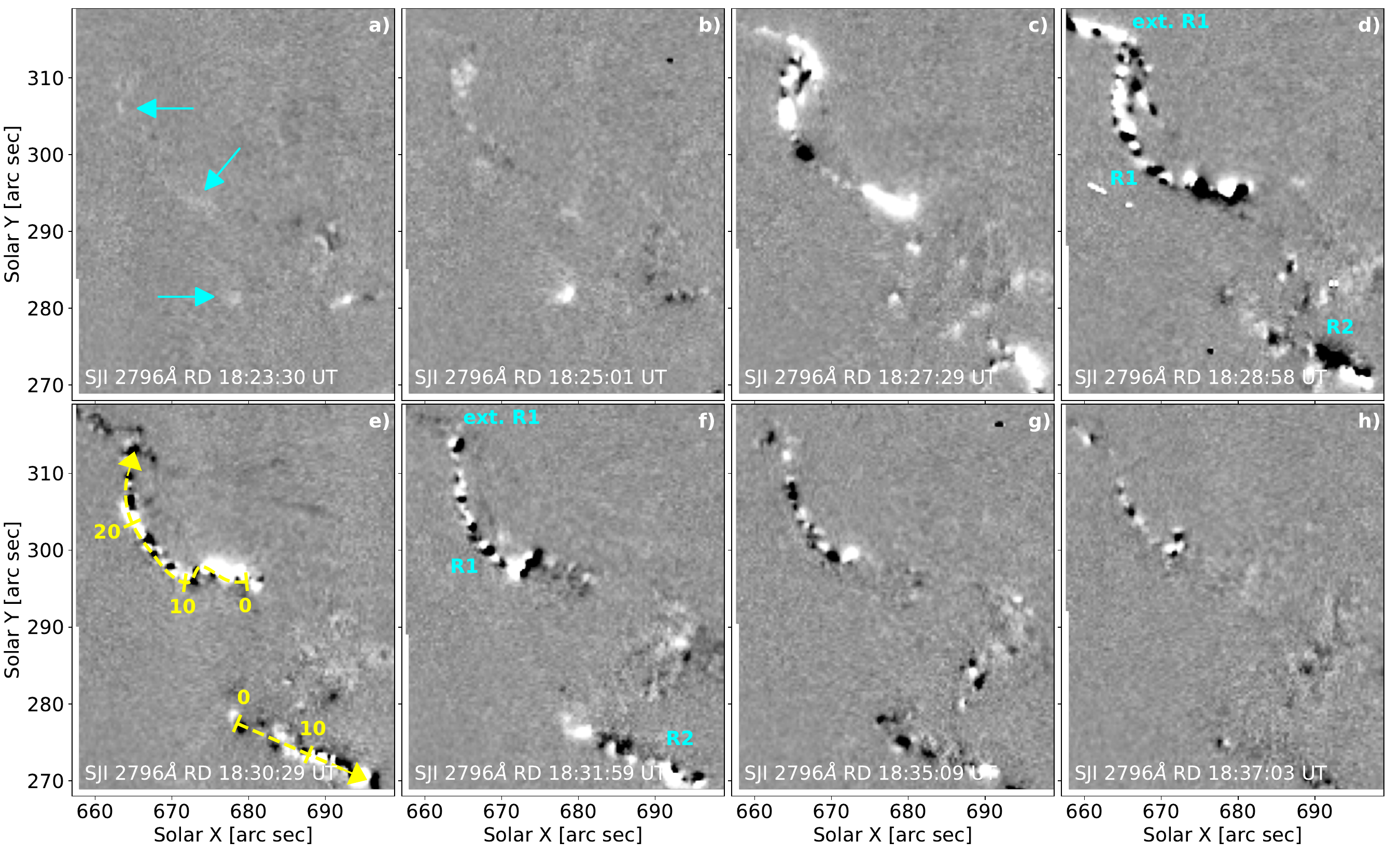}
    \caption{IRIS/SJI 2796\,\AA~running difference (RD) observations of the ribbons during the flare impulsive (panels a -- f) and peak (panels g \& h) phases. The images are saturated to $\pm 100$ DN\,s$^{-1}$, and the white color corresponds to intensity increase. The cyan arrows in panel a highlight weak, incipient flare ribbon emission at the flare onset. The yellow dashed arrows in panel e indicate artificial cuts used for producing the time-distance diagrams plotted in Figure \ref{fig:sji_rd_stackplots}. \\Animated version of the {SJI 2796\,\AA~RD} observations is available online. {For reference, the animation also shows level-2 SJI 2796\,\AA~channel observations used to produce the RD time series. The animations cover the time period between 18:15 -- 18:45 UT.}}
    \label{fig:sji_overview}
\end{figure}

The flare ribbons were composed of numerous flare kernels primarily exhibiting formation and dynamics during the flare's impulsive and peak phases. To highlight the dynamic substructure of the ribbons and suppress the slowly-varying background, in Figure \ref{fig:sji_overview} we plot running-difference (RD) 2796\,\AA~SJI observations with the time difference of 1 frame (8\,s, see also the animated version of the figure). The earliest signatures of ribbon emission were detected at $\approx$ 18:23:30 UT, soon after the flare onset, in the form of faint kernels in R1 and R2 highlighted using the cyan arrows plotted in panel a. The kernel intensities were slightly increasing in the course of the following two minutes (panel b), while R1 and R2 started to elongate after 18:27:30 UT, at which their intensity rapidly increased (panels c \& d). As R1 elongated, a bright portion of its extension (ext. R1) briefly reached beyond the SJI FOV in the north-east (c.f. Figure \ref{fig:sji_overview}d, Figure \ref{fig:overview}d). The brightest parts of R2, on the other hand, did not elongate past the extent of the SJI FOV (c.f. Figure \ref{fig:sji_overview}e, Figure \ref{fig:overview}d). 

The adjacent black and white blobs, clearly seen in Figure \ref{fig:sji_overview}d -- g, are representative of intensity increase and decrease between the neighboring locations and time frames. This pattern is indicative of apparent kernel slipping motions along the ribbons and R1 in particular. To study the kernel dynamics, we employ time-distance diagrams \citep[stackplots, e.g.,][]{Dudik14, LiT15, Lorincik19a} produced using artificial cuts plotted as the yellow dashed arrows along R1 and R2 (panel e). The cuts were chosen to match the brightest portions of the ribbons for as long as possible during the flare impulsive phase. Due to the difference in ribbon lengths discussed above, the curved cut along R1 is longer than that for R2, which is straight. The RD intensities were averaged across 2\arcsec~half-width of the cuts. This average was used in order to account for the slow ribbon separation and small deviations between the ribbon fine structure from the central cut location{, at the expense of a slight redution in spatial resolution and a decreased sensitivity to the finest resolvable ribbon structures.}

\begin{figure}[h]
    \centering
    \includegraphics[width=9.3cm, clip, viewport = 05 00 298 150]{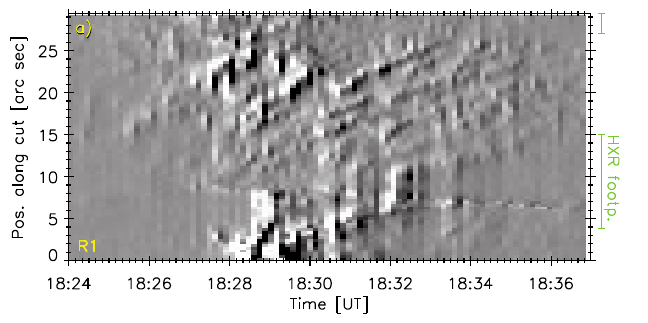}
    \includegraphics[width=8.475cm, clip, viewport = 18 00 285 150]{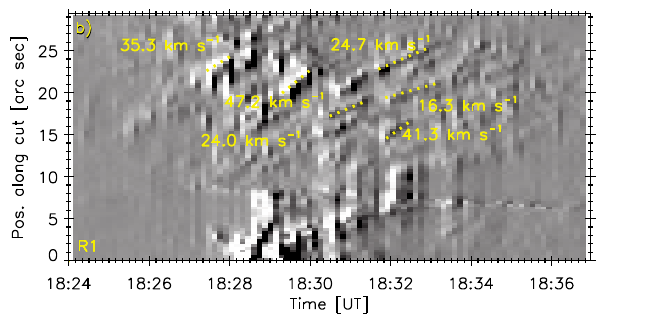}
    \\
    \includegraphics[width=9.3cm, clip, viewport = 05 00 298 150]{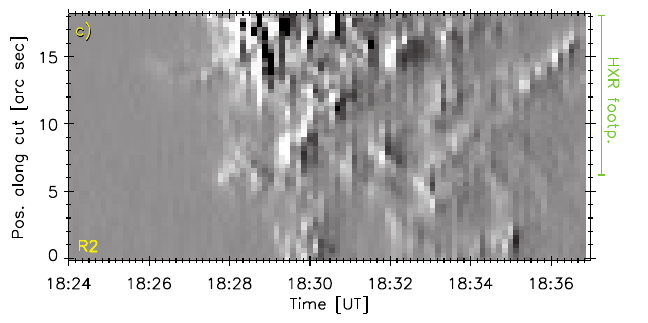}
    \includegraphics[width=8.475cm, clip, viewport = 18 00 285 150]{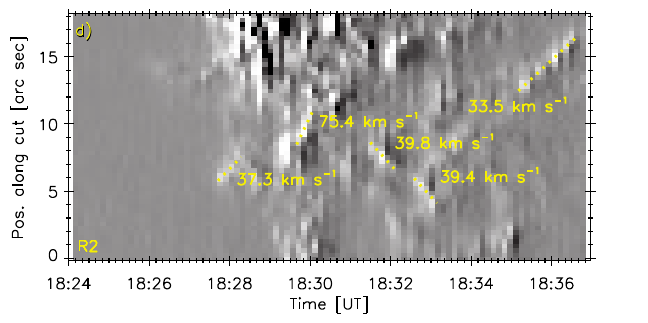}
    \caption{Time-distance diagrams (stackplots) used to study kernel motions along R1 (panels a \& b) and R2 (panels c \& d). Yellow dotted lines represent linear fits to kernel-imprinted traces in the stackplots, along with corresponding velocities. The green bars between the stackplots roughly indicate the extent of the 50\% STIX $32 - 50$ keV contours in these ribbons. The HXR footpoint in R1 overlaps only slightly with the kernel traces therein.}
    \label{fig:sji_rd_stackplots}
\end{figure}

The resulting stackplots for R1 and R2 are plotted in Figure \ref{fig:sji_rd_stackplots}a \& c. The inclined stripes visible therein represent the slipping kernels. The motions of the kernels intensify after about 18:27 UT, at the same time as the rapid rise of the GOES SXR flux (Section \ref{sec:overview}) and STIX QPPs (Section \ref{sec:stix_qpps}). The stackplots indicate a higher number of kernels along R1 (panel a) compared to R2 (panel c). Within R1, the kernels were numerous the most between locations $\approx 12 - 27\arcsec$ of the cut, corresponding to the curved part of R1 (c.f., Figure \ref{fig:sji_rd_stackplots}a and Figure \ref{fig:sji_overview}). Most of the slipping motions were directed towards the hooked ext. R1 in the north-east (Section \ref{sec:overview}). We estimate the kernel slipping velocities $v_\textrm{slip}$ parallel to the PIL by using linear fits to selected kernels traces. The fits are plotted using the yellow dotted lines in panels b \& d for the stackplots plotted in panels a \& c, respectively. The fits are indicative of relatively-low $v_\textrm{slip}$ between $16 - 75$\,\kps~in both ribbons. These measurements are similar to those reported in previous analyses of eruptive flares \citep[e.g.][]{Dudik14, LiT14, Dudik16}, but are lower (by up to two orders of magnitude) than the $v_\textrm{slip}$ values recently measured in confined flares \citep[][]{Lorincik25a, ZhangY25}, interpreted as evidence of slip-running reconnection. Prominent kernels traces disappear after $\approx$18:37 UT. Kernel formation and motions were no longer discernible after the flare peak, as seen in Figure \ref{fig:sji_overview}h and its animated version.

{It ought to be noted that, while still high, the 8\,s cadence of the SJI observations employed in this dataset might have hindered the identification of faster kernels \citep{Lorincik25a}, if any. The contrast between ribbon emission and surrounding structures as observed in the SJI 2796\,\AA~channel can also be lower compared to the 1330\,\AA~and 1400\,\AA~channels \citep[][]{Lorincik25a}, further decreasing the probability of detection of faster kernel motions or even finer ribbon substructure. Nevertheless, this limitation does not significantly affect our analysis, as we do not focus on fast kernel dynamics consistent with slip-running reconnection.}

\section{Quasi-periodic pulsations} \label{sec:qpps}

\subsection{STIX and GOES timeseries} \label{sec:stix_qpps}

\begin{figure}[h]
    \centering
    \includegraphics[width=18.00cm]{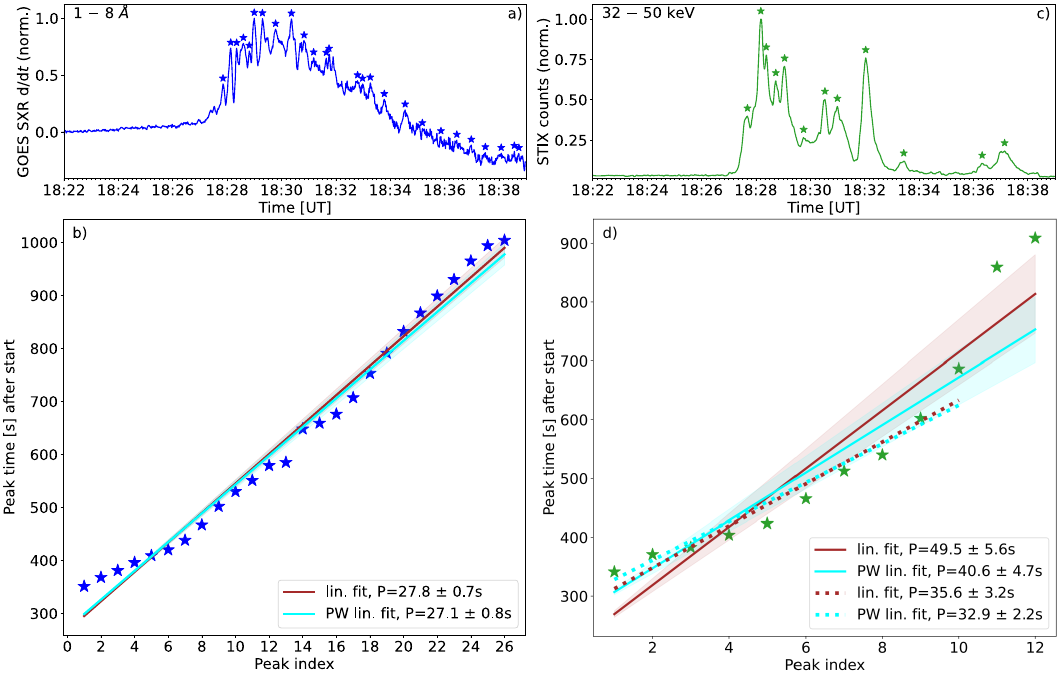}
    \caption{GOES $1 - 8$\,\AA~flux time derivative (panel a) and STIX $32 - 50$\,keV (panel c) lightcurves. The stars mark automatically-identified QPPs (see Section \ref{sec:stix_qpps} for details). Panels b and d present peak index vs. peak time scatterplots used to infer the QPP periods in signals plotted above.}
    \label{fig:goes_stix_qpps}
\end{figure}

Figure \ref{fig:goes_stix_qpps} presents GOES and STIX lightcurves exhibiting signatures of QPPs. Panel a shows the time derivative of the GOES $1-8$\,\AA~SXR flux in the time period starting from before the flare onset until the early gradual phase. The lightcurve was smoothed using the Savitzky-Golay filter with the window duration of 10\,s and a 3rd order polynomial in order to suppress the noise, while retaining both short- and long-period variations in time. The lightcurve shows QPPs starting at about 18:27 UT and lasting throughout the entire period plotted therein. The stars indicate peaks identified automatically using the \texttt{find\_peaks} routine in \texttt{SciPy} \citep[][]{Scipy20}. The peak selection satisfies two criteria: 1) their prominence must be $> 3 \sigma$ estimated in the pre-flare period (17:43 -- 18:20 UT), and 2) the peak duration must be at least three time bins.

Inspired by \citet[][]{Clarke21}, here we estimate the QPP period using a simple scatterplot showing the peak index on the horizontal axis and peak time on the vertical axis (panel b). The QPP period $P$ is then inferred from the slope of the linear fit to the datapoints. This approach provides dominant timescales of the pulsations, rather than a statistically robust periodicity determination. Simple linear fit (brown line) is indicative of $P = 27.8 \pm 0.7$\,s. A second linear fit plotted in cyan, weighted by the peak prominence (hereafter peak-weighted, `PW'), is indicative of $P = 27.1 \pm 0.8$\,s. Most of the datapoints are well represented by the linear trends, except for the peaks indexed $1 - 4$ and a few outliers among the peaks indexed $12 - 17$. Taking both measurements and their uncertainties into account, the resulting period of the GOES $1-8$\,\AA~flux time derivative QPPs is $26.3 < P_{\textrm{GOES}} < 28.5$\,s, aligning with previous measurements \citep[e.g.,][]{Hayes20}. Note however that the non-stationary nature of the QPPs underlines the fact that these period estimates should be regarded as an average timescale of recurrence, not a strictly constant oscillation period.

Figure \ref{fig:goes_stix_qpps}c shows the STIX time series in the $32 - 50$\,keV energy bands, summed across all detector pixels and smoothed using the same technique as the GOES lightcurve in panel a. The lower energy limit of 32\,keV was chosen to exclude contributions from the flare's thermal emission, which we verified by HXR spectral fitting (not shown). STIX time series at energies $> 50$\,keV exhibited the same pulsations but were noisier, which motivated the choice of the higher-energy limit. The STIX QPPs in these well-representative time series were again identified automatically, but we increased the peak prominence criterion to only flag the strongest QPPs during which the HXR source reconstruction was performed (see Section \ref{sec:qpp_sources}). The slope of the linear fit to the peak times, plotted using the brown line in panel d, is indicative of $P = 49.5 \pm 5.6$\,s. The peak-weighted linear fit (cyan) corresponds to a lower period $P = 40.6 \pm 4.7$\,s. The disparity between the measurements stems from the fact that the strongest QPPs were visible between 18:28 -- 18:32 UT (panel c) during the impulsive and early peak flare phases at which the QPPs were more frequent. Combining both period estimates results in $35.9 < P_{\textrm{STIX}} < 55.1$\,s for the QPPs in the $32 - 50$\,keV bands in the entire time period of interest. The increased spacing between the datapoints corresponding to the peaks indexed $5 - 11$ suggests an overall increase in the QPP period over time. This is either due to decay in QPP recurrence, or reduced detectability. In particular, the last two pulsations in the time series occurred after a period of $\approx 2$ minutes between 18:34 and 18:36 UT when no QPPs were detected. Limiting the linear fits to the peaks observed between 18:27 and 18:34 UT (dotted fits) leads to the decrease of the QPP period to $30.7 < P_{\textrm{STIX}} < 38.8$\,s, closer to the periods of the GOES (see above) and IRIS QPPs (Section \ref{sec:sji_qpps}). This result means that the frequency of the strongest non-thermal QPPs was the highest during the flare impulsive phase. 

\subsection{HXR footpoint sources} \label{sec:qpp_sources}

Next, we performed the HXR source reconstruction during the time STIX detected QPPs in the $32-50$\,keV bands, as discussed in the previous section. For this task, we employed the MEM\_GE algorithm \citep{Massa20} implemented in \texttt{stixpy}. We considered 10\,s short integration windows during the QPPs indicated using the star symbols in Figure \ref{fig:goes_stix_qpps}c. For the purposes of the source reconstruction, we chose appropriate background observations taken the day after the flare, during a period with no signs of increased activity. 

\begin{figure}[h]
    \centering
    \includegraphics[width=9.00cm]{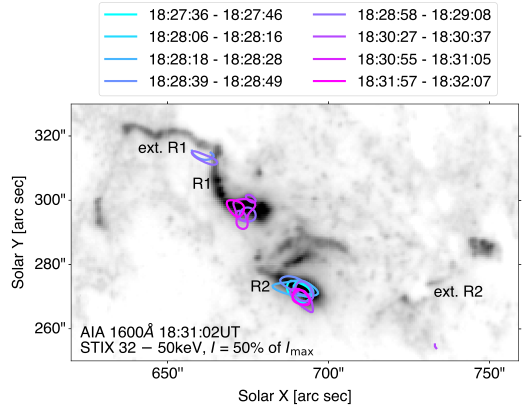}
    \caption{STIX footpoint sources during the strongest QPPs in the $32 - 50$\,keV energy bands reprojected to the Earth’s view. The colored contours correspond to 50\% of the maximal intensity in each STIX map (see legend for integration times).}
    \label{fig:qpp_sources}
\end{figure}

Figure \ref{fig:qpp_sources} presents the STIX $32 - 50$\,keV sources as contours, plotted on top of AIA 1600\,\AA~snapshots which detail the structure of the ribbons during the impulsive phase. The contours correspond to 50\% of the maximal intensity ($I_\textrm{max}$) in each frame and were colored to distinguish between the integration times (see legend for details). Since the emission above 32\,keV was non-thermal, the contours represent HXR footpoints as sites of non-thermal energy deposition from the reconnection site (Section \ref{sec:qsls}). HXR footpoint sources were reconstructed for 8 out of 12 QPPs under consideration, yielding $7\times10^3 - 2.3\times10^4$ counts in each STIX map where the source reconstruction was successful. The reconstruction was not reliable in four integration windows, corresponding to the QPP at $\approx$18:29:45 UT as well as the last three QPPs in the STIX time series (Figure \ref{fig:goes_stix_qpps}c), due to low counting statistics. Because of the limited dynamic range of the indirect imaging (see Section \ref{sec:disc_deposition}) and the short 10\,s integration windows, image reconstruction artifacts appear in the images at intensity fractions below 50\% of the $I_\textrm{max}$. To inspect the appearance of the signal sources at relatively-lower fractions of the $I_\textrm{max}$, we also increased the integration time to 20\,s. This increased the signal-to-noise ratio in the reconstructed images, yielding reliable reconstruction down to 30\% of the $I_\textrm{max}$ (not shown). Since the contours appeared qualitatively similar to those corresponding to 50\% of the $I_\textrm{max}$ assuming 10\,s integration windows, we opted to focus on the original reconstruction. 

The reconstructed sources are compact, spatially coincident with bright sections of R1 in the north and R2 in the south as well as flare loop footpoints (Section \ref{sec:overview}). Except for a slow motion of the source in R2, corresponding to the ribbon's separation (Section \ref{sec:overview}), the sources did not exhibit any organized or continuous motions. Interestingly, we found two secondary footpoint sources in the north-east (green contours in Figure \ref{fig:qpp_sources}) during two integration windows starting at 18:28:39 UT and 18:28:58 UT. This is roughly 30\,s after R1 started to elongate and form ext. R1 just to the north-east of the contours, visible in Figure \ref{fig:overview}d and the animated version of Figure \ref{fig:sji_overview}. These secondary sources disappear when plotting contours corresponding to $> 50$\% of the $I_\textrm{max}$.

We also studied how the locations of the HXR footpoints relate to the signatures of kernel slipping motions (Section \ref{sec:slipreco}). The green bars in Figure \ref{fig:sji_rd_stackplots} indicate the relative position of the HXR footpoints, as indicated by the 50\% STIX contours, to the cuts used for producing the stackplots. The HXR footpoints' locations overlap with the sites of the strongest UV intensity change: roughly at $4 - 8\arcsec$ of the cut in R1 and $\gtrsim 12\arcsec$ in R2. Because the inclination of the kernel traces (fitted lines) in these regions was low, the RD variation must have stemmed from slow (or even stationary) but pulsating kernels. Because the footpoint in R2 spans most of the ribbon extent as observed by SJI, it also overlaps with traces of the few slipping kernels therein (Figure \ref{fig:sji_rd_stackplots}c, d). 

\begin{figure}[h]
    \centering
    \includegraphics[width=0.99\linewidth, clip, viewport = 00 70 620 730]{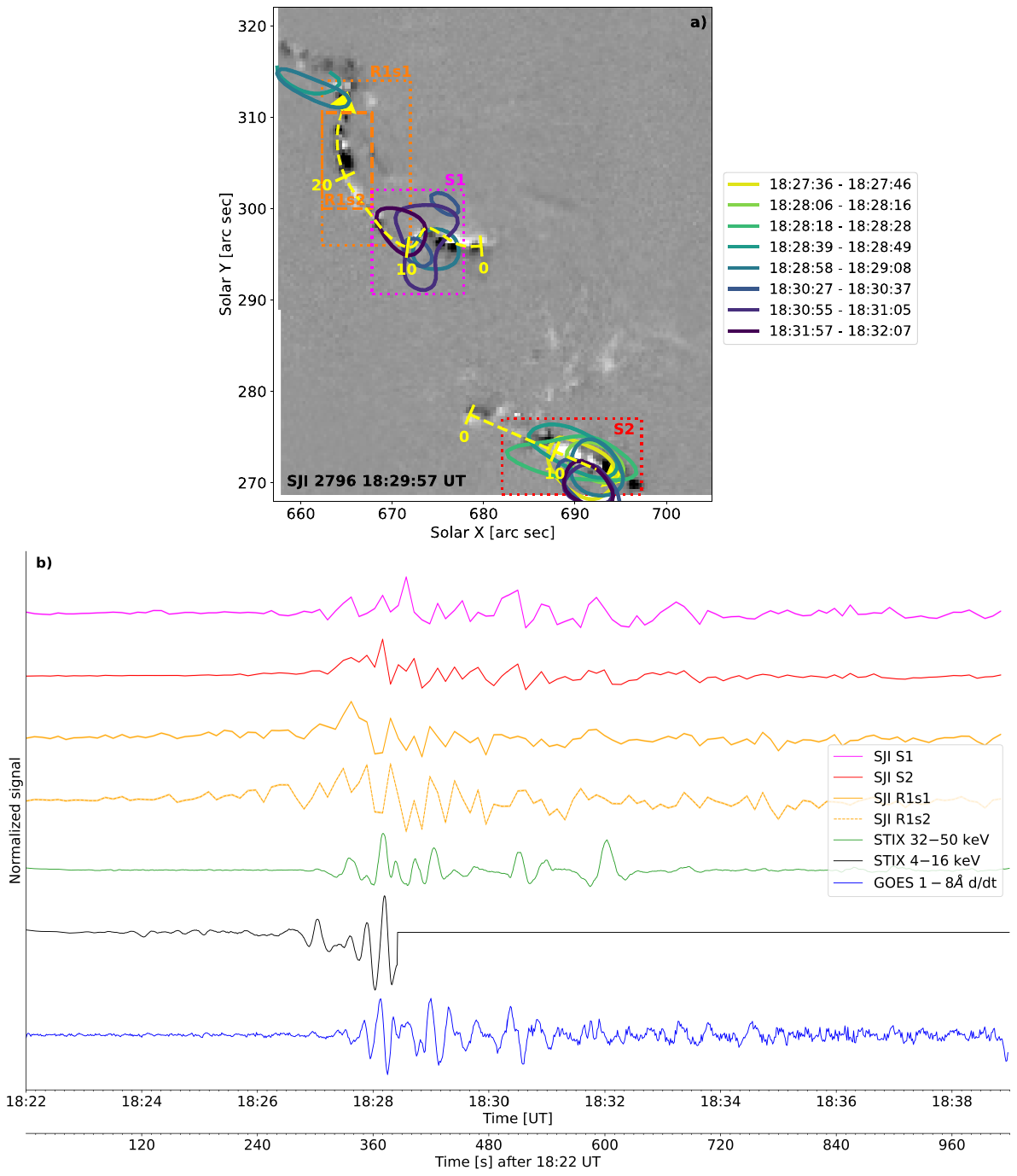}
    \caption{Panel a presents SJI 2796\,\AA~RD observations of the flare overplotted with 32 -- 50\,keV footpoint sources reconstructed during the QPPs (same as Figure \ref{fig:qpp_sources}). The dashed yellow arrows are artificial cuts used for producing the time-distance diagrams. The colored boxes mark different regions where IRIS/SJI intensity variations were studied. Panel b presents normalized lightcurves in different instruments and filter channels during the flare impulsive phase (see legend for details). The IRIS/SJI lightcurves were obtained in the boxes of same colors plotted in panel a.}
    \label{fig:sji_rd_lightcurves}
\end{figure}

A more peculiar result concerns the relative position of the kernels and the source extent in R1. Figure \ref{fig:sji_rd_stackplots}a, b shows a majority of kernel traces between the primary and secondary HXR footpoints (the latter observed only around 18:29 UT). Several faint kernels emerged at locations co-spatial with the stationary primary source, but they slipped away towards the north-east, where HXR footpoints were not identified. In those portions of R1, we did not find signatures of significant ($I >$ 50\% of $I_\textrm{max}$, see Figure \ref{fig:sji_rd_lightcurves}a) or even weaker ($I >$ 30\% of $I_\textrm{max}$) HXR emission, in the latter case using the longer 20\,s integration (Section \ref{sec:qpp_sources}). This absence of the reconstructed sources might reflect that the HXR emission was weaker or absent. While we cannot exclude the possibility that some HXR signal was still present, albeit below the dynamic range of STIX, the result that the non-thermal electron energy deposition into the primary footpoint dominated the deposition elsewhere along R1 still holds. This lack of strong non-thermal electron energy deposition into slipping kernels is another important result of our study. Further discussion regarding its significance as well the as limitations of indirect imaging is left for Section \ref{sec:disc_deposition}. 

\subsection{Pulsations of flare ribbon intensities} \label{sec:sji_qpps}

Having located the sources of strong HXR signals, we now proceed to study the intensity evolution in regions of interest along flare ribbons, as observed in the SJI 2796\,\AA~channel. Although the IRIS FOV did not contain the entire flare, its 8\,s cadence is higher than that of the AIA 304\,\AA~(12\,s) and 1600\,\AA~(24\,s) channel data. AIA 304\,\AA~observations are also contaminated by the emission of the overlying flare loop arcade (Section \ref{sec:overview}), which complicates the analysis of ribbon intensities. In analogy with Section \ref{sec:slipreco}, we use the RD data with the time difference of 1 snapshot in order to suppress contributions from the background emission while highlighting the intensity change along the ribbons. {Unlike the STIX \& GOES lightcurves, the RD SJI lightcurves were neither further smoothed nor detrended.}

Figure \ref{fig:sji_rd_lightcurves}a presents a SJI 2796\,\AA~RD snapshot of the ribbons overplotted with the HXR footpoints (Section \ref{sec:qpp_sources}). The yellow arrows represent the artificial cuts used to produce the stackplots for studying the kernel dynamics (Section \ref{sec:slipreco}). The RD intensity change was summed in four subregions designated using the colored boxes S1, S2, R1s1, and R1s2 in the same figure. S1 (purple) and S2 (red) were chosen to study the intensity variation within the extent of the stationary HXR footpoints in R1 and R2, respectively. The orange boxes R1s1 (dotted) and R1s2 (dashed) were used to study the intensity variation due to the kernel dynamics in R1, where the slippage was the most prominent (Section \ref{sec:slipreco}). R1s2 box represents a subregion within R1s1, which, however, does not overlap with the HXR footpoints (Figure \ref{fig:sji_rd_lightcurves}a). Lightcurves summed across the entire ribbons R1 and R2, as observed by IRIS/SJI, are discussed in Appendix \ref{sec:app_R1R2}.

Normalized lightcurves obtained in these four boxes are plotted using the same colors in Figure \ref{fig:sji_rd_lightcurves}b. The lightcurves are composed of numerous non-stationary pulsations, which are either positive, representing an overall intensity increase, or negative due to an intensity decrease. Because the ribbon substructure was dynamic, the RD time evolution is a net effect of 1) the intensity change due to slipping kernels, 2) the intensity change in stationary kernels, and, to a small extent, 3) the slow ribbon separation in the case of R2. The SJI lightcurves are plotted with respective error bars. Since the propagated photon noise was negligible for these SJI observations, the error bars correspond to $\pm \sigma$, where $\sigma$ is the standard deviation calculated in the pre-flare phase between the start of the IRIS dataset ($\approx$17:43 UT) and 18:05 UT. These noise estimates are further used in Section \ref{sec:heatmaps}. {Note that because no further processing was applied to the RD time series, these lightcurves are equivalent to the forward time differentiation of SJI lightcurves obtained by summing level-2 data}.

The purple (S1) and red (S2) lightcurves exhibited strong variation roughly between 18:27 and 18:34 UT during the flare impulsive phase. The intensity evolution is very similar, as both lightcurves exhibited similar simultaneous or near-simultaneous pulsations. This similarity is indicative of a coherent UV emission response in the stationary HXR footpoints. The intensity oscillation in the orange R1s1 (solid) and R1s2 (dashed) lightcurves started soon after the flare onset and continued beyond the flare peak. The amplitude of the variation before $\approx$18:27 UT is relatively low, but we verified that even the lowest SJI peaks can still be associated with weak kernel emission (see Figure \ref{fig:sji_overview}a and its animated version). 

The analysis of SJI QPP periods using the same method as employed in Section \ref{sec:stix_qpps} is presented in Appendix Figure \ref{fig:app_SJI_P} and discussed in Appendix \ref{sec:app_SJI_P}. We found that peaks stronger than $3\sigma$ appeared at timescales ranging roughly {between $25 < P_\textrm{SJI} < 34$\,s, depending} on the box and the fit used to infer the quasi-period. The longest period oscillation {of $31 < P_\textrm{SJI} < 34$\,s was} found in the box S1. Pulsations in {S2 and R1s2 exhibited $25 < P_\textrm{SJI} < 26$\,s. The latter implies} that the peaks induced by the slipping kernels away from the HXR footpoints occurred at relatively-higher frequency. %

\subsection{Correspondence between the SJI and STIX lightcurves}  

\subsubsection{General considerations} \label{sec:correspondence}

In Figure \ref{fig:sji_rd_lightcurves}b, we also plot the time evolution in the $32 - 50$\,keV (green) and $4-16$\,keV (black) energy bands of STIX as well as the time derivative of the $1-8$\,\AA~SXR flux detected by GOES (blue). The black lightcurve was padded roughly after 18:28 UT, when the lower-energy STIX bands started to be attenuated (Figure \ref{fig:overview}b). For plotting purposes, the smoothed (Section \ref{sec:stix_qpps}) GOES and STIX lightcurves were additionally detrended using a moving average window with a duration of 30\,s. By visual inspection of the lightcurves we verified that the detrending did not produce any artificial peaks (or troughs) along the lightcurves. We did not estimate the QPP period in the $4-16$\,keV STIX lightcurve as it was attenuated during most of the flare impulsive phase. Nevertheless, it is worth noting that the black lightcurve shows a discernible modulation after $\approx$18:24 UT, roughly three minutes before the QPPs were detected in the higher-energy $32-50$\,keV bands (green). While the amplitude of the initial modulation is much lower compared to the pulsations visible after $\approx$18:26:30 UT, it is still statistically significant, as we found $\sigma$ of these data to be negligible. 

Comparison of the UV SJI and HXR STIX lightcurves in Figure \ref{fig:sji_rd_lightcurves}b merits several remarks. The early $4-16$\,keV modulation was co-temporal with the signal variation along the R1s1, R1s2 lightcurves (orange) in the same period of time, suggestive of a physical link between the time series. However, we verified that the majority of the $4-16$\,keV emission at those times originated in the southern ribbon R2 and not R1 in the north, where these kernel-driven UV signals were measured. The oscillatory patterns along R1s1, R1s2 were also observed beyond the strongest $32-50$\,keV QPPs, co-temporal with discernible, albeit noisy variation captured by GOES (c.f., orange and blue lightcurves). These observables evidence that the UV intensity signature of kernel formation and slippage persisted for a longer period of time than the $32-50$\,keV HXR QPPs, as well as the UV pulsations captured in the HXR footpoints. This result is further discussed in Section \ref{sec:disc_burstyrec}. Finally, some of the higher-energy STIX QPPs seem to be synchronous (within the 8\,s SJI cadence) with the pulsations in the SJI lightcurves. This likeness is particularly the case for the S1 and S2 lightcurves that are, between 18:27 -- 18:34 UT, very similar to the $32-50$\,keV time evolution plotted in green. 

\begin{figure}[t]
    \centering
    \includegraphics[width=9.20cm, clip, viewport = 00 45 720 325]{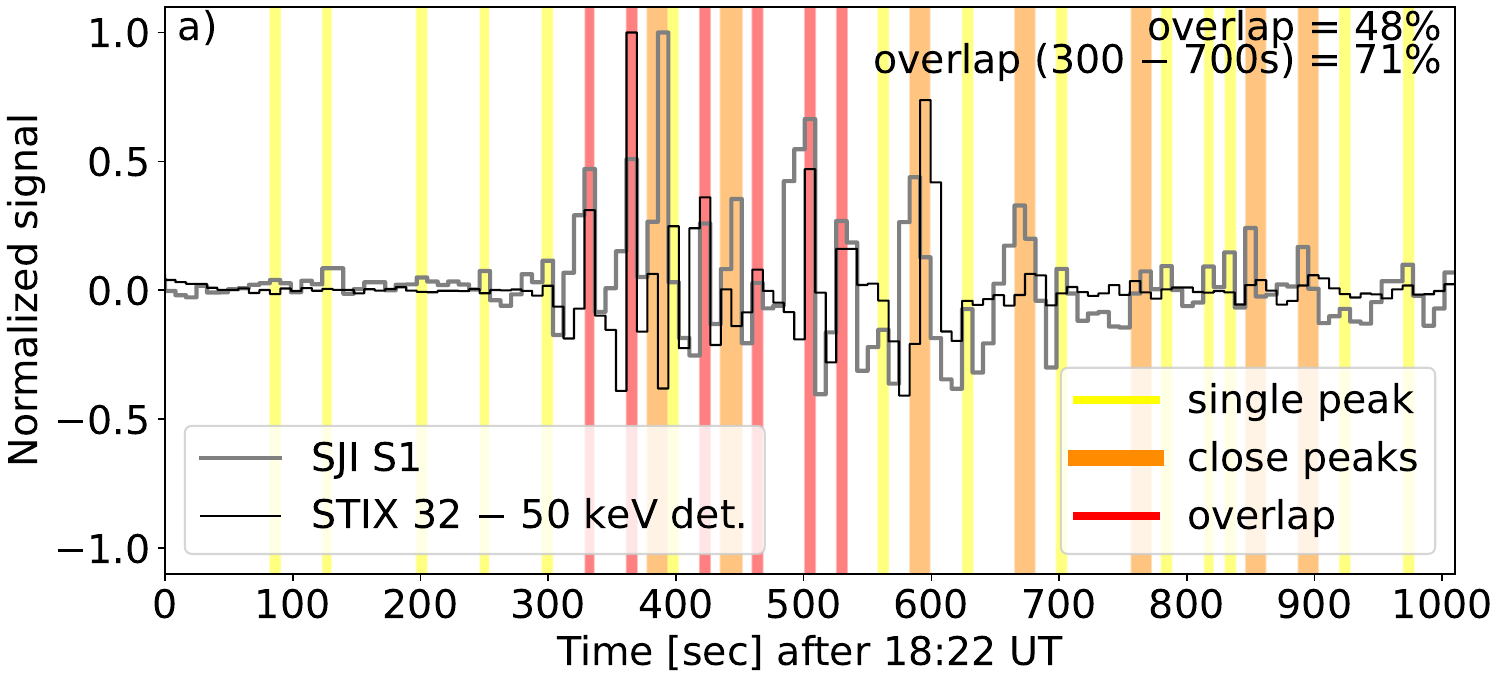}
    \includegraphics[width=8.254cm, clip, viewport = 74 45 720 325]{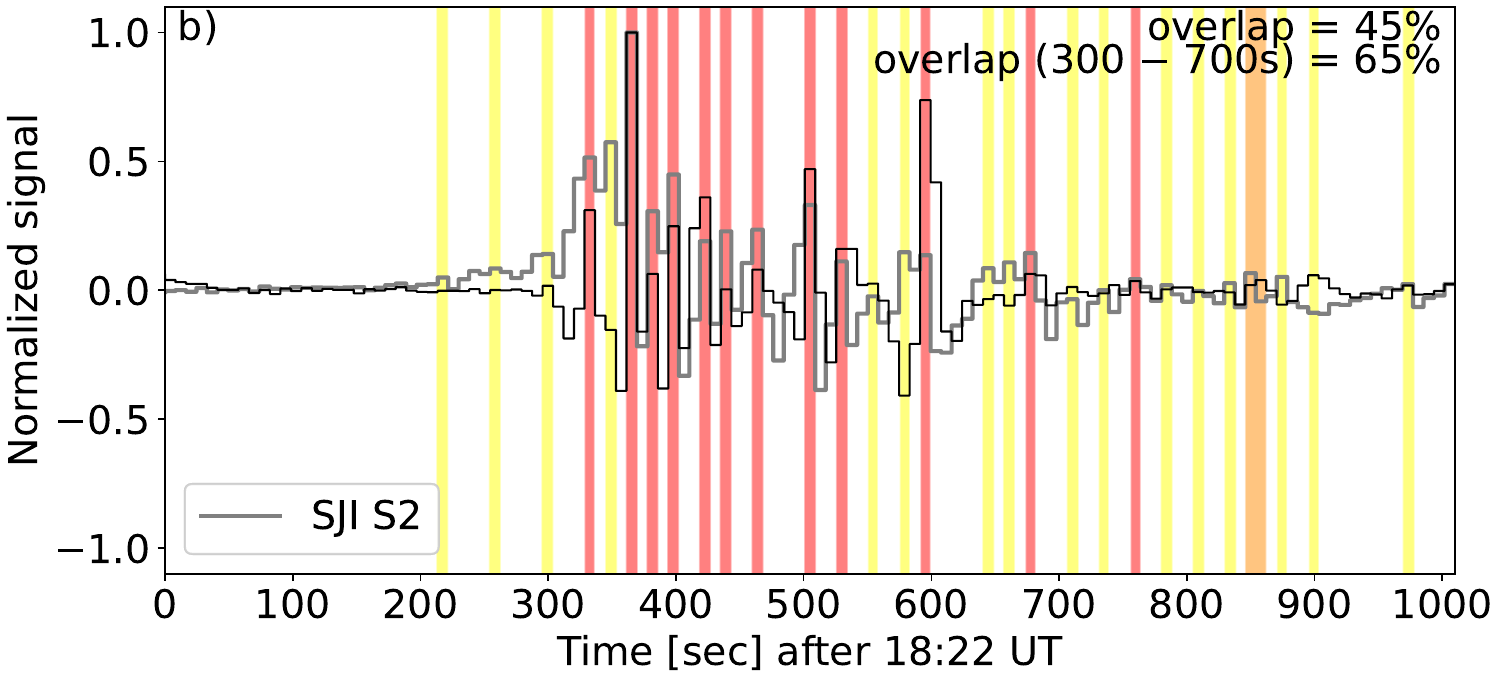}
    \\
    \includegraphics[width=9.20cm, clip, viewport = 00 02 720 325]{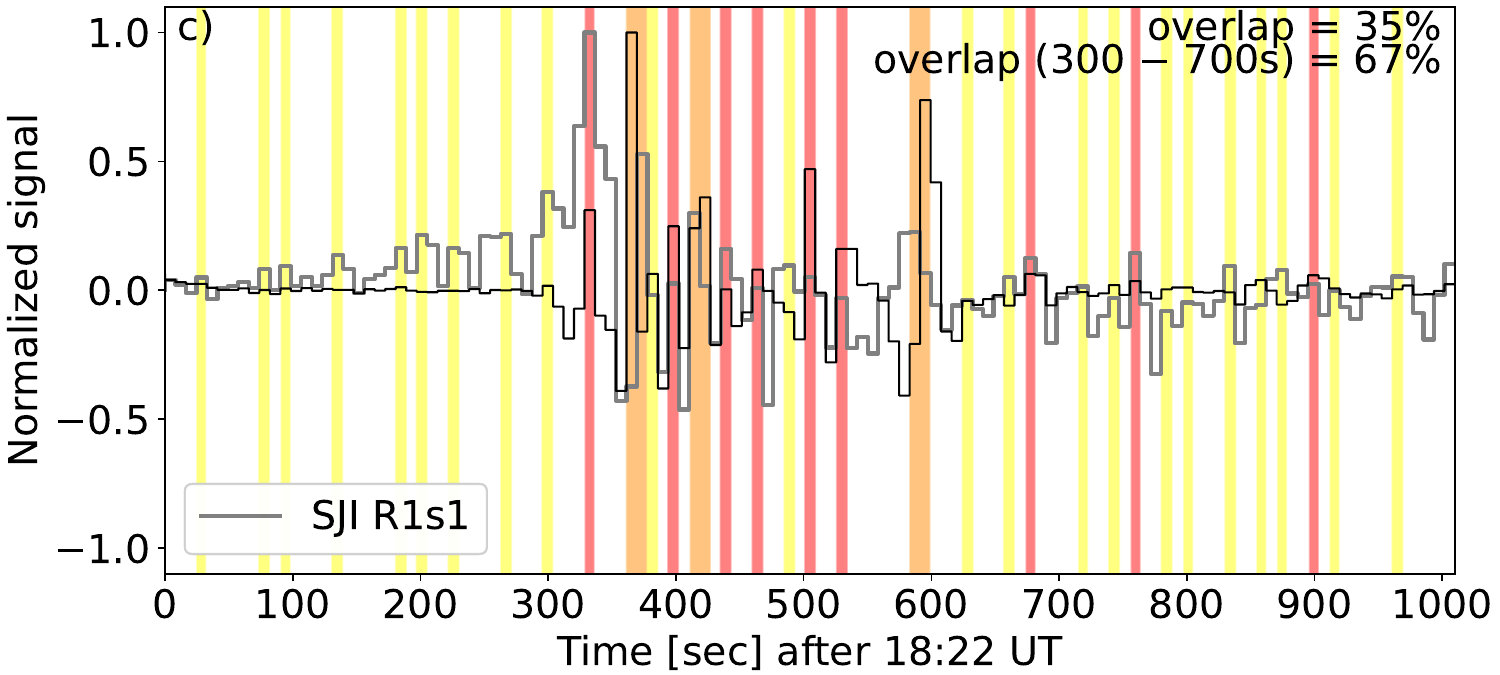}
    \includegraphics[width=8.254cm, clip, viewport = 74 02 720 325]{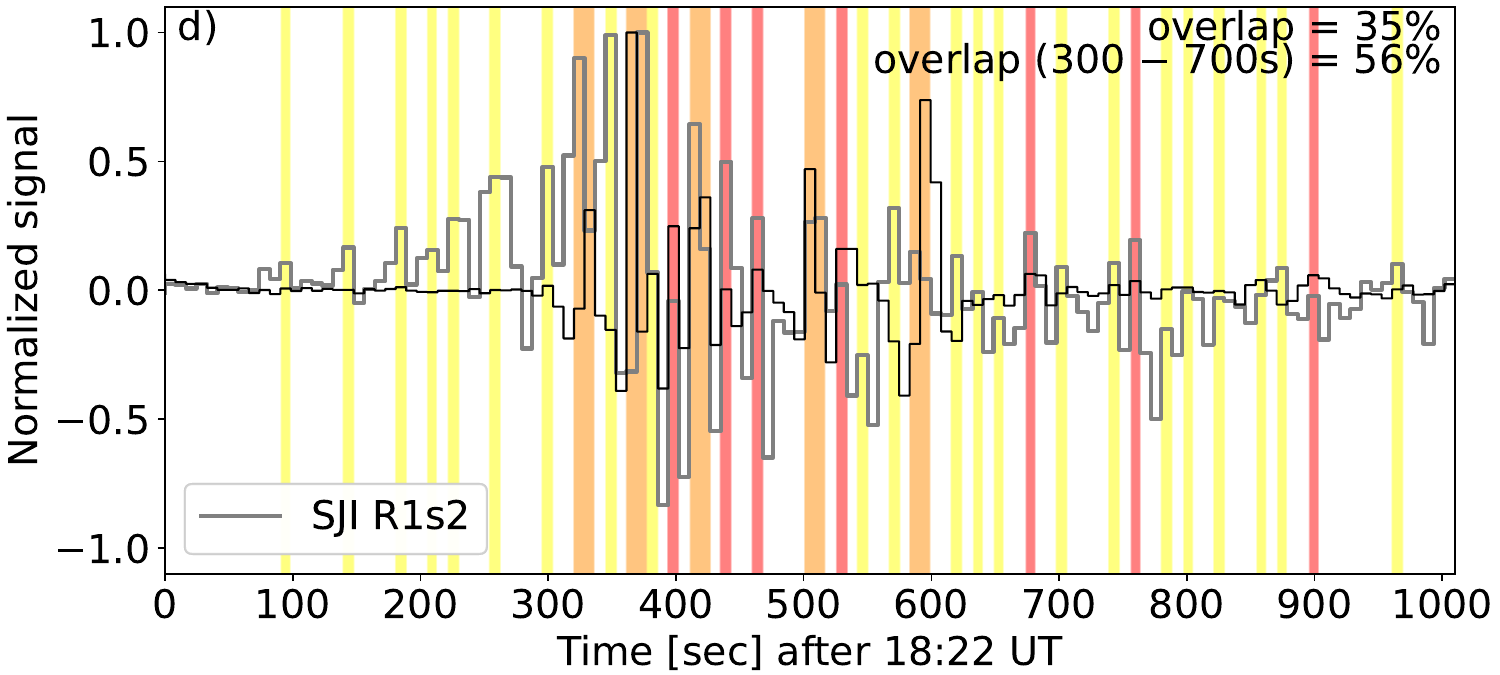}
    \caption{Peak heatmaps for the detrended STIX $32 - 50$\,keV lightcurve and RD intensity in different IRIS/SJI boxes, see Section \ref{sec:heatmaps} for the method. The yellow bars mark times containing at least one peak in either of the lightcurves. The orange bars flag instants (`close peaks') where each lightcurve contains a peak in adjacent time bins. The red bars flag times containing `overlapping' peaks occurring along both lightcurves simultaneously. Ratio between the total number of peaks to either close or overlapping peaks is indicated in the top-right of each panel, for both the entire time interval and a portion of it (300--700~s).} \label{fig:heatmaps}
\end{figure}

\subsubsection{SJI and STIX peak heatmaps}  \label{sec:heatmaps}

The correspondence between the SJI and STIX $32-50$\,keV pulsations (Section \ref{sec:correspondence}) was studied using simple peak heatmaps. The heatmaps (Figure \ref{fig:heatmaps}) compare the detrended STIX $32-50$\,keV lightcurve (black) with the SJI RD lightcurves (grey) in the four boxes S1 (panel a), S2 (b), R1s1 (c), and R1s2 (d). Because the data of the two instruments have different time resolution, 8\,s for SJI and 0.5\,s for STIX, the lightcurves were interpolated to a fixed time binning of 8\,s. Colored bars plotted therein indicate instances containing peaks in the pair of lightcurves. A $3\sigma$ peak detection criterion was employed for the SJI lightcurves (see Appendix \ref{sec:app_SJI_P} for details). For the STIX lightcurve, we focused on the strongest QPPs discussed in Section \ref{sec:stix_qpps}. The yellow vertical bars mark peaks with no corresponding pair in the second lightcurve. The orange bars flag peaks with a pair in a neighboring time bin. Finally, the magenta bars indicate times containing perfectly overlapping peaks in the lightcurve pair. In the top-right of each panel, `overlap' values are listed. These are defined as the ratio between the number of close and perfectly overlapping peaks to the total number of peaks in both lightcurves. The two overlap values correspond to either the entire time range plotted in Figure \ref{fig:heatmaps}, or the overlap between 300\,s -- 700\,s after the start of the plotted time range, when most of the HXR QPPs occurred. Note that the overlap simply quantifies temporal co-occurrence within the 8\,s heatmap time resolution and should not be interpreted as a statistical confidence level.

The overlap between the pulsations in the STIX and SJI S1 and S2 lightcurves is {up to} $\approx 70$\% when the shorter time range is considered. Strikingly, as indicated by the numerous red bars between $t $ = 360 -- {530}\,s in panel b, the STIX and SJI S2 pulsations (panel b) were coherent over a period of more than 2 minutes. The overlap {is 67}\% in the box R1s1 (panel c) and {decreases to} 56\% in R1s2 (panel d) in the same time frame. The {similarity between overlaps in S1 and R1s1} stems from the fact that R1s1 still extends to the HXR footpoints in the ribbon R1 (Figure \ref{fig:sji_rd_lightcurves}). When the entire time period is considered, the overlaps {decrease to only} 35\% in the boxes R1s1 and R1s2. This decrease is caused by the lack of the $32-50$\,keV HXR QPPs during the first $\approx 300$\,s of the timeseries. 

The heatmap analysis indicates that most of the HXR QPPs can be associated with an RD intensity pulsation in UV captured by IRIS/SJI in source regions of said QPPs. This \textit{spatio-temporal} correspondence between the HXR and UV pulsations implies that both HXR bremsstrahlung and UV emission, originating in the lower atmosphere, respond to the flare energy release close in time and space (Section \ref{sec:disc}). The correlation between the intensity variations induced by the slipping kernels and the HXR QPPs is lower. The UV intensity variation set on minutes before the HXR QPPs, and the slippage was the most prominent outside of the strongest HXR footpoints (Section \ref{sec:qpp_sources}). It is worth noting that the thermal $4-16$\,keV STIX lightcurve does show variations in the first $\approx 300$\,s of the timeseries, which are absent in the high-energy STIX bands (Section \ref{sec:correspondence}). We also investigated peak heatmaps comparing the lower-energy lightcurve to the SJI R1s1 and R1s2 lightcurves, but we did not find overlap exceeding 33\% (not shown). Therefore, even though the lower-energy HXR emission exhibited variation during the period of early kernel formation and slippage, it was only weakly correlated with the intensity variation induced by these early kernels. This weak correlation reflects the fact that R1s1 \& R1s2 trace intensity changes induced by the slipping kernels, not a direct energization into the HXR footpoint. Furthermore, the source reconstruction identified dominant sources of this emission in R2, rather than R1 which was distinguished by prominent kernel slippage.

\begin{figure}[h]
    \centering
    \includegraphics[width=18.00cm, clip, viewport = 00 10 510 430]{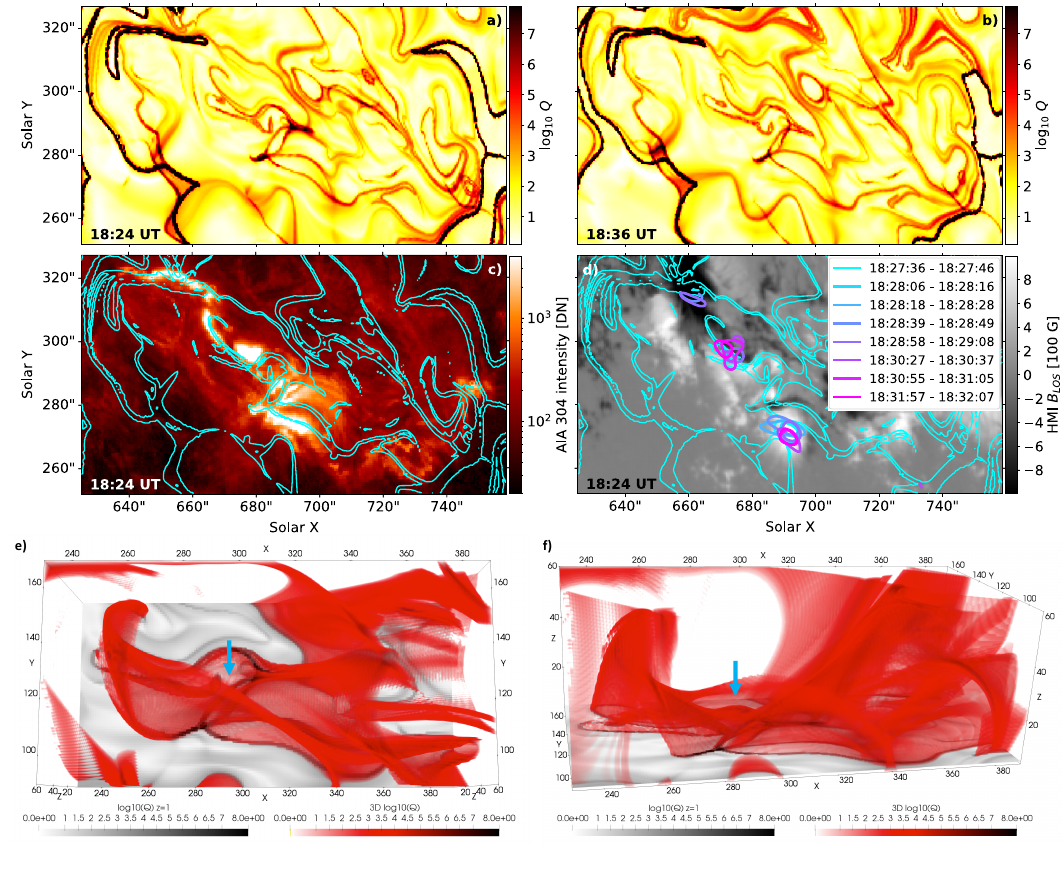}
    \caption{Panels a \& b show the $Q$ maps at $z \approx 1$\,Mm above the photosphere at 18:24 and 18:36 UT, respectively, during the impulsive and peak flare phases. Panel c presents the distribution of the photospheric $B_\mathrm{LOS}$ saturated to $\pm 1000$\,G. Ribbon observations from the AIA 304\,\AA~channel at 18:30 UT are presented in panel d. Cyan contours in panels c and d correspond to log$_{10}$ ($Q$) = 3 and delineate QSL footprints. The colored contours in panel c delineate HXR footpoints (see Section \ref{sec:qpp_sources} for details). Panels e and f show 3D rendering of $Q$ at 18:24 UT from two different vantage points of the flare central region. The blue arrows indicate possible reconnection site at the intersection of QSL volumes.}
    \label{fig:qsls}
\end{figure}

\section{Three-dimensional magnetic reconnection and topology} \label{sec:qsls}

The indirect imaging of STIX (Section \ref{sec:qpp_sources}) showed that $32-50$\,keV QPPs received dominant signal contribution from a pair of stationary HXR footpoints corresponding to portions of R1 and R2 in the central flare region. Here, we investigate the footpoint locations and ribbon morphology in the context of the magnetic environment in which the flare occurred. In particular, we analyze the spatio-temporal distribution of the squashing factor $Q$, whose high values ($Q >> 1$) indicate the presence of QSLs as preferred locations for 3D reconnection (Section \ref{sec:introduction}).

In order to calculate $Q$, we first performed nonlinear force-free field (NLFFF) extrapolations of four SHARP magnetograms acquired by HMI between 18:12 -- 18:48 UT at a cadence of 12 minutes. The \texttt{NF2} package \citep{Jarolim23} was employed for this task, while enforcing the divergence-free condition. The credibility of the model was verified by integrating field lines in the AR12975, which were found to closely resemble coronal loops observed in the AIA 171\,\AA~channel. In addition, the quality evaluation metrics of the extrapolation were comparable to those discussed in \citet[][see Appendix A therein]{Jarolim24b}. 3D distribution of $Q$ was calculated using the \texttt{FastQSL} code \citep{ZhangP22}, implemented as an optional output metric in \texttt{NF2}. To validate our results, we also obtained $Q$ using \texttt{QSLSquasher} \citep{Tassev17}. This was performed in one of the extrapolated magnetograms only, as photospheric magnetograms showed little evolution during the flare. The distribution of $Q$ was found to qualitatively agree with that resulting from \texttt{FastQSL}, in agreement with the analysis presented in \citet{ZhangP22}. 

The spatial distribution of $Q$ at $z = 1 $\,Mm above the photosphere during the flare {impulsive and peak phases} is plotted in Figure \ref{fig:qsls}a \& b, respectively. QSL footprints associated with the flare are located in the center of the FOV plotted therein. We observed little evolution of $Q_\textrm{z=1}$ between the two snapshots, with the key features visible already at 18:12 UT (not shown), roughly 20\, minutes before the flare peak \citep[c.f. Figure 5 of][]{Zhao16}. The QSL footprints at 18:24 UT, delineated using the cyan log$_{10}$ ($Q$) = 3 contours, qualitatively correspond to the location and shape of the ribbons observed 6 minutes later in the AIA 304\,\AA~channel (panel c), in agreement with previous literature \cite[see Section 3.3. in review][]{Dudik25}. They follow the complex PIL separating the positive- and the negative-polarity flux distributions observed by the HMI (panel d). {Comparison of panels c and d reveals that the short ribbon R2 was spatially coincident with a strong and compact positive-polarity flux concentration with $B_{\textrm{LOS}}$ up to $\approx 1200$\,G. The longer R1 formed in spatially-inhomogeneous negative-polarity flux regions with strengths typically |$B_{\textrm{LOS}}$| < 500\,G, but increasing to $|B_{\textrm{LOS}}| > 1000$\,G above $Y$ = 309\arcsec.} The hooked extension of R1, located in the north-east, follows a high-$Q$ (log$_{10}$ ($Q$) $> 7$, black) footprint in the same region (c.f., panels a \& c). $Q \rightarrow \infty$ values are indicative of true separatrices. It is, however, likely that the precision of the $Q$ calculations at the peripheries of the FOV was here limited by the size of the domain. The bottom row of Figure \ref{fig:qsls} presents a 3D rendering of $Q$ at 18:24 UT from two different vantage points zoomed into the central flare region. The blue arrow in panels e and f points to the intersection of QSLs volumes. Located at an altitude of about 8\,Mm between R1 and R2, this intersection of high-$Q$ volumes is a likely candidate location for connectivity changes associated with reconnection. It projects to an elongated ellipse-shaped QSL footprint at $X \approx 690\arcsec$, $Y \approx 285 \arcsec$ visible in Figure \ref{fig:qsls}a. 

The colored contours in Figure \ref{fig:qsls} delineate the HXR footpoint sources during the QPPs detected by STIX in the $32-50$\,keV bands, the same as those plotted in Figure \ref{fig:qpp_sources} and discussed in Section \ref{sec:qpp_sources}. {The two sources were rooted in regions where the line-of-sight field strength differed by up to two orders of magnitude. The southern footpoint was located in the strong-field region with $B_{\textrm{LOS}}$ of around and exceeding 1000\,G (analogous to R2). The northern source was observed in comparatively moderate fields, with $B_{\textrm{LOS}}$ typically below 500\,G.} Consistent with the morphology of R1 and R2, the HXR footpoints' locations correspond to QSL footprints. Non-thermal electrons released by reconnection thus precipitated into locations with high connectivity gradients, i.e., hosting field lines likely to partake in reconnection, in agreement with the standard flare model and its 3D extensions. The comparison between the spatial extent of the HXR footpoints and the QSL footprints suggests that recurrent non-thermal energy injection from the reconnection site was dominant within a relatively-compact loop system anchored roughly 20\,Mm on either side of the reconnection site in both R1 and R2. As evidenced by the spatial distribution of $Q$ as well as that of the ribbons, this loop system was embedded in a large-scale QSL where energy deposition was weaker. The spatial inhomogeneity in the distribution of the reconnected energy is discussed in Section \ref{sec:disc_deposition}. 

\section{IRIS spectroscopy} \label{sec:spectroscopy}

The IRIS slit crossed the compact ribbon R2 in a region coincident with the HXR footpoint therein (c.f., Figure \ref{fig:overview}d, e). Here we briefly discuss the temporal evolution of the \ion{Si}{4} 1402.77\,\AA~spectra captured at a very high cadence of $\approx0.8$\,s. For analyses of high-cadence observations of different events, see e.g., \citet{Lorincik22, Lorincik25b, Ashfield25}. In the flare under study, the \ion{Si}{4} 1402.77\,\AA~line exhibited complex multi-component profiles, similar to those recently reported in another M-class flare \citep[][]{Ashfield25}. The profiles often consisted of at least three components in emission, corresponding to the rest wavelength, a weakly-redshifted ($v_\mathrm{D} \lessapprox 70$\,\kps) but strong component, and another highly-redshifted ($v_\mathrm{D} \lessapprox 200$\,\kps) weaker component. In certain locations, we also observed weak traces of absorption features due to \ion{Fe}{2} lines as well as weak self-absorption at the rest wavelength. Unlike \citet{Ashfield25}, who employed six-Gaussian fits to assess the components' behavior in great detail, here we, for simplicity, study the spectra using moments. Because the profiles were broad, redshifted, and the respective spectral window can contain several weak intercombination lines \citep{Polito16, Young18}, the moments were calculated over the wavelength range of $-0.9$\,\AA~to $+1.5$\,\AA~from the laboratory rest wavelength (1402.77\,\AA). Prior to the moment calculation, we performed subtraction of the FUV continuum averaged between 1405.01\,\AA $-$ 1405.71\,\AA. The first moment (centroid) was used to obtain the Doppler velocity, calculated using the reference wavelength of 1402.77\,\AA~\citep[][]{Sandlin86}. The line broadening is here expressed in terms of the standard deviation ($\sigma$) calculated as the square root of the second moment (variance).

\begin{figure}[h]
    \centering
    \includegraphics[width=18.00cm, clip, viewport = 00 00 500 255]{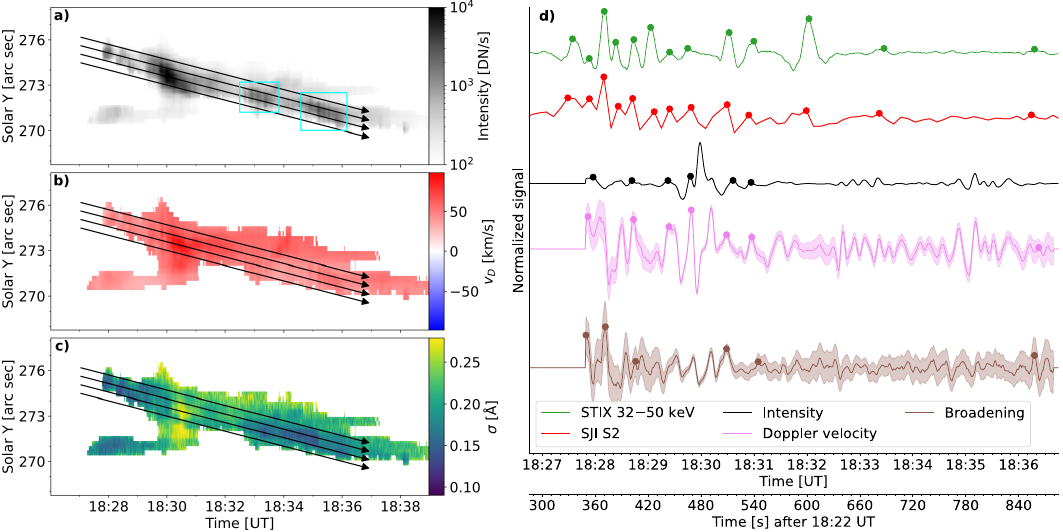}
    \caption{Maps of the intensity (panel a), Doppler velocity (panel b), and broadening of the \ion{Si}{4} 1402.77\,\AA~line inferred using the moment analysis as a function of time during the flare impulsive and peak phases. The black arrows denote cuts used to track the moments in the ribbon's emission under slit in time. Panel d compared the HXR $32 - 50$\,keV (green) and the SJI S2 (red) lightcurves with those of the moments averaged across the four cuts in time. The black, magenta, and brown lightcurves correspond to the \ion{Si}{4} intensity, Doppler speed, and broadening, respectively. The dots along the green lightcurve indicates major HXR QPPs. The colored dots along the remaining lightcurves mark pulsations associated with HXR QPPs.}
    \label{fig:spec_moments}
\end{figure}

Figure \ref{fig:spec_moments} presents the spatio-temporal distribution of the \ion{Si}{4} intensity (panel a), Doppler velocity (panel b), and broadening (panel c) along the IRIS slit in time. The \ion{Si}{4} intensity increase set on after about 18:28 UT, at which the kernels composing R2 appeared in the location scanned by the slit. The total line intensity exhibited a strong maximum at about 18:30 UT. Interestingly, the intensity exhibited oscillatory patterns visible as adjacent blobs around 18:33 UT and 18:35 UT, highlighted using the cyan frames in panel a. The line was consistently redshifted due to chromospheric condensation, exhibiting downflows at speeds up to $v_\mathrm{D} \lessapprox 100$\,\kps~as indicated by the moments (panel b). As seen in panel c, the line broadening peaked roughly 30\,s after the intensity maximum at $\approx$18:30:35 UT. 

The slanted arrows in panels a -- c track the evolution of the moments in different pixels within the separating ribbon (Section \ref{sec:overview}). The line intensity, Doppler speed, and broadening, averaged across the four arrows as a function of time, are plotted in Figure \ref{fig:spec_moments}d (see also Appendix Figure \ref{fig:app_spec_P}). To maintain consistency with the analysis of the STIX and GOES signals, the lightcurves were first detrended to remove the slowly-varying components and subsequently smoothed using the same procedure and detrending windows as discussed in Section \ref{sec:correspondence}. For comparison, this panel also shows two lightcurves from Figure \ref{fig:sji_rd_lightcurves}; the detrended $32-50$\,keV lightcurve observed by STIX (green) as well as the SJI RD intensity variation in the box S2 (red).

These spectral properties exhibited clear temporal variation. The line intensity evolution (black) aligns with that inferred from the inspection of the intensity map in Figure \ref{fig:spec_moments}a, indicative of the strong pulsation at 18:30 UT and the two series of lower-amplitude pulsations at 18:33 UT and 18:35 UT. The pulsation period (Appendix \ref{sec:app_spec_P}) estimated over the entire time interval is $P_{\textrm{Si IV} I} \approx 30$\,s, similar to the roughly 27\,s period estimated using the SJI data in the same region (Appendix \ref{sec:app_SJI_P}). The intensity oscillation period, however, lowers to about $P_{\textrm{Si IV} I} \approx 19$\,s for the pulsations detected after a brief period of weak signal around 18:32 UT. The oscillations were most prominent in the Doppler velocity lightcurve plotted in violet, with the estimated period $P_{\textrm{Si IV} v_\mathrm{D}}$ of about 22\,s. Note that the Doppler velocity also exhibited pulsations at about 18:31:30 UT and 18:34 UT, at which no corresponding intensity pulsations were detected. The broadening lightcurve (brown), although encumbered by higher uncertainties, exhibited several pulsations coincident with those of the Doppler velocity at $P_{\textrm{Si IV} \sigma} \approx$ 20\,s. This relationship could be due to the fact that the broadening, inferred from the moment analysis, is intimately related to the Doppler velocity, whose variability can be dominated by the secondary redshifted component \citep{Lorincik22}. At the same time, a broadening pulsation at 18:28:30 UT appears to be in anti-phase with corresponding Doppler velocity pulsations. Studying the degree of coherence between the pulsations of different spectral properties was outside the scope of this work, but merits the attention of future studies. 

Here, we instead focus on the correlation between the \ion{Si}{4} spectral pulsations and the QPPs captured by STIX in the $32-50$\,keV bands. In Figure \ref{fig:spec_moments}d, prominent HXR QPPs are marked for clarity using the colored dots along the green lightcurve. In the remaining lightcurves, we only highlight pulsations occurring during or close in time to the STIX QPPs. As we already established in Sections \ref{sec:correspondence} and \ref{sec:heatmaps}, the HXR lightcurve was well-correlated with RD intensity pulsations captured by the SJI in ribbon subregions corresponding to the HXR footpoints. Indeed, the green and red lightcurves exhibit many peaks occurring within 8\,s from each other. The \ion{Si}{4} intensity (black) and Doppler velocity (violet) exhibit 6 pulsations, low-amplitude in the former case, which occurred within 5\,s from the corresponding HXR QPPs. Six HXR QPPs also overlap with the broadening pulsations (brown), although this number should be interpreted cautiously due to the presence of the noise. There are also several noteworthy discrepancies between the HXR and \ion{Si}{4} lightcurves (see also Section \ref{sec:disc_burstyrec}). The strongest \ion{Si}{4} intensity pulsation at 18:30 UT, as well as prominent Doppler velocity and broadening pulsations with low error bars observed ca. 20\,s later, were not associated with any HXR QPP. Similarly, the strong HXR QPP at 18:32 UT did coincide with a pulsation in SJI S2, but at the same time the \ion{Si}{4} line was weak and did not exhibit any discernible pulsations. After this pulsation, the HXR lightcurve exhibited only small-amplitude variation, whereas the \ion{Si}{4} variability continued for at least 4 more minutes until 18:36 UT. The two series of intensity pulsations at 18:33 UT and 18:35 UT, as well as the persistent oscillatory redshifts, thus cannot be associated with HXR oscillations at these energies. Some of the small-amplitude peaks observed in $32 - 50$\,keV in this period appeared more prominently at lower energies (e.g., $15 - 22$\,keV), exhibiting a clear non-thermal component as indicated by the fitting (not shown). Unlike the strong QPPs in $32 - 50$\,keV, these were, however, difficult to match to peaks in visible raw lightcurves, and we did not analyze them further. It is also worth noting that the shorter ($\approx 18$\,s) quasi-periods of the \ion{Si}{4} spectral properties are close to twice the 8\,s cadence of the SJI observations. Had oscillations with similar periods manifested themselves in the imaging observations, they likely wouldn't be resolved due to the Nyquist criterion. 

\section{Discussion} \label{sec:disc}

\subsection{Overview of key observational results}

\begin{figure}[h]
    \centering
    \includegraphics[width=9.00cm, clip, viewport = 00 80 500 450]{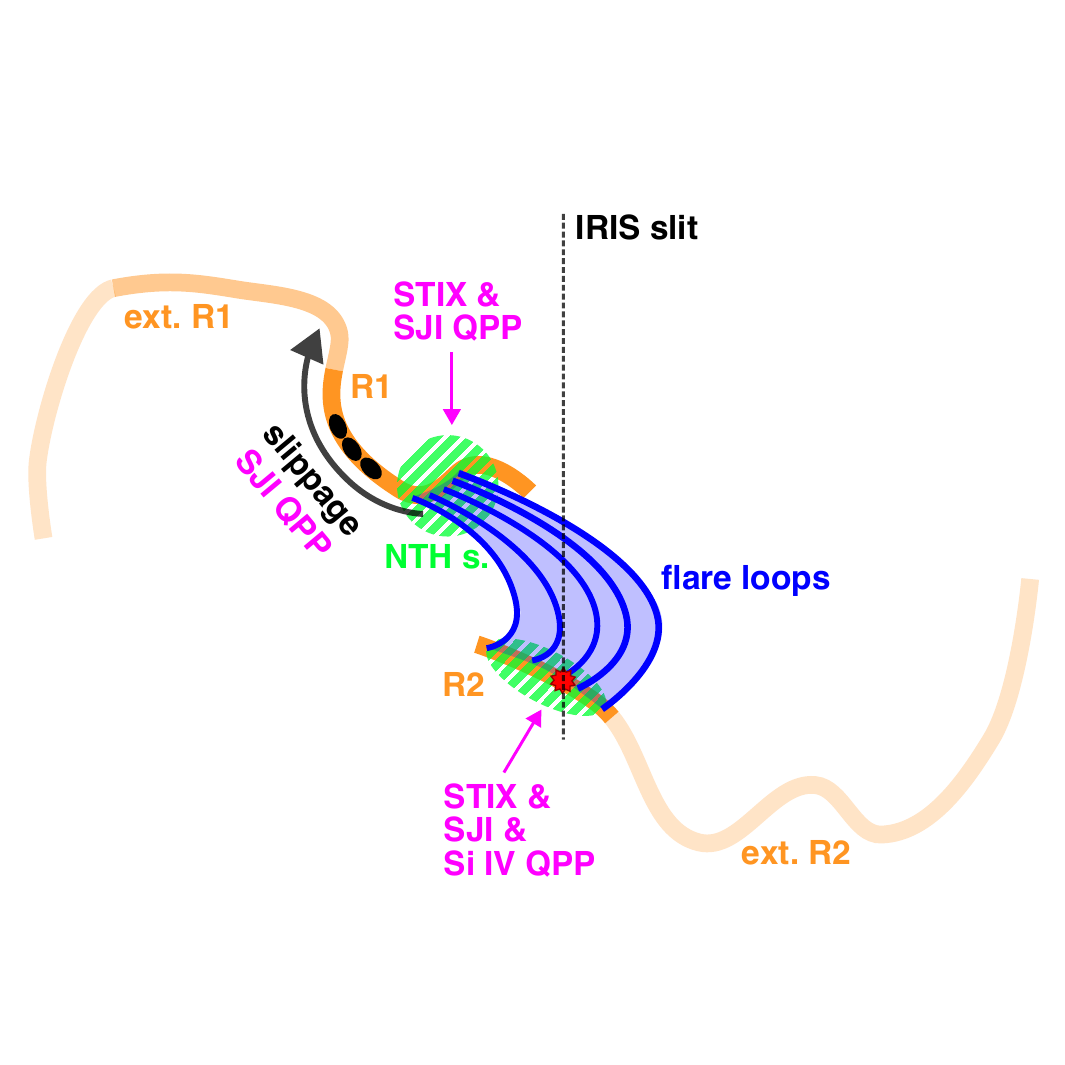}
    \caption{Cartoon summarizing IRIS and STIX observations of the 2022 March 31 flare.}
    \label{fig:cartoon}
\end{figure}

Prior to discussing our findings and their implications, we briefly reiterate key results from our observational analysis summarized in a cartoon sketch shown in Figure \ref{fig:cartoon}. The ribbons R1 and R2, as well as their extensions ext. R1 and ext. R2, are indicated in orange. The ribbons correspond to QSL footprints (Section~\ref{sec:qsls}) within a large-scale 3D reconnection geometry. The primary HXR footpoints, drawn in green, are found in the central flare region at the footpoints of the flare loop arcade (blue) in the vicinity of the likely reconnection region. Kernels are indicated using the black ellipses in R1, and the black curved arrow indicates the dominant direction of slippage. The magenta labels indicate regions where QPPs were captured by IRIS, STIX, or both instruments. The short-period ($\approx 30$\,s) QPPs were found in HXR emission, UV ribbon intensities, and transition region \ion{Si}{4} spectra during the impulsive phase. The HXR and UV pulsations were often co-spatial and co-temporal within the time resolution of the IRIS data ($\approx$~8s). Slipping reconnection manifested itself in the appearance and apparent motions of kernels along flare ribbons, in particular along the curved part R1. The region distinguished by prominent kernel slippage was not associated with \textit{strong} HXR footpoint emission.

\subsection{Temporal coherence and bursty reconnection}  \label{sec:disc_burstyrec}

The strong temporal overlap between UV and HXR pulsations observed during the impulsive phase provides compelling evidence that these signals were of a common physical origin. The period of the pulsations closely matches the $\approx 32$\,s periods recently reported by \citet{Ashfield25}. In that study, coherent UV and HXR pulsations were attributed to bursty reconnection driving episodic particle acceleration and chromospheric condensation. Viewed in this context, the overlap between the HXR QPPs, UV intensity patterns in the HXR footpoints, and several \ion{Si}{4} spectroscopic pulsations in our observations suggests that they were likewise driven by bursty reconnection. The formation and subsequent kernel motions observed in the conjugate ribbon R1, governed by slipping reconnection, also exhibited a bursty character in agreement with previous results \citep{ZhangY25}. {Despite growing observational evidence (Section \ref{sec:introduction}), the mechanisms responsible for the bursty character of slipping reconnection remain unknown. Bursty reconnection is often associated with formation of plasmoids in tearing current sheets. Using a 3D MHD eruption simulation, \citet{Dahlin25} demonstrated that self-consistently forming plasmoids map to transient corrugations of ribbon fronts, so-called `spirals' or `breaking waves' \citep[see e.g.,][]{Wyper21, French25}. According to \citet{Dahlin25}, these corrugations develop predominantly in the direction perpendicular to the ribbon, unlike the field line slippage occurring along ribbons. The connection between the plasmoid-imprinted corrugations and apparent slipping motions remains an open question.}

While UV imaging and spectroscopic pulsations persisted into the flare peak, they had no detectable counterpart in STIX HXR observations during that phase (Section~\ref{sec:spectroscopy}). This suggests that, during the flare peak, the energization of the lower atmosphere may have been dominated by other heating mechanisms, such as thermal conduction \citep[e.g.,][]{Macniece86, Longcope14, Polito18, Ashfield22, Lorincik25b} or Alfv\'{e}n waves \citep[e.g.,][]{Russell13, Reep16, Kerr16, Lorincik25b}. A similar disparity between UV and HXR signals was also observed immediately after flare onset, when the UV intensity variations induced by the slipping kernels preceded the onset of $32-50$\,keV HXR QPPs by several minutes (Section~\ref{sec:sji_qpps}). This earlier UV onset suggests that non-thermal energization into the slipping kernels was initially weaker, or that alternative heating mechanisms dominated during the early phase of the flare. We also cannot exclude the possibility that weak HXR counterparts to the UV pulsations were present but remained below what can be detected by STIX.

The degree of temporal coherence between UV and HXR pulsations warrants a brief discussion. The HXR \& UV ribbon intensity pulsations, observed over a period of {more than} 2 minutes in the ribbon R2 (Section \ref{sec:heatmaps}), were synchronous within the 8\,s time resolution of the 2796\,\AA~SJI observations. When assessing the overlap between the pulsations in the peak heatmaps (Figure \ref{fig:heatmaps}), we considered peaks occurring in neighboring time bins to be also concurrent. By doing so, we relaxed the UV \& HXR pulsation synchronicity criterion down to 16\,s of each other. Under this definition, we identified several instances when the UV pulsation preceded the corresponding HXR QPP (by 1 time step, 8\,s). We verified that this behavior cannot {be fully attributed to artifacts introduced by the differentiation of the SJI observations, or the interpolation of the STIX time series to the cadence of the SJI data. While minor timing shifts could have arisen from the post-processing procedures, they do not systematically account for the observed offsets.} While the time resolution of SJI data precluded a more precise timing comparison, these delays could have been of a physical nature \citep[see e.g.,][]{Miklenic07, Jeffrey18}.

\subsection{Spatial distribution of energy deposition during slipping reconnection} \label{sec:disc_deposition}

While the temporal correspondence between UV and HXR pulsations supports bursty reconnection as their common driver, the spatial distribution of non-thermal energy deposition during slipping reconnection appears non-uniform. The majority of HXR QPPs detected by STIX coincided with the SJI UV intensity pulsations measured in the same footpoint locations. In contrast, the portion of ribbon R1 where slipping motions were most prominent did not coincide with strong ($> 30$\% of the $I_\textrm{max}$) HXR emission sources (Section \ref{sec:qpp_sources}). The spatial separation between the region distinguished by prominent kernel slippage and HXR footpoint along R1 is apparent from Figure~\ref{fig:cartoon}. This behavior contrasts with the results reported by \citet{Purkhart25}, a recent study allowing for a particularly relevant comparison with ours as they also focused on HXR emission in an event exhibiting signatures of slipping reconnection (Section \ref{sec:introduction}). In particular, the authors found the location of UV kernels to be sometimes consistent with HXR footpoints. Although the dynamics of kernel slippage reported therein were comparable to those found in our study, the correlated UV and HXR pulsations occurred at significantly longer periods ($> 3$\,min). The comparison between the two events suggests that similar reconnection dynamics can be associated with different QPP periods, but significantly different patterns of non-thermal energy deposition.

A possible explanation for the absence of strong HXR sources coincident with slipping kernels in our event is the limited dynamic range of indirect HXR imaging. Relatively weak HXR emission associated with slipping kernels could have simply remained below the detection threshold of STIX. As indicated by the peak heatmap {analysis, $\approx 35$\% of} the pulsations driven by slipping kernels did coincide with HXR QPPs, although the UV and HXR intensities were measured in different (but adjacent) regions. This suggests that at least a fraction of non-thermal energy was deposited into the slipping kernels, producing bright UV emission (Appendix~\ref{sec:app_R1R2}), but the corresponding HXR signal was too weak for the sources to be detected. Nevertheless, even when accounting for this limitation, our observations indicate a spatial offset between the location of strong HXR emission and majority of the slipping kernels. This observable suggests that the rate of non-thermal energy deposition into the slipping kernels was locally reduced compared to where the HXR footpoint was found. This hypothesis has been already proposed based on previous observations of slipping reconnection \citep{Dudik16, Lorincik19a}. 

The local decrease of the energy deposition rate could have been related to the asymmetric magnetic structure of the event. Only some of the QSL footprints exhibited a visible ribbon counterpart, and only certain ribbon sections brightened significantly during the flare (Figure~\ref{fig:qsls}c). As described in Section~\ref{sec:overview}, the brightest portions of R1 were at least twice as long as R2. Non-thermal electrons traveling along field lines slipping along R1, exciting kernel emission, would likely deposit their energy over a longer ribbon portion compared to those rooted in compact R2. {Our examination of the photospheric magnetic field strength (Section \ref{sec:qsls}) supports this notion. Compared to R2, R1 formed in spatially more inhomogeneous and extended concentrations of surface magnetic field. Prominent kernel slippage along R1 occurred in regions of comparatively weaker field strengths, suggesting a more spatially-distributed energization. A clear correspondence between strong-field regions and localized HXR emission was observed in R2, where the southern footpoint overlapped with the majority of this compact ribbon. In contrast, the primary northern source mapped to a region of comparatively weaker fields. R1 however briefly hosted a secondary source separated from the primary one by $> 20\arcsec$, further supporting the scenario of the non-thermal energy deposition being more spread-out along this ribbon.} 

In agreement with previous literature \citep[see e.g.,][and references therein]{Fletcher04}, the HXR observations indicate that strong non-thermal energy input was highly localized. The energy of a majority of $32-50$\,keV electrons was injected into a loop subsystem \citep[see Section \ref{sec:qsls} and e.g.,][]{Temmer07} close to the reconnection site within the large-scale QSL. The relative position of the flare loop and HXR footpoints suggests that the energy deposition therein has driven efficient chromospheric evaporation (Figure \ref{fig:overview}g, \ref{fig:cartoon}), unlike elsewhere in the ribbons. Although based on several assumptions and simplifications, this reasoning provides a plausible explanation for the strong spatial confinement of the non-thermal energy deposition into R2, the stronger intensity of 131\AA~loops anchored in R2 (e.g. due to possibly stronger evaporation) and the comparatively weaker HXR signatures associated with slipping kernels along R1. 

To conclude, these results indicate that while slipping reconnection can proceed in a bursty manner and drive coherent temporal variability across multiple diagnostics, strong non-thermal energy deposition can remain highly localized in space. The contrast between our results and those of \citet{Purkhart25} highlights the need for a larger statistical sample of flares observed simultaneously in UV and HXR in order to address which factors, such as magnetic topology, flare magnitude, size, duration, or eruptivity, control the energy deposition during slipping reconnection.

\section{Summary} \label{sec:summary}

In this manuscript, we investigated the 2022 March 31 M9.6-class flare observed at high cadence by IRIS and Solar Orbiter/STIX. The flare exhibited quasi-periodic pulsations (QPPs) at a broad range of periods, roughly between $26 - 55$\,s, detected in X-rays by GOES and STIX, as well as UV imaging and spectroscopy by IRIS. We primarily focused on the correlation between HXR QPPs and UV pulsations captured in different ribbon subregions, including regions exhibiting signatures of magnetic slipping reconnection. Our results can be summarized as follows:

\begin{enumerate}
    \item The flare produced a pair of flare ribbons R1 and R2 composed of flare kernels, well-resolved in the IRIS/SJI 2796\,\AA~channel data. The kernels were exhibiting apparent slipping motions at speeds $v_\textrm{slip}$ up to 76\,\kps, best seen along the longer R1. Only a few slipping kernels were detected in the compact R2.
    \item The time derivative of the GOES $1-8$\,\AA~SXR flux and the non-thermal $32-50$\,keV STIX lightcurve exhibited QPPs with periods of $26 < P_{\textrm{GOES}} < 29$\,s and $36 < P_{\textrm{STIX}} < 55$\,s during the flare impulsive and peak phases. The STIX QPP period estimated during the flare impulsive phase, when the pulsations were more frequent, lowers to $31 < P_{\textrm{STIX}} < 39$\,s. Thermal $4-16$\,keV STIX emission also exhibited signal variability starting $\approx 3$ minutes before the non-thermal bands. 
    \item The non-thermal STIX QPPs originated mostly from a pair of HXR footpoints located in the central flare region and nearly stationary in time. The sources were spatially coincident with bright and stationary portions of R1 and R2 as well as bright high-temperature flare loop emission, as seen in AIA 131\,\AA, presumably due to stronger chromospheric evaporation .
    \item IRIS/SJI 2796\,\AA~(dominated by chromospheric emission) captured intensity pulsations in several ribbon subregions at quasi-periods {of $25 < P_\textrm{SJI} < 34$\,s. The regions} of interest included the portions of R1 and R2 corresponding to the HXR footpoints, as well as the portion of R1 distinguished by prominent kernel slipping motions. This result provides further evidence that slipping reconnection can be of a bursty nature.
    \item {Roughly} 70\% of the pulsations detected in the $32-50$\,keV bands of STIX and SJI, in ribbon portions corresponding to the HXR footpoints, were either synchronous or occurred close in time. Most of the high-energy STIX QPPs were thus associated with a UV intensity response captured by SJI. 
    \item This correlation was {lower ($\approx 35$\%) for} UV intensity pulsations in the region distinguished by prominent slipping motions. The kernel formation and slippage set on nearly 5 minutes before STIX detected QPPs in the $32-50$\,keV band. While some of the kernels formed close to the footpoint in R1 from where they slipped, we did not find dominant sources of HXR emission along the trajectory of the slipping kernels. This is indicative of a relatively lower rate of non-thermal energy deposition into the slipping kernels compared to the pair of bright HXR sources.
    \item The ribbons were found to correspond to quasi-separatrix layer (QSL) footprints calculated using extrapolated HMI magnetograms. Prominent non-thermal energy deposition, manifested in the formation of the HXR footpoints, was highly localized, likely restricted to a compact system of flare loops in the vicinity of the reconnection region within an extensive 3D reconnection structure.
    \item IRIS also provided useful spectroscopic diagnostics of the transition region \ion{Si}{4} emission observed in the ribbon R2 at high sub-second cadence. The line exhibited pulsations in the intensity, Doppler speed, and broadening at periods roughly between $ P_\textrm{Si IV} \approx 19 - 30$\,s. 
    \item Several peaks of the line intensity and Doppler speed occurred concurrently with $32-50$\,keV HXR QPPs, suggesting they were driven by bursty reconnection, in line with the results of \citet{Ashfield25}. The variability of spectral characteristics, however, lasted for longer than the HXR QPPs, suggestive of a change in the energy transport agent as the flare progressed.
\end{enumerate}

These results support a scenario in which two distinct behaviors can be identified within the flare ribbons. In the part of the ribbon with the strongest HXR emission, connected to the core of the hot loops seen in AIA 131\,\AA, bursty reconnection produces the multi-wavelength QPPs observed in different atmospheric layers and domains of the electromagnetic spectrum, aligning with the recent study of \citet{Ashfield25}. This suggests that a more sustained energy release occurs therein, driving evaporation, strong HXR footpoint emission, and QPPs. In the part of the ribbon further away from the primary HXR footpoint, we observe prominent signatures of slipping reconnection, but insignificant or weaker HXR emission compared to the previous region, possibly due to STIX's dynamic range. This different behavior might be due to the fact that, because of the slipping motions, the non-thermal energy is not consistently deposited in an individual area of the ribbon for a long enough time to drive strong evaporation and HXR emission. 

The detection of short ($\approx$30\,s) oscillations periods keeps emphasizing  the need for flare observations at high cadence. This capability is in HXR now offered by STIX onboard Solar Orbiter \citep{Muller20} and the Hard X-ray Imager (HXI) onboard ASO-S \citep{Gan19}. In UV and EUV, these diagnostics are made possible thanks to IRIS (uniquely for UV spectroscopy), the Extreme Ultraviolet Imager \citep[EUI;][]{Rochus20} onboard Solar Orbiter, and will be made available by the upcoming Multi-Slit Solar Explorer \citep[MUSE;][]{Depontieu22, Cheung22} mission. 

Because the dynamic range of indirect HXR imaging poses limitations in terms of the identification of weak HXR footpoints in slipping kernels, it remains to be determined whether apparently-slipping flare kernels can be associated to HXR footpoints exhibiting similar dynamics, and what are the properties of non-thermal electrons deposited therein. Further comparison between IRIS spectroscopic observations, radiative hydrodynamic models and HXR emission will be needed to investigate those properties in more detail \citep{Polito16a, Polito23b, Lorincik25b} and thus advance the understanding of flare energetics during 3D slipping reconnection. Future direct HXR imaging missions, such as FIERCE \citep{Shih20} or the Focusing Optics X-ray Solar Imager \citep[FOXSI;][]{Christe16} sounding rocket reflights, will offer promising diagnostics of where and how much energy is deposited in complex 3D magnetic topologies.

\section*{Acknowledgments}

The authors thank {the anonymous referee for insightful suggestions and} Jana Ka\v{s}parov\'{a}, Jaroslav Dud\'{i}k, Robert Jarolim, Shane Maloney, and Peijin Zhang for useful discussions. This work was funded by the Heliophysics Guest Investigator (H-GI) Open grant 80NSSC24K0553. J.L., V.P., and N.F. acknowledge support from NASA under the contract NNG09FA40C (IRIS). L.A.H is supported by a Royal Society-Research Ireland University Research Fellowship (URF$\backslash$R1$\backslash$241775). N.F. acknowledges support from NASA under the contract NNG04EA00C (SDO/AIA), NNM07AA01C (Hinode/SOT) and 80GSFC21C0011 (MUSE). IRIS is a NASA small explorer mission developed and operated by LMSAL with mission operations executed at NASA Ames Research Center and major contributions to downlink communications funded by ESA and the Norwegian Space Agency. Solar Orbiter is a space mission of international collaboration between ESA and NASA, operated by ESA. The STIX instrument is an international collaboration between Switzerland, Poland, France, Czech Republic, Germany, Austria, Ireland, and Italy. SDO data were obtained courtesy of NASA/SDO and the AIA and HMI science teams. 

\appendix 
\newcounter{appendix}
\renewcommand{\theappendix}{Appendix}
\refstepcounter{appendix}

\setcounter{figure}{0}    
\renewcommand{\figurename}{Appendix Figure}

\begin{figure}[h]
    \centering
    \includegraphics[width=0.4\linewidth]{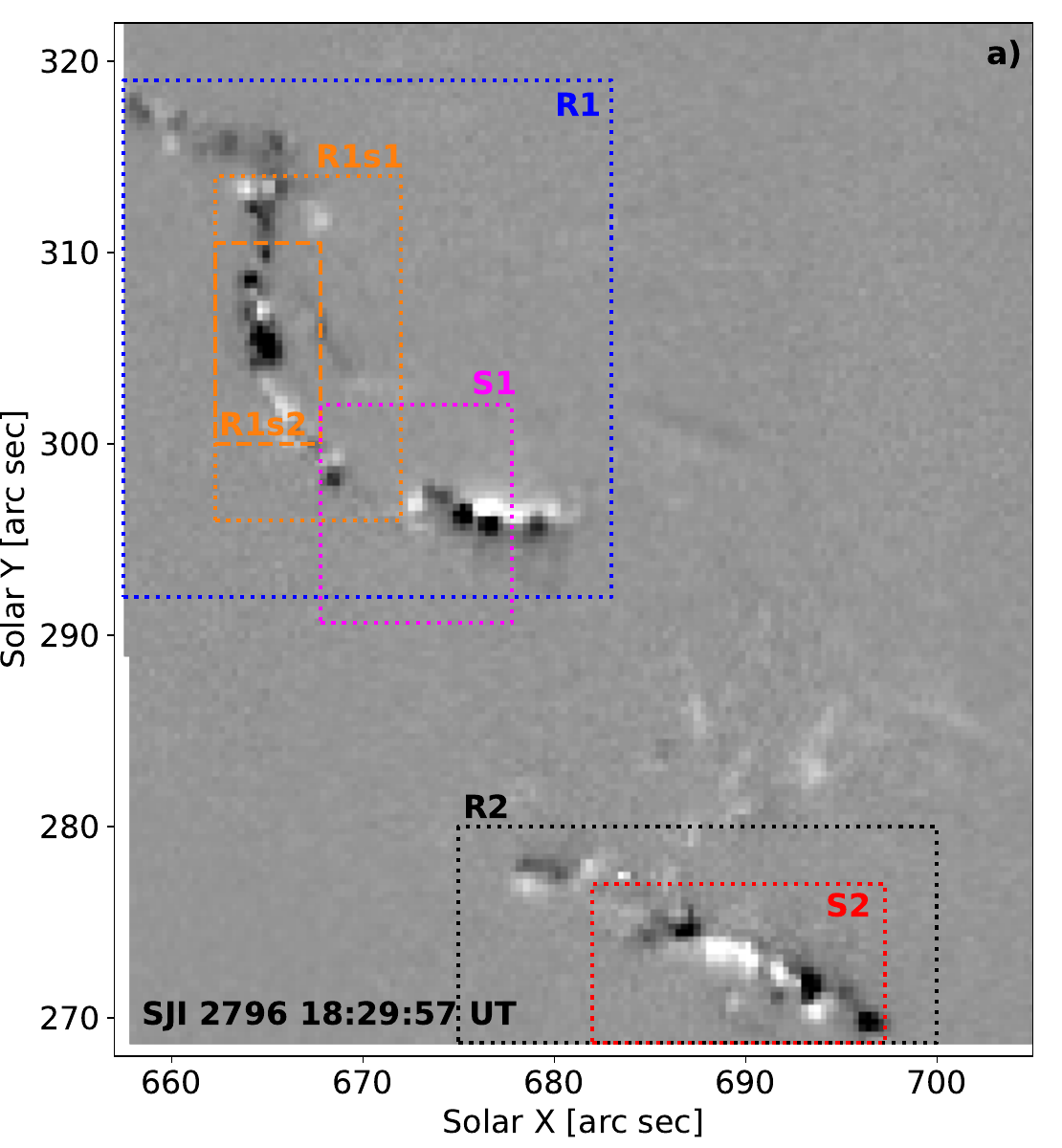}
    \includegraphics[width=0.99\linewidth]{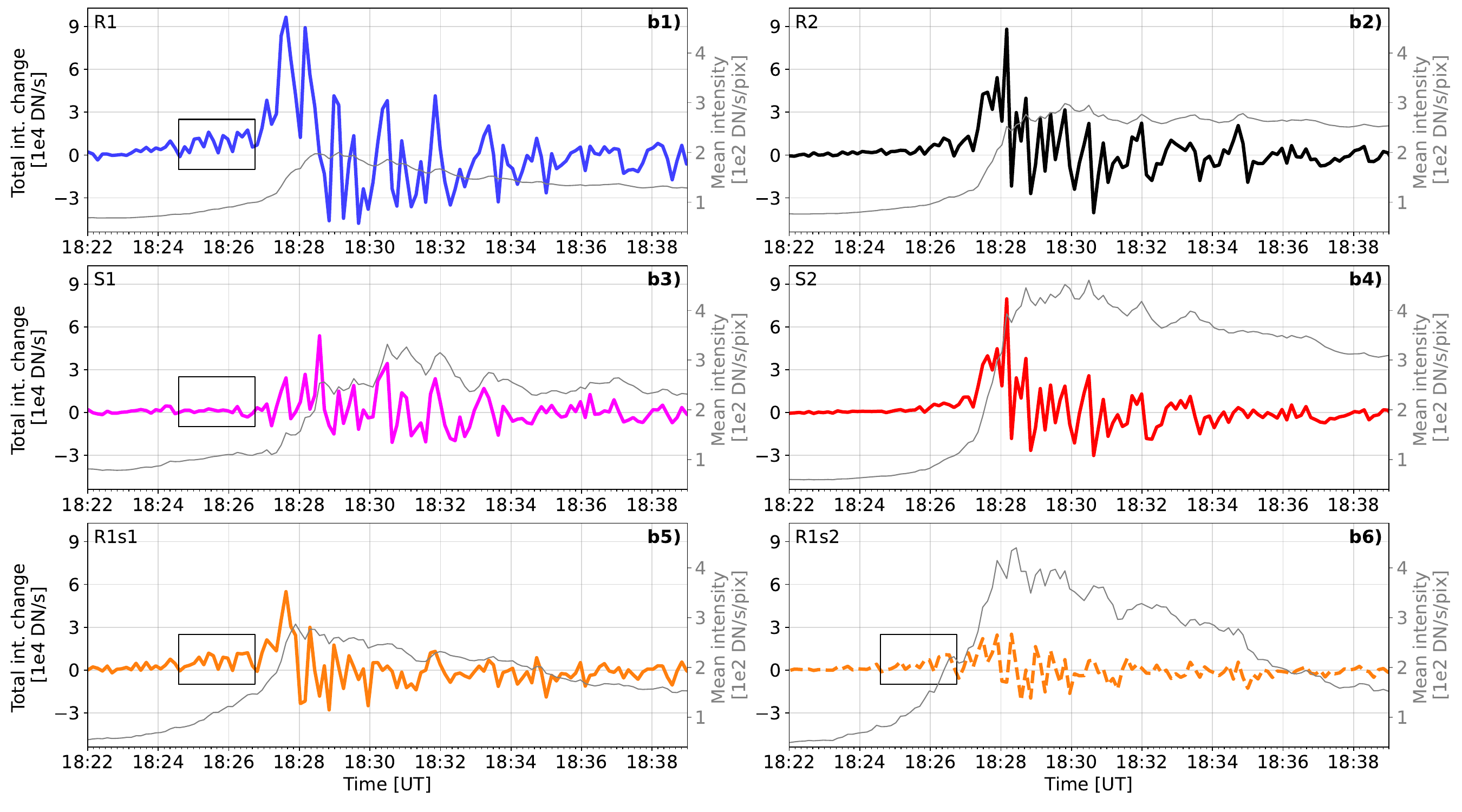}
    \caption{Boxes focused on different ribbons and ribbon subregions (panel a) for studying the intensity variation as observed in SJI 2796\,\AA~(panels b1 -- b6). The thick lines delineate the total intensity change inferred from the RD SJI observations in different boxes. The thin gray lines represent the mean intensity in each of the boxes.}
    \label{fig:app_R1_R2}
\end{figure}

\section{Ribbon UV intensities}
\label{sec:app_R1R2}

Appendix Figure \ref{fig:app_R1_R2}a presents SJI 2796\,\AA~snapshot of the ribbons with boxes used to study the evolution of the intensity in different regions. In addition to the R1s1, R1s2, S1, and S2 boxes discussed in Section \ref{sec:sji_qpps}, here we also study the boxes R1 (blue) and R2 (black) capturing the two ribbons in their entirety as observed in the SJI FOV. Panels b1 -- b6 present the lightcurves corresponding to the mean intensity, inferred from normalizing the total intensity by the area of each box (grey) and the total intensity change (same as Figure \ref{fig:sji_rd_lightcurves}b, colored), the latter inferred from the RD data. Note that the lightcurves are plotted without normalization and the plotting ranges on the vertical axes are fixed to facilitate their comparison.

Comparison between panels b1 and b2 reveals that the mean intensity in the box R2 was roughly 50\% higher than that in R1. Similarly, the maximum mean intensity in the HXR footpoint source box S2 (panel b4) exceeded that measured in S1 (b3) by up to 40\%. The mean intensity of the compact ribbon R2{, located in a strong-field region, was consistently higher than that of its longer counterpart R1 { located in a more fragmented flux (Section \ref{sec:qsls})}}. Interestingly, the mean intensity in the segment of R1 where the slippage was prominent (R1s1) pars with that in the ribbon portion corresponding to the primary HXR footprint (S1; c.f., panels b3 \& b5). The peak mean intensity in the smallest box R1s2 focused on the slipping kernels, exceeds that measured in the HXR footpoint by $\approx 35$\% (c.f., panels b3 \& b6). The portion of the ribbon distinguished by prominent kernel slippage therefore was not, on average, fainter than that corresponding to the HXR footprints. 

\begin{figure}[h]
    \centering
    \includegraphics[width=0.99\linewidth, clip, viewport = 00 80 620 720]{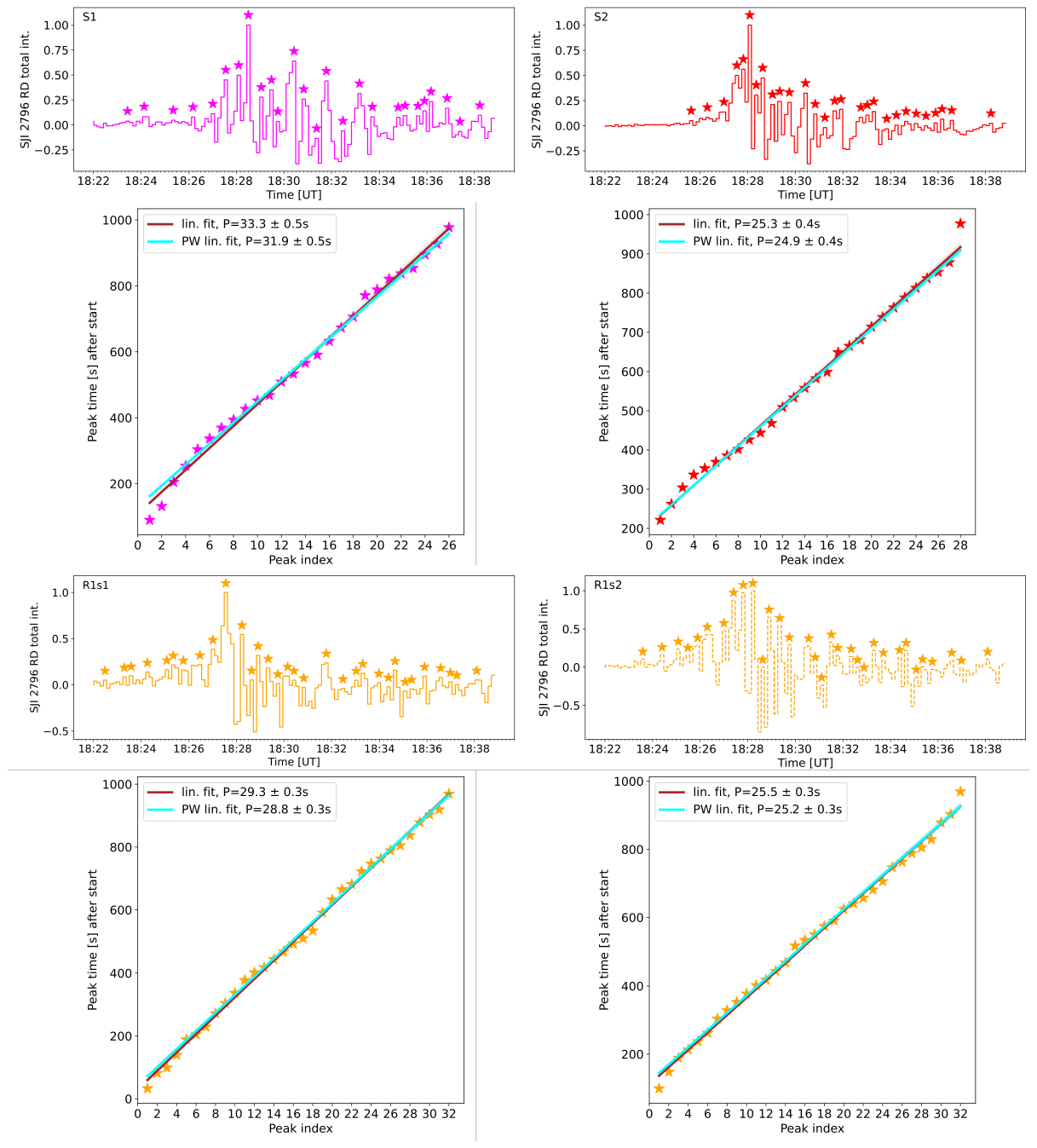}
    \caption{Analysis of SJI QPP periods using linear fits to peak index vs. peak time data for the RD intensities in the four boxes: S1 (purple, top left), S2 (red, top right), R1s1 (orange solid, bottom left), R1s2 (orange dashed, bottom right). The stars mark automatically-identified peaks along the lightcurves indicated above the respective period analysis panel (see Section \ref{sec:stix_qpps}) for the method.}
    \label{fig:app_SJI_P}
\end{figure}

As for the total intensity change, the strongest pulsation in R1, peaking at 18:27:35 UT (panel b1), received $< 30$\% intensity change contribution from the S1 (panel b3) but nearly $60$\% contribution from a pulsation occurring in R1s1. This means that there was a strong kernel in the region contained within both boxes R1s1 and S1 (panel a). Kernels appearing within S1 were dominating the contributions into the pulsations between 18:27 -- 18:34 UT (c.f., panels b1 \& b3). Of particular interest are several low-amplitude ($< 1.7 \times 10^4$\,DN\,s$^{-1}$) peaks visible in panel b1 around 18:26 UT highlighted using the black frame. Because only negligible variation occurred at the same time in S1 (panel b3), these pulsations stemmed from the slipping kernels whose intensity change is well captured in panels b5 and b6. Ribbon emission and dynamics within the R1s2 (b6) had little ($< 30$\%) contribution into the total intensity change of R1. This is because the box captures the smallest segment of the ribbon, chosen to be outside of the strongest HXR footpoint therein (Section \ref{sec:sji_qpps}). The time evolution in R2 (black; panel b2) and S2 (red; panel b4) boxes is nearly identical, with the exception of a single QPP at $\approx$18:34:30 UT only present in the black lightcurve. This demonstrates that increasing the box size (R2), beyond the ribbon subregion with the strongest HXR emission (S2), has little overall effect on the lightcurves. And finally, comparing the blue and black lightcurves in panels b1 and b2, respectively, indicates that R1 intensities showed more variability before 18:27 UT compared to R2. This is a slight limitation of the relatively-smaller SJI FOV, which did not capture the ribbons in their entirety (Section \ref{sec:overview}). Ext. R2 exhibited brightening and signatures of kernel motions at about [720\arcsec, 260\arcsec] in the same time frame, visible e.g. in Figure \ref{fig:overview}c and the animated version of the 304\,\AA~observations. 

\section{Analysis of SJI pulsation periods} \label{sec:app_SJI_P}

Appendix Figure \ref{fig:app_SJI_P} presents the analysis of pulsation periods in the four SJI 2796\,\AA~RD lightcurves investigated in Section \ref{sec:sji_qpps}. Below each normalized lightcurve, the peak index versus time plot for inferring pulsation periods is presented (see Section \ref{sec:stix_qpps} for the method). The stars above the lightcurves indicate automatically-detected peaks whose amplitude is $> 3\sigma$ obtained in the pre-flare period. The period estimated using the linear fits (see Section \ref{sec:stix_qpps} for the method) is the lowest for {S2, $24.5 < P_\textrm{SJI} < 25.7$\,s, and R1s2, $24.9 < P_\textrm{SJI} < 25.8$\,s. Slightly longer period $28.5 < P_\textrm{SJI} < 29.6$\,s is found in R1s1, and the longest, of $31.4 < P_\textrm{SJI} < 33.8$\,s, in S1}. 

\section{Analysis of \ion{Si}{4} pulsation periods} \label{sec:app_spec_P}

The analysis of approximate periods of pulsations exhibited by the \ion{Si}{4} 1402.77\,\AA~line in the ribbon R2 is presented in Appendix Figure \ref{fig:app_spec_P}. The time evolution of the total intensity, Doppler speed, and broadening ($\sigma$), averaged along the four cuts indicated in Figure \ref{fig:spec_moments}a -- c, is presented in the top row. The stars plotted therein mark peaks detected in the lightcurves. For the line intensity (black, left) we focused on the peaks whose prominence exceeded 30\,DN\,s$^{-1}$, roughly equivalent $3 \times \sigma_\mathrm{dc}$, where $\sigma_\mathrm{dc} = 3.1$\,DN is the dark current subtraction uncertainty. In the case of the Doppler velocity (violet, middle) and $\sigma$ (brown, right) we lowered the prominence to flag the peaks higher than the local noise level. Due to the presence of the noise in the broadening lightcurve, only four peaks were selected therein. 

\begin{figure}[h]
    \centering
    \includegraphics[width=0.99\linewidth]{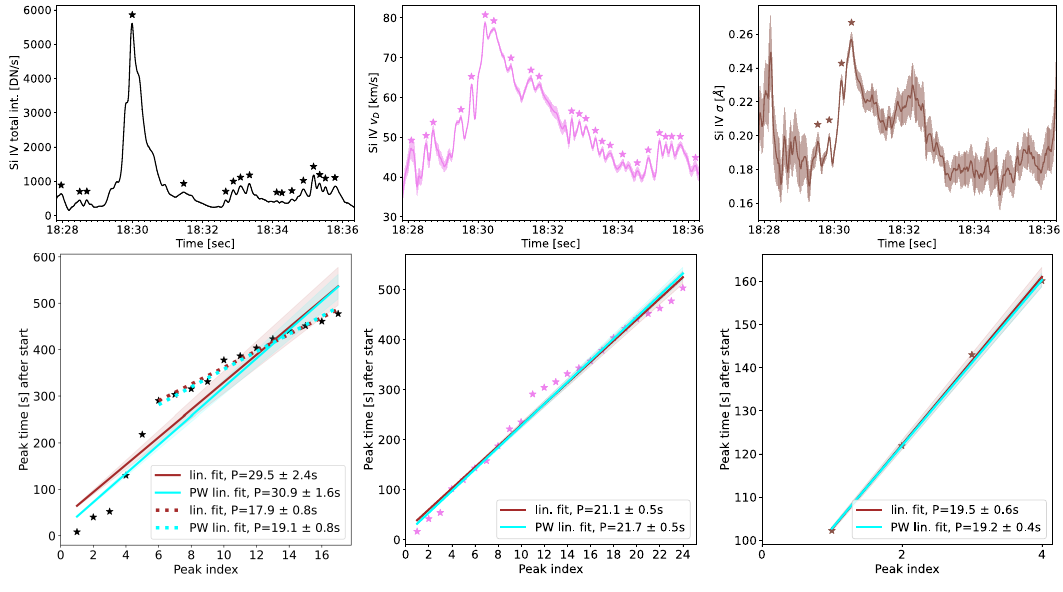}
    \caption{Same as Appendix Figure \ref{fig:app_SJI_P}, but for the \ion{Si}{4} intensity (black, left), Doppler speed (violet, middle), and broadening (brown, right) averaged across the four cuts in Figure \ref{fig:spec_moments}a -- c.}
    \label{fig:app_spec_P}
\end{figure}

In analogy with the analysis of the STIX, GOES, and SJI quasi-periodic signals, we estimate pulsation periods using the linear fits to the peak index versus time data plotted below respective lightcurves. The total intensity exhibited the period of $27.1 < P_{\textrm{Si IV} I} < 32.5$\,s for all peaks detected in the time range indicated in the top row. Note that several data points, in particular those corresponding to the peaks indexed 1 -- 3 and 6 -- 8, diverge significantly from the fits. This is due to the fact that the pulsations were occurring in bursts separated by intervals when no peaks were detected. Limiting the linear fit to the peaks indexed 6 -- 17 detected after 18:32 UT (dotted fits), results in the period decrease down to $17.1 < P_{\textrm{Si IV} I} < 19.9$\,s and a better fit. The period of the Doppler velocity pulsations, occurring throughout the time interval plotted in the top panel, is $20.7 < P_{\textrm{Si IV} v_\mathrm{D}} < 22.2$\,s. The linear fits do not approximate well the peaks indexed 11 -- 13, which is caused by the lack of pulsations around 18:32 UT. The pulsations identified in the broadening lightcurve appeared at the period of roughly $18.8 < P_{\textrm{Si IV} \sigma} < 20.1$\,s, although note that this estimation is based on four data points only.

Based on the quality of the fits discussed above we conclude that the estimation of the periods of the pulsations in the spectral characteristics of the \ion{Si}{4} line was less precise compared to the rest of the lightcurves analyzed in this study. This fact however does not limit the main conclusions of this work focused primarily on HXR and UV imaging pulsations rather than flare ribbon spectroscopy. 

\bibliography{bibliography}

@ARTICLE{Asai04,
       author = {{Asai}, Ayumi and {Yokoyama}, Takaaki and {Shimojo}, Masumi and {Masuda}, Satoshi and {Kurokawa}, Hiroki and {Shibata}, Kazunari},
        title = "{Flare Ribbon Expansion and Energy Release Rate}",
      journal = {\apj},
         year = 2004,
        month = aug,
       volume = {611},
       number = {1},
        pages = {557-567},
          doi = {10.1086/422159},
       adsurl = {https://ui.adsabs.harvard.edu/abs/2004ApJ...611..557A}
}

@ARTICLE{Ashfield22,
       author = {{Ashfield}, IV, William H. and {Longcope}, Dana W. and {Zhu}, Chunming and {Qiu}, Jiong},
        title = "{Connecting Chromospheric Condensation Signatures to Reconnection-driven Heating Rates in an Observed Flare}",
      journal = {\apj},
         year = 2022,
        month = feb,
       volume = {926},
       number = {2},
          eid = {164},
        pages = {164},
          doi = {10.3847/1538-4357/ac402d},
archivePrefix = {arXiv},
       eprint = {2112.02150},
 primaryClass = {astro-ph.SR},
       adsurl = {https://ui.adsabs.harvard.edu/abs/2022ApJ...926..164A}
}

@ARTICLE{Ashfield25,
       author = {{Ashfield}, William and {Polito}, Vanessa and {L{\"o}rin{\v{c}}{\'\i}k}, Juraj and {De Pontieu}, Bart and {Chintzoglou}, Georgios and {Bose}, Souvik and {Freij}, Nabil and {Rouppe van der Voort}, Luc and {Joshi}, Reetika and {Thoen Faber}, Jonas},
        title = "{Spectroscopic Observations of Solar Flare Pulsations Driven by Oscillatory Reconnection}",
       journal = {Research Square},
         year = 2025,
         note = {preprint}
}

@ARTICLE{Aulanier06,
       author = {{Aulanier}, G. and {Pariat}, E. and {D{\'e}moulin}, P. and {Devore}, C.~R.},
        title = "{Slip-Running Reconnection in Quasi-Separatrix Layers}",
      journal = {\solphys},
         year = 2006,
        month = nov,
       volume = {238},
       number = {2},
        pages = {347-376},
          doi = {10.1007/s11207-006-0230-2},
       adsurl = {https://ui.adsabs.harvard.edu/abs/2006SoPh..238..347A}
}

@ARTICLE{Benz17,
       author = {{Benz}, Arnold O.},
        title = "{Flare Observations}",
      journal = {Living Reviews in Solar Physics},
         year = 2017,
        month = dec,
       volume = {14},
       number = {1},
          eid = {2},
        pages = {2},
          doi = {10.1007/s41116-016-0004-3},
       adsurl = {https://ui.adsabs.harvard.edu/abs/2017LRSP...14....2B}
}

@ARTICLE{Bogachev05,
       author = {{Bogachev}, Sergey A. and {Somov}, Boris V. and {Kosugi}, Takeo and {Sakao}, Taro},
        title = "{The Motions of the Hard X-Ray Sources in Solar Flares: Images and Statistics}",
      journal = {\apj},
         year = 2005,
        month = sep,
       volume = {630},
       number = {1},
        pages = {561-572},
          doi = {10.1086/431918},
       adsurl = {https://ui.adsabs.harvard.edu/abs/2005ApJ...630..561B}
}

@ARTICLE{Brueckner95,
       author = {{Brueckner}, G.~E. and {Howard}, R.~A. and {Koomen}, M.~J. and {Korendyke}, C.~M. and {Michels}, D.~J. and {Moses}, J.~D. and {Socker}, D.~G. and {Dere}, K.~P. and {Lamy}, P.~L. and {Llebaria}, A. and {Bout}, M.~V. and {Schwenn}, R. and {Simnett}, G.~M. and {Bedford}, D.~K. and {Eyles}, C.~J.},
        title = "{The Large Angle Spectroscopic Coronagraph (LASCO)}",
      journal = {\solphys},
         year = 1995,
        month = dec,
       volume = {162},
       number = {1-2},
        pages = {357-402},
          doi = {10.1007/BF00733434},
       adsurl = {https://ui.adsabs.harvard.edu/abs/1995SoPh..162..357B}
}

@ARTICLE{Carmichael64,
   author = {{Carmichael}, H.},
    title = "{A Process for Flares}",
  journal = {NASA Special Publication},
     year = 1964,
   volume = 50,
    pages = {451},
   adsurl = {http://adsabs.harvard.edu/abs/1964NASSP..50..451C}
}

@ARTICLE{Cheung22,
       author = {{Cheung}, Mark C.~M. and {Mart{\'\i}nez-Sykora}, Juan and {Testa}, Paola and {De Pontieu}, Bart and {Chintzoglou}, Georgios and {Rempel}, Matthias and {Polito}, Vanessa and {Kerr}, Graham S. and {Reeves}, Katharine K. and {Fletcher}, Lyndsay and {Jin}, Meng and {N{\'o}brega-Siverio}, Daniel and {Danilovic}, Sanja and {Antolin}, Patrick and {Allred}, Joel and {Hansteen}, Viggo and {Ugarte-Urra}, Ignacio and {DeLuca}, Edward and {Longcope}, Dana and {Takasao}, Shinsuke and {DeRosa}, Marc L. and {Boerner}, Paul and {Jaeggli}, Sarah and {Nitta}, Nariaki V. and {Daw}, Adrian and {Carlsson}, Mats and {Golub}, Leon and {The}},
        title = "{Probing the Physics of the Solar Atmosphere with the Multi-slit Solar Explorer (MUSE). II. Flares and Eruptions}",
      journal = {\apj},
         year = 2022,
        month = feb,
       volume = {926},
       number = {1},
          eid = {53},
        pages = {53},
          doi = {10.3847/1538-4357/ac4223},
archivePrefix = {arXiv},
       eprint = {2106.15591},
 primaryClass = {astro-ph.SR},
       adsurl = {https://ui.adsabs.harvard.edu/abs/2022ApJ...926...53C}
}

@ARTICLE{Christe16,
       author = {{Christe}, Steven and {Glesener}, Lindsay and {Buitrago-Casas}, Camilo and {Ishikawa}, Shin-Nosuke and {Ramsey}, Brian and {Gubarev}, Mikhail and {Kilaru}, Kiranmayee and {Kolodziejczak}, Jeffery J. and {Watanabe}, Shin and {Takahashi}, Tadayuki and {Tajima}, Hiroyasu and {Turin}, Paul and {Shourt}, Van and {Foster}, Natalie and {Krucker}, Sam},
        title = "{FOXSI-2: Upgrades of the Focusing Optics X-ray Solar Imager for its Second Flight}",
      journal = {Journal of Astronomical Instrumentation},
         year = 2016,
        month = mar,
       volume = {5},
       number = {1},
          eid = {1640005-625},
        pages = {1640005-625},
          doi = {10.1142/S2251171716400055},
archivePrefix = {arXiv},
       eprint = {2403.07610},
 primaryClass = {astro-ph.IM},
       adsurl = {https://ui.adsabs.harvard.edu/abs/2016JAI.....540005C}
}

@ARTICLE{Clarke21,
       author = {{Clarke}, Brendan P. and {Hayes}, Laura A. and {Gallagher}, Peter T. and {Maloney}, Shane A. and {Carley}, Eoin P.},
        title = "{Quasi-periodic Particle Acceleration in a Solar Flare}",
      journal = {\apj},
         year = 2021,
        month = apr,
       volume = {910},
       number = {2},
          eid = {123},
        pages = {123},
          doi = {10.3847/1538-4357/abe463},
archivePrefix = {arXiv},
       eprint = {2102.04267},
 primaryClass = {astro-ph.SR},
       adsurl = {https://ui.adsabs.harvard.edu/abs/2021ApJ...910..123C}
}

@ARTICLE{Collier23,
       author = {{Collier}, Hannah and {Hayes}, Laura A. and {Battaglia}, Andrea F. and {Harra}, Louise K. and {Krucker}, S{\"a}m},
        title = "{Characterising fast-time variations in the hard X-ray time profiles of solar flares using Solar Orbiter's STIX}",
      journal = {\aap},
         year = 2023,
        month = mar,
       volume = {671},
          eid = {A79},
        pages = {A79},
          doi = {10.1051/0004-6361/202245293},
archivePrefix = {arXiv},
       eprint = {2301.08040},
 primaryClass = {astro-ph.SR},
       adsurl = {https://ui.adsabs.harvard.edu/abs/2023A&A...671A..79C}
}

@ARTICLE{Collier24,
       author = {{Collier}, Hannah and {Hayes}, Laura A. and {Yu}, Sijie and {Battaglia}, Andrea F. and {Ashfield}, William and {Polito}, Vanessa and {Harra}, Louise K. and {Krucker}, S{\"a}m},
        title = "{Localising pulsations in the hard X-ray and microwave emission of an X-class flare}",
      journal = {\aap},
         year = 2024,
        month = apr,
       volume = {684},
          eid = {A215},
        pages = {A215},
          doi = {10.1051/0004-6361/202348652},
archivePrefix = {arXiv},
       eprint = {2402.10546},
 primaryClass = {astro-ph.SR},
       adsurl = {https://ui.adsabs.harvard.edu/abs/2024A&A...684A.215C}
}

@ARTICLE{Dahlin25,
       author = {{Dahlin}, Joel T. and {Antiochos}, Spiro K. and {Wyper}, Peter F. and {Qiu}, Jiong and {DeVore}, C. Richard},
        title = "{Determining the 3D Dynamics of Solar Flare Magnetic Reconnection}",
      journal = {\apj},
         year = 2025,
        month = nov,
       volume = {993},
       number = {1},
          eid = {31},
        pages = {31},
          doi = {10.3847/1538-4357/ae03c5},
archivePrefix = {arXiv},
       eprint = {2504.00913},
 primaryClass = {astro-ph.SR},
       adsurl = {https://ui.adsabs.harvard.edu/abs/2025ApJ...993...31D}
}

@ARTICLE{Demoulin96a,
       author = {{D{\'e}moulin}, P. and {Henoux}, J.~C. and {Priest}, E.~R. and {Mandrini}, C.~H.},
        title = "{Quasi-Separatrix layers in solar flares. I. Method.}",
      journal = {\aap},
         year = 1996,
        month = apr,
       volume = {308},
        pages = {643-655},
       adsurl = {https://ui.adsabs.harvard.edu/abs/1996A&A...308..643D}
}

@ARTICLE{Demoulin06,
       author = {{D{\'e}moulin}, P.},
        title = "{Extending the concept of separatrices to QSLs for magnetic reconnection}",
      journal = {Advances in Space Research},
         year = 2006,
        month = jan,
       volume = {37},
       number = {7},
        pages = {1269-1282},
          doi = {10.1016/j.asr.2005.03.085},
       adsurl = {https://ui.adsabs.harvard.edu/abs/2006AdSpR..37.1269D}
}

@ARTICLE{Depontieu14,
       author = {{De Pontieu}, B. and {Title}, A.~M. and {Lemen}, J.~R. and {Kushner}, G.~D. and {Akin}, D.~J. and {Allard}, B. and {Berger}, T. and {Boerner}, P. and {Cheung}, M. and {Chou}, C. and {Drake}, J.~F. and {Duncan}, D.~W. and {Freeland}, S. and {Heyman}, G.~F. and {Hoffman}, C. and {Hurlburt}, N.~E. and {Lindgren}, R.~W. and {Mathur}, D. and {Rehse}, R. and {Sabolish}, D. and {Seguin}, R. and {Schrijver}, C.~J. and {Tarbell}, T.~D. and {W{\"u}lser}, J. -P. and {Wolfson}, C.~J. and {Yanari}, C. and {Mudge}, J. and {Nguyen-Phuc}, N. and {Timmons}, R. and {van Bezooijen}, R. and {Weingrod}, I. and {Brookner}, R. and {Butcher}, G. and {Dougherty}, B. and {Eder}, J. and {Knagenhjelm}, V. and {Larsen}, S. and {Mansir}, D. and {Phan}, L. and {Boyle}, P. and {Cheimets}, P.~N. and {DeLuca}, E.~E. and {Golub}, L. and {Gates}, R. and {Hertz}, E. and {McKillop}, S. and {Park}, S. and {Perry}, T. and {Podgorski}, W.~A. and {Reeves}, K. and {Saar}, S. and {Testa}, P. and {Tian}, H. and {Weber}, M. and {Dunn}, C. and {Eccles}, S. and {Jaeggli}, S.~A. and {Kankelborg}, C.~C. and {Mashburn}, K. and {Pust}, N. and {Springer}, L. and {Carvalho}, R. and {Kleint}, L. and {Marmie}, J. and {Mazmanian}, E. and {Pereira}, T.~M.~D. and {Sawyer}, S. and {Strong}, J. and {Worden}, S.~P. and {Carlsson}, M. and {Hansteen}, V.~H. and {Leenaarts}, J. and {Wiesmann}, M. and {Aloise}, J. and {Chu}, K. -C. and {Bush}, R.~I. and {Scherrer}, P.~H. and {Brekke}, P. and {Martinez-Sykora}, J. and {Lites}, B.~W. and {McIntosh}, S.~W. and {Uitenbroek}, H. and {Okamoto}, T.~J. and {Gummin}, M.~A. and {Auker}, G. and {Jerram}, P. and {Pool}, P. and {Waltham}, N.},
        title = "{The Interface Region Imaging Spectrograph (IRIS)}",
      journal = {Solar Physics},
         year = 2014,
        month = jul,
       volume = {289},
       number = {7},
        pages = {2733-2779},
          doi = {10.1007/s11207-014-0485-y},
archivePrefix = {arXiv},
       eprint = {1401.2491},
 primaryClass = {astro-ph.SR},
       adsurl = {https://ui.adsabs.harvard.edu/abs/2014SoPh..289.2733D}
}

@ARTICLE{Depontieu21,
       author = {{De Pontieu}, Bart and {Polito}, Vanessa and {Hansteen}, Viggo and {Testa}, Paola and {Reeves}, Katharine K. and {Antolin}, Patrick and {N{\'o}brega-Siverio}, Daniel Elias and {Kowalski}, Adam F. and {Martinez-Sykora}, Juan and {Carlsson}, Mats and {McIntosh}, Scott W. and {Liu}, Wei and {Daw}, Adrian and {Kankelborg}, Charles C.},
        title = "{A New View of the Solar Interface Region from the Interface Region Imaging Spectrograph (IRIS)}",
      journal = {\solphys},
         year = 2021,
        month = may,
       volume = {296},
       number = {5},
          eid = {84},
        pages = {84},
          doi = {10.1007/s11207-021-01826-0},
archivePrefix = {arXiv},
       eprint = {2103.16109},
 primaryClass = {astro-ph.SR},
       adsurl = {https://ui.adsabs.harvard.edu/abs/2021SoPh..296...84D}
}

@ARTICLE{Depontieu22,
       author = {{De Pontieu}, Bart and {Testa}, Paola and {Mart{\'\i}nez-Sykora}, Juan and {Antolin}, Patrick and {Karampelas}, Konstantinos and {Hansteen}, Viggo and {Rempel}, Matthias and {Cheung}, Mark C.~M. and {Reale}, Fabio and {Danilovic}, Sanja and {Pagano}, Paolo and {Polito}, Vanessa and {De Moortel}, Ineke and {N{\'o}brega-Siverio}, Daniel and {Van Doorsselaere}, Tom and {Petralia}, Antonino and {Asgari-Targhi}, Mahboubeh and {Boerner}, Paul and {Carlsson}, Mats and {Chintzoglou}, Georgios and {Daw}, Adrian and {DeLuca}, Edward and {Golub}, Leon and {Matsumoto}, Takuma and {Ugarte-Urra}, Ignacio and {McIntosh}, Scott W. and {the MUSE Team}},
        title = "{Probing the Physics of the Solar Atmosphere with the Multi-slit Solar Explorer (MUSE). I. Coronal Heating}",
      journal = {\apj},
         year = 2022,
        month = feb,
       volume = {926},
       number = {1},
          eid = {52},
        pages = {52},
          doi = {10.3847/1538-4357/ac4222},
archivePrefix = {arXiv},
       eprint = {2106.15584},
 primaryClass = {astro-ph.SR},
       adsurl = {https://ui.adsabs.harvard.edu/abs/2022ApJ...926...52D}
}

@ARTICLE{Dudik14,
       author = {{Dud{\'\i}k}, J. and {Janvier}, M. and {Aulanier}, G. and {Del Zanna}, G. and {Karlick{\'y}}, M. and {Mason}, H.~E. and {Schmieder}, B.},
        title = "{Slipping Magnetic Reconnection during an X-class Solar Flare Observed by SDO/AIA}",
      journal = {\apj},
         year = 2014,
        month = apr,
       volume = {784},
       number = {2},
          eid = {144},
        pages = {144},
          doi = {10.1088/0004-637X/784/2/144},
archivePrefix = {arXiv},
       eprint = {1401.7529},
 primaryClass = {astro-ph.SR},
       adsurl = {https://ui.adsabs.harvard.edu/abs/2014ApJ...784..144D}
}

@ARTICLE{Dudik16,
       author = {{Dud{\'\i}k}, Jaroslav and {Polito}, Vanessa and {Janvier}, Miho and {Mulay}, Sargam M. and {Karlick{\'y}}, Marian and {Aulanier}, Guillaume and {Del Zanna}, Giulio and {Dzif{\v{c}}{\'a}kov{\'a}}, Elena and {Mason}, Helen E. and {Schmieder}, Brigitte},
        title = "{Slipping Magnetic Reconnection, Chromospheric Evaporation, Implosion, and Precursors in the 2014 September 10 X1.6-Class Solar Flare}",
      journal = {\apj},
         year = 2016,
        month = may,
       volume = {823},
       number = {1},
          eid = {41},
        pages = {41},
          doi = {10.3847/0004-637X/823/1/41},
archivePrefix = {arXiv},
       eprint = {1603.06092},
 primaryClass = {astro-ph.SR},
       adsurl = {https://ui.adsabs.harvard.edu/abs/2016ApJ...823...41D}
}

@ARTICLE{Dudik25,
       author = {{Dud{\'\i}k}, Jaroslav and {Aulanier}, Guillaume and {L{\"o}rin{\v{c}}{\'\i}k}, Juraj and {Zemanov{\'a}}, Alena},
        title = "{Quasi-Separatrix Layers and Three-Dimensional Magnetic Reconnection: Theory and Observations of Solar Flares (Invited Review)}",
      journal = {\solphys},
         year = 2025,
        month = oct,
       volume = {300},
       number = {10},
          eid = {139},
        pages = {139},
          doi = {10.1007/s11207-025-02549-2},
       adsurl = {https://ui.adsabs.harvard.edu/abs/2025SoPh..300..139D}
}

@ARTICLE{Huang24,
       author = {{Huang}, Jing and {Tan}, Baolin and {Zhang}, Yin and {Zhu}, Xiaoshuai and {Yang}, Shangbin and {Deng}, Yuanyong},
        title = "{The Slipping Magnetic Reconnection and Damped Quasiperiodic Pulsations in a Circular Ribbon Flare}",
      journal = {\apj},
         year = 2024,
        month = apr,
       volume = {965},
       number = {2},
          eid = {137},
        pages = {137},
          doi = {10.3847/1538-4357/ad3353},
       adsurl = {https://ui.adsabs.harvard.edu/abs/2024ApJ...965..137H}
}

@ARTICLE{Fletcher02,
       author = {{Fletcher}, L. and {Hudson}, H.~S.},
        title = "{Spectral and Spatial Variations of Flare Hard X-ray Footpoints}",
      journal = {\solphys},
         year = 2002,
        month = nov,
       volume = {210},
       number = {1},
        pages = {307-321},
          doi = {10.1023/A:1022479610710},
       adsurl = {https://ui.adsabs.harvard.edu/abs/2002SoPh..210..307F}
}

@ARTICLE{Fletcher04,
       author = {{Fletcher}, Lyndsay and {Pollock}, Jennifer A. and {Potts}, Hugh E.},
        title = "{Tracking of TRACE Ultraviolet Flare Footpoints}",
      journal = {\solphys},
         year = 2004,
        month = aug,
       volume = {222},
       number = {2},
        pages = {279-298},
          doi = {10.1023/B:SOLA.0000043580.89730.4d},
       adsurl = {https://ui.adsabs.harvard.edu/abs/2004SoPh..222..279F}
}

@ARTICLE{Fletcher11,
       author = {{Fletcher}, L. and {Dennis}, B.~R. and {Hudson}, H.~S. and {Krucker}, S. and {Phillips}, K. and {Veronig}, A. and {Battaglia}, M. and {Bone}, L. and {Caspi}, A. and {Chen}, Q. and {Gallagher}, P. and {Grigis}, P.~T. and {Ji}, H. and {Liu}, W. and {Milligan}, R.~O. and {Temmer}, M.},
        title = "{An Observational Overview of Solar Flares}",
      journal = {\ssr},
         year = 2011,
        month = sep,
       volume = {159},
       number = {1-4},
        pages = {19-106},
          doi = {10.1007/s11214-010-9701-8},
archivePrefix = {arXiv},
       eprint = {1109.5932},
 primaryClass = {astro-ph.SR},
       adsurl = {https://ui.adsabs.harvard.edu/abs/2011SSRv..159...19F}
}

@ARTICLE{French24,
       author = {{French}, Ryan J. and {Hayes}, Laura A. and {Kazachenko}, Maria D. and {Reeves}, Katharine K. and {Shen}, Chengcai and {L{\"o}rin{\v{c}}{\'\i}k}, Juraj},
        title = "{X-Ray and Spectral Ultraviolet Observations of Periodic Pulsations in a Solar Flare Fan/Looptop}",
      journal = {\apj},
         year = 2024,
        month = dec,
       volume = {977},
       number = {2},
          eid = {207},
        pages = {207},
          doi = {10.3847/1538-4357/ad8ed1},
archivePrefix = {arXiv},
       eprint = {2411.02634},
 primaryClass = {astro-ph.SR},
       adsurl = {https://ui.adsabs.harvard.edu/abs/2024ApJ...977..207F}
}

@ARTICLE{French25,
       author = {{French}, Ryan J. and {Kazachenko}, Maria D. and {Berghmans}, David and {D'Huys}, Elke and {Dominique}, Marie and {Patel}, Ritesh and {Talpeanu}, Dana-Camelia and {Tamburri}, Cole A. and {Yadav}, Rahul},
        title = "{Evolution of Flare Ribbon Bead-like Structures in a Solar Flare}",
      journal = {\apjl},
         year = 2025,
        month = dec,
       volume = {995},
       number = {2},
          eid = {L54},
        pages = {L54},
          doi = {10.3847/2041-8213/ae2684},
archivePrefix = {arXiv},
       eprint = {2512.00710},
 primaryClass = {astro-ph.SR},
       adsurl = {https://ui.adsabs.harvard.edu/abs/2025ApJ...995L..54F}
}

@ARTICLE{Gan19,
       author = {{Gan}, Wei-Qun and {Zhu}, Cheng and {Deng}, Yuan-Yong and {Li}, Hui and {Su}, Yang and {Zhang}, Hai-Ying and {Chen}, Bo and {Zhang}, Zhe and {Wu}, Jian and {Deng}, Lei and {Huang}, Yu and {Yang}, Jian-Feng and {Cui}, Ji-Jun and {Chang}, Jin and {Wang}, Chi and {Wu}, Ji and {Yin}, Zeng-Shan and {Chen}, Wen and {Fang}, Cheng and {Yan}, Yi-Hua and {Lin}, Jun and {Xiong}, Wei-Ming and {Chen}, Bin and {Bao}, Hai-Chao and {Cao}, Cai-Xia and {Bai}, Yan-Ping and {Wang}, Tao and {Chen}, Bing-Long and {Li}, Xin-Yu and {Zhang}, Ye and {Feng}, Li and {Su}, Jiang-Tao and {Li}, Ying and {Chen}, Wei and {Li}, You-Ping and {Su}, Ying-Na and {Wu}, Hai-Yan and {Gu}, Mei and {Huang}, Lei and {Tang}, Xue-Jun},
        title = "{Advanced Space-based Solar Observatory (ASO-S): an overview}",
      journal = {Research in Astronomy and Astrophysics},
         year = 2019,
        month = nov,
       volume = {19},
       number = {11},
          eid = {156},
        pages = {156},
          doi = {10.1088/1674-4527/19/11/156},
       adsurl = {https://ui.adsabs.harvard.edu/abs/2019RAA....19..156G}
}

@ARTICLE{Graham15,
       author = {{Graham}, D.~R. and {Cauzzi}, G.},
        title = "{Temporal Evolution of Multiple Evaporating Ribbon Sources in a Solar Flare}",
      journal = {\apjl},
         year = 2015,
        month = jul,
       volume = {807},
       number = {2},
          eid = {L22},
        pages = {L22},
          doi = {10.1088/2041-8205/807/2/L22},
archivePrefix = {arXiv},
       eprint = {1506.03465},
 primaryClass = {astro-ph.SR},
       adsurl = {https://ui.adsabs.harvard.edu/abs/2015ApJ...807L..22G}
}

@ARTICLE{Graham20,
       author = {{Graham}, David R. and {Cauzzi}, Gianna and {Zangrilli}, Luca and {Kowalski}, Adam and {Sim{\~o}es}, Paulo and {Allred}, Joel},
        title = "{Spectral Signatures of Chromospheric Condensation in a Major Solar Flare}",
      journal = {\apj},
         year = 2020,
        month = may,
       volume = {895},
       number = {1},
          eid = {6},
        pages = {6},
          doi = {10.3847/1538-4357/ab88ad},
archivePrefix = {arXiv},
       eprint = {2004.05075},
 primaryClass = {astro-ph.SR},
       adsurl = {https://ui.adsabs.harvard.edu/abs/2020ApJ...895....6G}
}

@ARTICLE{Hayes20,
       author = {{Hayes}, Laura A. and {Inglis}, Andrew R. and {Christe}, Steven and {Dennis}, Brian and {Gallagher}, Peter T.},
        title = "{Statistical Study of GOES X-Ray Quasi-periodic Pulsations in Solar Flares}",
      journal = {\apj},
         year = 2020,
        month = may,
       volume = {895},
       number = {1},
          eid = {50},
        pages = {50},
          doi = {10.3847/1538-4357/ab8d40},
archivePrefix = {arXiv},
       eprint = {2004.11775},
 primaryClass = {astro-ph.SR},
       adsurl = {https://ui.adsabs.harvard.edu/abs/2020ApJ...895...50H}
}

@ARTICLE{Hirayama74,
   author = {{Hirayama}, T.},
    title = "{Theoretical Model of Flares and Prominences. I: Evaporating Flare Model}",
  journal = {Solar Physics},
     year = 1974,
    month = feb,
   volume = 34,
    pages = {323-338},
      doi = {10.1007/BF00153671},
   adsurl = {http://adsabs.harvard.edu/abs/1974SoPh...34..323H}
}

@ARTICLE{Inglis12,
       author = {{Inglis}, A.~R. and {Dennis}, B.~R.},
        title = "{The Relationship between Hard X-Ray Pulse Timings and the Locations of Footpoint Sources during Solar Flares}",
      journal = {\apj},
         year = 2012,
        month = apr,
       volume = {748},
       number = {2},
          eid = {139},
        pages = {139},
          doi = {10.1088/0004-637X/748/2/139},
archivePrefix = {arXiv},
       eprint = {1303.6309},
 primaryClass = {astro-ph.SR},
       adsurl = {https://ui.adsabs.harvard.edu/abs/2012ApJ...748..139I}
}

@ARTICLE{Inglis13,
       author = {{Inglis}, A.~R. and {Gilbert}, H.~R.},
        title = "{Hard X-Ray and Ultraviolet Emission during the 2011 June 7 Solar Flare}",
      journal = {\apj},
         year = 2013,
        month = nov,
       volume = {777},
       number = {1},
          eid = {30},
        pages = {30},
          doi = {10.1088/0004-637X/777/1/30},
archivePrefix = {arXiv},
       eprint = {1307.2874},
 primaryClass = {astro-ph.SR},
       adsurl = {https://ui.adsabs.harvard.edu/abs/2013ApJ...777...30I}
}

@ARTICLE{Janvier13,
       author = {{Janvier}, M. and {Aulanier}, G. and {Pariat}, E. and {D{\'e}moulin}, P.},
        title = "{The standard flare model in three dimensions. III. Slip-running reconnection properties}",
      journal = {\aap},
         year = 2013,
        month = jul,
       volume = {555},
          eid = {A77},
        pages = {A77},
          doi = {10.1051/0004-6361/201321164},
archivePrefix = {arXiv},
       eprint = {1305.4053},
 primaryClass = {astro-ph.SR},
       adsurl = {https://ui.adsabs.harvard.edu/abs/2013A&A...555A..77J}
}

@ARTICLE{Janvier17,
       author = {{Janvier}, Miho},
        title = "{Three-dimensional magnetic reconnection and its application to solar flares}",
      journal = {Journal of Plasma Physics},
         year = 2017,
        month = feb,
       volume = {83},
       number = {1},
          eid = {535830101},
        pages = {535830101},
          doi = {10.1017/S0022377817000034},
archivePrefix = {arXiv},
       eprint = {1612.06513},
 primaryClass = {astro-ph.SR},
       adsurl = {https://ui.adsabs.harvard.edu/abs/2017JPlPh..83a5301J}
}

@ARTICLE{Jarolim23,
       author = {{Jarolim}, R. and {Thalmann}, J.~K. and {Veronig}, A.~M. and {Podladchikova}, T.},
        title = "{Probing the solar coronal magnetic field with physics-informed neural networks.}",
      journal = {Nature Astronomy},
         year = 2023,
        month = oct,
       volume = {7},
        pages = {1171-1179},
          doi = {10.1038/s41550-023-02030-9},
       adsurl = {https://ui.adsabs.harvard.edu/abs/2023NatAs...7.1171J}
}

@ARTICLE{Jarolim24b,
       author = {{Jarolim}, Robert and {Veronig}, Astrid M. and {Purkhart}, Stefan and {Zhang}, Peijin and {Rempel}, Matthias},
        title = "{Magnetic Field Evolution of the Solar Active Region 13664}",
      journal = {\apjl},
         year = 2024,
        month = nov,
       volume = {976},
       number = {1},
          eid = {L12},
        pages = {L12},
          doi = {10.3847/2041-8213/ad8914},
archivePrefix = {arXiv},
       eprint = {2409.08124},
 primaryClass = {astro-ph.SR},
       adsurl = {https://ui.adsabs.harvard.edu/abs/2024ApJ...976L..12J}
}

@ARTICLE{Jeffrey18,
       author = {{Jeffrey}, N.~L.~S. and {Fletcher}, L. and {Labrosse}, N. and {Sim{\~o}es}, P.~J.~A.},
        title = "{The development of lower-atmosphere turbulence early in a solar flare}",
      journal = {Science Advances},
         year = 2018,
        month = dec,
       volume = {4},
       number = {12},
        pages = {2794},
          doi = {10.1126/sciadv.aav2794},
archivePrefix = {arXiv},
       eprint = {1812.09906},
 primaryClass = {astro-ph.SR},
       adsurl = {https://ui.adsabs.harvard.edu/abs/2018SciA....4.2794J}
}

@ARTICLE{Jing17,
       author = {{Jing}, Ju and {Liu}, Rui and {Cheung}, Mark C.~M. and {Lee}, Jeongwoo and {Xu}, Yan and {Liu}, Chang and {Zhu}, Chunming and {Wang}, Haimin},
        title = "{Witnessing a Large-scale Slipping Magnetic Reconnection along a Dimming Channel during a Solar Flare}",
      journal = {\apjl},
         year = 2017,
        month = jun,
       volume = {842},
       number = {2},
          eid = {L18},
        pages = {L18},
          doi = {10.3847/2041-8213/aa774d},
archivePrefix = {arXiv},
       eprint = {1706.01355},
 primaryClass = {astro-ph.SR},
       adsurl = {https://ui.adsabs.harvard.edu/abs/2017ApJ...842L..18J}
}

@ARTICLE{Kashapova20,
       author = {{Kashapova}, L.~K. and {Kupriyanova}, E.~G. and {Xu}, Z. and {Reid}, H.~A.~S. and {Kolotkov}, D.~Y.},
        title = "{The origin of quasi-periodicities during circular ribbon flares}",
      journal = {\aap},
         year = 2020,
        month = oct,
       volume = {642},
          eid = {A195},
        pages = {A195},
          doi = {10.1051/0004-6361/201833947},
archivePrefix = {arXiv},
       eprint = {2008.02010},
 primaryClass = {astro-ph.SR},
       adsurl = {https://ui.adsabs.harvard.edu/abs/2020A&A...642A.195K}
}

@ARTICLE{Kerr16,
       author = {{Kerr}, Graham S. and {Fletcher}, Lyndsay. and {Russell}, Alexander J.~B. and {Allred}, Joel C.},
        title = "{Simulations of the Mg II k and Ca II 8542 lines from an Alfv{\'e}n Wave-heated Flare Chromosphere}",
      journal = {\apj},
         year = 2016,
        month = aug,
       volume = {827},
       number = {2},
          eid = {101},
        pages = {101},
          doi = {10.3847/0004-637X/827/2/101},
archivePrefix = {arXiv},
       eprint = {1605.05888},
 primaryClass = {astro-ph.SR},
       adsurl = {https://ui.adsabs.harvard.edu/abs/2016ApJ...827..101K}
}

@ARTICLE{Kopp76,
   author = {{Kopp}, R.~A. and {Pneuman}, G.~W.},
    title = "{Magnetic reconnection in the corona and the loop prominence phenomenon}",
  journal = {Solar Physics},
     year = 1976,
    month = oct,
   volume = 50,
    pages = {85-98},
      doi = {10.1007/BF00206193},
   adsurl = {http://adsabs.harvard.edu/abs/1976SoPh...50...85K}
}

@ARTICLE{Krucker05,
       author = {{Krucker}, S{\"a}m and {Fivian}, M.~D. and {Lin}, R.~P.},
        title = "{Hard X-ray footpoint motions in solar flares: Comparing magnetic reconnection models with observations}",
      journal = {Advances in Space Research},
         year = 2005,
        month = jan,
       volume = {35},
       number = {10},
        pages = {1707-1711},
          doi = {10.1016/j.asr.2005.05.054},
       adsurl = {https://ui.adsabs.harvard.edu/abs/2005AdSpR..35.1707K}
}

@ARTICLE{Krucker08,
       author = {{Krucker}, S. and {Battaglia}, M. and {Cargill}, P.~J. and {Fletcher}, L. and {Hudson}, H.~S. and {MacKinnon}, A.~L. and {Masuda}, S. and {Sui}, L. and {Tomczak}, M. and {Veronig}, A.~L. and {Vlahos}, L. and {White}, S.~M.},
        title = "{Hard X-ray emission from the solar corona}",
      journal = {\aapr},
         year = 2008,
        month = oct,
       volume = {16},
        pages = {155-208},
          doi = {10.1007/s00159-008-0014-9},
       adsurl = {https://ui.adsabs.harvard.edu/abs/2008A&ARv..16..155K}
}

@ARTICLE{Krucker20,
       author = {{Krucker}, S{\"a}m and {Hurford}, G.~J. and {Grimm}, O. and {K{\"o}gl}, S. and {Gr{\"o}belbauer}, H. -P. and {Etesi}, L. and {Casadei}, D. and {Csillaghy}, A. and {Benz}, A.~O. and {Arnold}, N.~G. and {Molendini}, F. and {Orleanski}, P. and {Schori}, D. and {Xiao}, H. and {Kuhar}, M. and {Hochmuth}, N. and {Felix}, S. and {Schramka}, F. and {Marcin}, S. and {Kobler}, S. and {Iseli}, L. and {Dreier}, M. and {Wiehl}, H.~J. and {Kleint}, L. and {Battaglia}, M. and {Lastufka}, E. and {Sathiapal}, H. and {Lapadula}, K. and {Bednarzik}, M. and {Birrer}, G. and {Stutz}, St. and {Wild}, Ch. and {Marone}, F. and {Skup}, K.~R. and {Cichocki}, A. and {Ber}, K. and {Rutkowski}, K. and {Bujwan}, W. and {Juchnikowski}, G. and {Winkler}, M. and {Darmetko}, M. and {Michalska}, M. and {Seweryn}, K. and {Bia{\l}ek}, A. and {Osica}, P. and {Sylwester}, J. and {Kowalinski}, M. and {{\'S}cis{\l}owski}, D. and {Siarkowski}, M. and {St{\k{e}}{\'s}licki}, M. and {Mrozek}, T. and {Podg{\'o}rski}, P. and {Meuris}, A. and {Limousin}, O. and {Gevin}, O. and {Le Mer}, I. and {Brun}, S. and {Strugarek}, A. and {Vilmer}, N. and {Musset}, S. and {Maksimovi{\'c}}, M. and {F{\'a}rn{\'\i}k}, F. and {Koz{\'a}{\v{c}}ek}, Z. and {Ka{\v{s}}parov{\'a}}, J. and {Mann}, G. and {{\"O}nel}, H. and {Warmuth}, A. and {Rendtel}, J. and {Anderson}, J. and {Bauer}, S. and {Dionies}, F. and {Paschke}, J. and {Pl{\"u}schke}, D. and {Woche}, M. and {Schuller}, F. and {Veronig}, A.~M. and {Dickson}, E.~C.~M. and {Gallagher}, P.~T. and {Maloney}, S.~A. and {Bloomfield}, D.~S. and {Piana}, M. and {Massone}, A.~M. and {Benvenuto}, F. and {Massa}, P. and {Schwartz}, R.~A. and {Dennis}, B.~R. and {van Beek}, H.~F. and {Rodr{\'\i}guez-Pacheco}, J. and {Lin}, R.~P.},
        title = "{The Spectrometer/Telescope for Imaging X-rays (STIX)}",
      journal = {\aap},
         year = 2020,
        month = oct,
       volume = {642},
          eid = {A15},
        pages = {A15},
          doi = {10.1051/0004-6361/201937362},
       adsurl = {https://ui.adsabs.harvard.edu/abs/2020A&A...642A..15K}
}

@ARTICLE{Lemen12,
   author = {{Lemen}, J.~R. and {Title}, A.~M. and {Akin}, D.~J. and {Boerner}, P.~F. and 
	{Chou}, C. and {Drake}, J.~F. and {Duncan}, D.~W. and {Edwards}, C.~G. and 
	{Friedlaender}, F.~M. and {Heyman}, G.~F. and {Hurlburt}, N.~E. and 
	{Katz}, N.~L. and {Kushner}, G.~D. and {Levay}, M. and {Lindgren}, R.~W. and 
	{Mathur}, D.~P. and {McFeaters}, E.~L. and {Mitchell}, S. and 
	{Rehse}, R.~A. and {Schrijver}, C.~J. and {Springer}, L.~A. and 
	{Stern}, R.~A. and {Tarbell}, T.~D. and {Wuelser}, J.-P. and 
	{Wolfson}, C.~J. and {Yanari}, C. and {Bookbinder}, J.~A. and 
	{Cheimets}, P.~N. and {Caldwell}, D. and {Deluca}, E.~E. and 
	{Gates}, R. and {Golub}, L. and {Park}, S. and {Podgorski}, W.~A. and 
	{Bush}, R.~I. and {Scherrer}, P.~H. and {Gummin}, M.~A. and 
	{Smith}, P. and {Auker}, G. and {Jerram}, P. and {Pool}, P. and 
	{Soufli}, R. and {Windt}, D.~L. and {Beardsley}, S. and {Clapp}, M. and 
	{Lang}, J. and {Waltham}, N.},
    title = "{The Atmospheric Imaging Assembly (AIA) on the Solar Dynamics Observatory (SDO)}",
  journal = {Solar Physics},
     year = 2012,
    month = jan,
   volume = 275,
    pages = {17-40},
      doi = {10.1007/s11207-011-9776-8},
   adsurl = {http://adsabs.harvard.edu/abs/2012SoPh..275...17L}
}

@ARTICLE{LiT14,
       author = {{Li}, Ting and {Zhang}, Jun},
        title = "{Slipping Magnetic Reconnection Triggering a Solar Eruption of a Triangle-shaped Flag Flux Rope}",
      journal = {\apjl},
         year = 2014,
        month = aug,
       volume = {791},
       number = {1},
          eid = {L13},
        pages = {L13},
          doi = {10.1088/2041-8205/791/1/L13},
archivePrefix = {arXiv},
       eprint = {1407.4180},
 primaryClass = {astro-ph.SR},
       adsurl = {https://ui.adsabs.harvard.edu/abs/2014ApJ...791L..13L}
}

@ARTICLE{LiT15,
       author = {{Li}, Ting and {Zhang}, Jun},
        title = "{Quasi-periodic Slipping Magnetic Reconnection During an X-class Solar Flare Observed by the Solar Dynamics Observatory and Interface Region Imaging Spectrograph}",
      journal = {\apjl},
         year = 2015,
        month = may,
       volume = {804},
       number = {1},
          eid = {L8},
        pages = {L8},
          doi = {10.1088/2041-8205/804/1/L8},
archivePrefix = {arXiv},
       eprint = {1504.01111},
 primaryClass = {astro-ph.SR},
       adsurl = {https://ui.adsabs.harvard.edu/abs/2015ApJ...804L...8L}
}

@ARTICLE{LiT16,
       author = {{Li}, Ting and {Yang}, Kai and {Hou}, Yijun and {Zhang}, Jun},
        title = "{Slipping Magnetic Reconnection of Flux-rope Structures as a Precursor to an Eruptive X-class Solar Flare}",
      journal = {\apj},
         year = 2016,
        month = oct,
       volume = {830},
       number = {2},
          eid = {152},
        pages = {152},
          doi = {10.3847/0004-637X/830/2/152},
archivePrefix = {arXiv},
       eprint = {1608.02057},
 primaryClass = {astro-ph.SR},
       adsurl = {https://ui.adsabs.harvard.edu/abs/2016ApJ...830..152L}
}

@ARTICLE{Longcope14,
       author = {{Longcope}, D.~W.},
        title = "{A Simple Model of Chromospheric Evaporation and Condensation Driven Conductively in a Solar Flare}",
      journal = {\apj},
         year = 2014,
        month = nov,
       volume = {795},
       number = {1},
          eid = {10},
        pages = {10},
          doi = {10.1088/0004-637X/795/1/10},
archivePrefix = {arXiv},
       eprint = {1409.1886},
 primaryClass = {astro-ph.SR},
       adsurl = {https://ui.adsabs.harvard.edu/abs/2014ApJ...795...10L}
}

@ARTICLE{Lorincik19a,
       author = {{L{\"o}rin{\v{c}}{\'\i}k}, Juraj and {Aulanier}, Guillaume and {Dud{\'\i}k}, Jaroslav and {Zemanov{\'a}}, Alena and {Dzif{\v{c}}{\'a}kov{\'a}}, Elena},
        title = "{Velocities of Flare Kernels and the Mapping Norm of Field Line Connectivity}",
      journal = {\apj},
         year = 2019,
        month = aug,
       volume = {881},
       number = {1},
          eid = {68},
        pages = {68},
          doi = {10.3847/1538-4357/ab298f},
archivePrefix = {arXiv},
       eprint = {1906.01880},
 primaryClass = {astro-ph.SR},
       adsurl = {https://ui.adsabs.harvard.edu/abs/2019ApJ...881...68L}
}

@ARTICLE{Lorincik22,
       author = {{L{\"o}rin{\v{c}}{\'\i}k}, Juraj and {Polito}, Vanessa and {De Pontieu}, Bart and {Yu}, Sijie and {Freij}, Nabil},
        title = "{Rapid variations of Si IV spectra in a flare observed by interface region imaging spectrograph at a sub-second cadence}",
      journal = {Frontiers in Astronomy and Space Sciences},
         year = 2022,
        month = nov,
       volume = {9},
          eid = {334},
        pages = {334},
          doi = {10.3389/fspas.2022.1040945},
archivePrefix = {arXiv},
       eprint = {2210.12205},
 primaryClass = {astro-ph.SR},
       adsurl = {https://ui.adsabs.harvard.edu/abs/2022FrASS...940945L}
}

@ARTICLE{Lorincik25a,
       author = {{L{\"o}rin{\v{c}}{\'\i}k}, Juraj and {Dud{\'\i}k}, Jaroslav and {Sainz Dalda}, Alberto and {Aulanier}, Guillaume and {Polito}, Vanessa and {De Pontieu}, Bart},
        title = "{Observation of super-Alfv{\'e}nic slippage of reconnecting magnetic field lines on the Sun}",
      journal = {Nature Astronomy},
         year = 2025,
        month = jan,
       volume = {9},
        pages = {45-54},
          doi = {10.1038/s41550-024-02396-4},
       adsurl = {https://ui.adsabs.harvard.edu/abs/2025NatAs...9...45L}
}

@ARTICLE{Lorincik25b,
       author = {{L{\"o}rin{\v{c}}{\'\i}k}, Juraj and {Polito}, Vanessa and {Kerr}, Graham S. and {Hayes}, Laura A. and {Russell}, Alexander J.~B.},
        title = "{Probing Progression of Heating Through the Lower Flare Atmosphere via High-cadence IRIS Spectroscopy}",
      journal = {\apj},
         year = 2025,
        month = jun,
       volume = {986},
       number = {1},
          eid = {73},
        pages = {73},
          doi = {10.3847/1538-4357/adccc8},
archivePrefix = {arXiv},
       eprint = {2504.10619},
 primaryClass = {astro-ph.SR},
       adsurl = {https://ui.adsabs.harvard.edu/abs/2025ApJ...986...73L}
}

@ARTICLE{Macniece86,
       author = {{MacNeice}, P.},
        title = "{A Numerical Hydrodynamic Model of a Heated Coronal Loop}",
      journal = {\solphys},
         year = 1986,
        month = jan,
       volume = {103},
       number = {1},
        pages = {47-66},
          doi = {10.1007/BF00154858},
       adsurl = {https://ui.adsabs.harvard.edu/abs/1986SoPh..103...47M}
}

@ARTICLE{Massa20,
       author = {{Massa}, Paolo and {Schwartz}, Richard and {Tolbert}, A. Kim and {Massone}, Anna Maria and {Dennis}, Brian R. and {Piana}, Michele and {Benvenuto}, Federico},
        title = "{MEM\_GE: A New Maximum Entropy Method for Image Reconstruction from Solar X-Ray Visibilities}",
      journal = {\apj},
         year = 2020,
        month = may,
       volume = {894},
       number = {1},
          eid = {46},
        pages = {46},
          doi = {10.3847/1538-4357/ab8637},
archivePrefix = {arXiv},
       eprint = {2002.07921},
 primaryClass = {astro-ph.SR},
       adsurl = {https://ui.adsabs.harvard.edu/abs/2020ApJ...894...46M}
}

@ARTICLE{McLaughlin18,
       author = {{McLaughlin}, J.~A. and {Nakariakov}, V.~M. and {Dominique}, M. and {Jel{\'\i}nek}, P. and {Takasao}, S.},
        title = "{Modelling Quasi-Periodic Pulsations in Solar and Stellar Flares}",
      journal = {\ssr},
         year = 2018,
        month = feb,
       volume = {214},
       number = {1},
          eid = {45},
        pages = {45},
          doi = {10.1007/s11214-018-0478-5},
archivePrefix = {arXiv},
       eprint = {1802.04180},
 primaryClass = {astro-ph.SR},
       adsurl = {https://ui.adsabs.harvard.edu/abs/2018SSRv..214...45M}
}

@ARTICLE{Miklenic07,
       author = {{Miklenic}, C.~H. and {Veronig}, A.~M. and {Vr{\v{s}}nak}, B. and {Hanslmeier}, A.},
        title = "{Reconnection and energy release rates in a two-ribbon flare}",
      journal = {\aap},
         year = 2007,
        month = jan,
       volume = {461},
       number = {2},
        pages = {697-706},
          doi = {10.1051/0004-6361:20065751},
       adsurl = {https://ui.adsabs.harvard.edu/abs/2007A&A...461..697M}
}

@ARTICLE{Muller20,
       author = {{M{\"u}ller}, D. and {St. Cyr}, O.~C. and {Zouganelis}, I. and {Gilbert}, H.~R. and {Marsden}, R. and {Nieves-Chinchilla}, T. and {Antonucci}, E. and {Auch{\`e}re}, F. and {Berghmans}, D. and {Horbury}, T.~S. and {Howard}, R.~A. and {Krucker}, S. and {Maksimovic}, M. and {Owen}, C.~J. and {Rochus}, P. and {Rodriguez-Pacheco}, J. and {Romoli}, M. and {Solanki}, S.~K. and {Bruno}, R. and {Carlsson}, M. and {Fludra}, A. and {Harra}, L. and {Hassler}, D.~M. and {Livi}, S. and {Louarn}, P. and {Peter}, H. and {Sch{\"u}hle}, U. and {Teriaca}, L. and {del Toro Iniesta}, J.~C. and {Wimmer-Schweingruber}, R.~F. and {Marsch}, E. and {Velli}, M. and {De Groof}, A. and {Walsh}, A. and {Williams}, D.},
        title = "{The Solar Orbiter mission. Science overview}",
      journal = {\aap},
         year = 2020,
        month = oct,
       volume = {642},
          eid = {A1},
        pages = {A1},
          doi = {10.1051/0004-6361/202038467},
archivePrefix = {arXiv},
       eprint = {2009.00861},
 primaryClass = {astro-ph.SR},
       adsurl = {https://ui.adsabs.harvard.edu/abs/2020A&A...642A...1M}
}

@ARTICLE{Nakariakov09,
       author = {{Nakariakov}, V.~M. and {Melnikov}, V.~F.},
        title = "{Quasi-Periodic Pulsations in Solar Flares}",
      journal = {\ssr},
         year = 2009,
        month = dec,
       volume = {149},
       number = {1-4},
        pages = {119-151},
          doi = {10.1007/s11214-009-9536-3},
       adsurl = {https://ui.adsabs.harvard.edu/abs/2009SSRv..149..119N}
}

@ARTICLE{Odwyer10,
       author = {{O'Dwyer}, B. and {Del Zanna}, G. and {Mason}, H.~E. and {Weber}, M.~A. and {Tripathi}, D.},
        title = "{SDO/AIA response to coronal hole, quiet Sun, active region, and flare plasma}",
      journal = {\aap},
         year = 2010,
        month = oct,
       volume = {521},
          eid = {A21},
        pages = {A21},
          doi = {10.1051/0004-6361/201014872},
       adsurl = {https://ui.adsabs.harvard.edu/abs/2010A&A...521A..21O}
}

@ARTICLE{Pesnell12,
       author = {{Pesnell}, W. Dean and {Thompson}, B.~J. and {Chamberlin}, P.~C.},
        title = "{The Solar Dynamics Observatory (SDO)}",
      journal = {Solar Physics},
         year = 2012,
        month = jan,
       volume = {275},
       number = {1-2},
        pages = {3-15},
          doi = {10.1007/s11207-011-9841-3},
       adsurl = {https://ui.adsabs.harvard.edu/abs/2012SoPh..275....3P}
}

@ARTICLE{Poland82,
       author = {{Poland}, A.~I. and {Machado}, M.~E. and {Wolfson}, C.~J. and {Frost}, K.~J. and {Woodgate}, B.~E. and {Shine}, R.~A. and {Kenny}, P.~J. and {Cheng}, C.~C. and {Tandberg-Hanssen}, E.~A. and {Bruner}, E.~C. and {Henze}, W.},
        title = "{The Impulsive and Gradual Phases of a Solar Limb Flare as Observed from the Solar Maximum Mission Satellite}",
      journal = {\solphys},
         year = 1982,
        month = jun,
       volume = {78},
       number = {2},
        pages = {201-213},
          doi = {10.1007/BF00151603},
       adsurl = {https://ui.adsabs.harvard.edu/abs/1982SoPh...78..201P}
}

@ARTICLE{Polito16a,
       author = {{Polito}, V. and {Reep}, J.~W. and {Reeves}, K.~K. and {Sim{\~o}es}, P.~J.~A. and {Dud{\'\i}k}, J. and {Del Zanna}, G. and {Mason}, H.~E. and {Golub}, L.},
        title = "{Simultaneous IRIS and Hinode/EIS Observations and Modelling of the 2014 October 27 X2.0 Class Flare}",
      journal = {\apj},
         year = 2016,
        month = jan,
       volume = {816},
       number = {2},
          eid = {89},
        pages = {89},
          doi = {10.3847/0004-637X/816/2/89},
archivePrefix = {arXiv},
       eprint = {1512.06378},
 primaryClass = {astro-ph.SR},
       adsurl = {https://ui.adsabs.harvard.edu/abs/2016ApJ...816...89P}
}

@ARTICLE{Polito16,
       author = {{Polito}, V. and {Del Zanna}, G. and {Dud{\'\i}k}, J. and {Mason}, H.~E. and {Giunta}, A. and {Reeves}, K.~K.},
        title = "{Density diagnostics derived from the O iv and S iv intercombination lines observed by IRIS}",
      journal = {\aap},
         year = 2016,
        month = oct,
       volume = {594},
          eid = {A64},
        pages = {A64},
          doi = {10.1051/0004-6361/201628965},
archivePrefix = {arXiv},
       eprint = {1607.05072},
 primaryClass = {astro-ph.SR},
       adsurl = {https://ui.adsabs.harvard.edu/abs/2016A&A...594A..64P}
}

@ARTICLE{Polito23b,
       author = {{Polito}, Vanessa and {Peterson}, Marianne and {Glesener}, Lindsay and {Testa}, Paola and {Yu}, Sijie and {Reeves}, Katharine K. and {Sun}, Xudong and {Duncan}, Jessie},
        title = "{Multi-wavelength observations and modeling of a microflare: constraining non-thermal particle acceleration}",
      journal = {Frontiers in Astronomy and Space Sciences},
         year = 2023,
        month = sep,
       volume = {10},
          eid = {1214901},
        pages = {1214901},
          doi = {10.3389/fspas.2023.1214901},
       adsurl = {https://ui.adsabs.harvard.edu/abs/2023FrASS..1014901P}
}

@ARTICLE{Priest95,
       author = {{Priest}, E.~R. and {D{\'e}moulin}, P.},
        title = "{Three-dimensional magnetic reconnection without null points. 1. Basic theory of magnetic flipping}",
      journal = {\jgr},
         year = 1995,
        month = dec,
       volume = {100},
       number = {A12},
        pages = {23443-23464},
          doi = {10.1029/95JA02740},
       adsurl = {https://ui.adsabs.harvard.edu/abs/1995JGR...10023443P}
}

@ARTICLE{Purkhart25,
       author = {{Purkhart}, Stefan and {Collier}, Hannah and {Hayes}, Laura A. and {Veronig}, Astrid M. and {Janvier}, Miho and {Krucker}, S{\"a}m},
        title = "{Spatiotemporal evolution of UV pulsations and their connection to 3D magnetic reconnection and particle acceleration}",
      journal = {\aap},
         year = 2025,
        month = jun,
       volume = {698},
          eid = {A318},
        pages = {A318},
          doi = {10.1051/0004-6361/202554475},
archivePrefix = {arXiv},
       eprint = {2506.03825},
 primaryClass = {astro-ph.SR},
       adsurl = {https://ui.adsabs.harvard.edu/abs/2025A&A...698A.318P}
}

@ARTICLE{Qiu12,
       author = {{Qiu}, J. and {Cheng}, J.~X. and {Hurford}, G.~J. and {Xu}, Y. and {Wang}, H.},
        title = "{Solar flare hard X-ray spikes observed by RHESSI: a case study}",
      journal = {\aap},
         year = 2012,
        month = nov,
       volume = {547},
          eid = {A72},
        pages = {A72},
          doi = {10.1051/0004-6361/201118609},
archivePrefix = {arXiv},
       eprint = {1210.7040},
 primaryClass = {astro-ph.SR},
       adsurl = {https://ui.adsabs.harvard.edu/abs/2012A&A...547A..72Q}
}

@ARTICLE{Reep16,
       author = {{Reep}, J.~W. and {Russell}, A.~J.~B.},
        title = "{Alfv{\'e}nic Wave Heating of the Upper Chromosphere in Flares}",
      journal = {\apjl},
         year = 2016,
        month = feb,
       volume = {818},
       number = {1},
          eid = {L20},
        pages = {L20},
          doi = {10.3847/2041-8205/818/1/L20},
archivePrefix = {arXiv},
       eprint = {1601.01969},
 primaryClass = {astro-ph.SR},
       adsurl = {https://ui.adsabs.harvard.edu/abs/2016ApJ...818L..20R}
}

@ARTICLE{Rochus20,
       author = {{Rochus}, P. and {Auch{\`e}re}, F. and {Berghmans}, D. and {Harra}, L. and {Schmutz}, W. and {Sch{\"u}hle}, U. and {Addison}, P. and {Appourchaux}, T. and {Aznar Cuadrado}, R. and {Baker}, D. and {Barbay}, J. and {Bates}, D. and {BenMoussa}, A. and {Bergmann}, M. and {Beurthe}, C. and {Borgo}, B. and {Bonte}, K. and {Bouzit}, M. and {Bradley}, L. and {B{\"u}chel}, V. and {Buchlin}, E. and {B{\"u}chner}, J. and {Cab{\'e}}, F. and {Cadiergues}, L. and {Chaigneau}, M. and {Chares}, B. and {Choque Cortez}, C. and {Coker}, P. and {Condamin}, M. and {Coumar}, S. and {Curdt}, W. and {Cutler}, J. and {Davies}, D. and {Davison}, G. and {Defise}, J.-M. and {Del Zanna}, G. and {Delmotte}, F. and {Delouille}, V. and {Dolla}, L. and {Dumesnil}, C. and {D{\"u}rig}, F. and {Enge}, R. and {Fran{\c{c}}ois}, S. and {Fourmond}, J.-J. and {Gillis}, J.-M. and {Giordanengo}, B. and {Gissot}, S. and {Green}, L.~M. and {Guerreiro}, N. and {Guilbaud}, A. and {Gyo}, M. and {Haberreiter}, M. and {Hafiz}, A. and {Hailey}, M. and {Halain}, J.-P. and {Hansotte}, J. and {Hecquet}, C. and {Heerlein}, K. and {Hellin}, M.-L. and {Hemsley}, S. and {Hermans}, A. and {Hervier}, V. and {Hochedez}, J.-F. and {Houbrechts}, Y. and {Ihsan}, K. and {Jacques}, L. and {J{\'e}r{\^o}me}, A. and {Jones}, J. and {Kahle}, M. and {Kennedy}, T. and {Klaproth}, M. and {Kolleck}, M. and {Koller}, S. and {Kotsialos}, E. and {Kraaikamp}, E. and {Langer}, P. and {Lawrenson}, A. and {Le Clech'}, J.-C. and {Lenaerts}, C. and {Liebecq}, S. and {Linder}, D. and {Long}, D.~M. and {Mampaey}, B. and {Markiewicz-Innes}, D. and {Marquet}, B. and {Marsch}, E. and {Matthews}, S. and {Mazy}, E. and {Mazzoli}, A. and {Meining}, S. and {Meltchakov}, E. and {Mercier}, R. and {Meyer}, S. and {Monecke}, M. and {Monfort}, F. and {Morinaud}, G. and {Moron}, F. and {Mountney}, L. and {M{\"u}ller}, R. and {Nicula}, B. and {Parenti}, S. and {Peter}, H. and {Pfiffner}, D. and {Philippon}, A. and {Phillips}, I. and {Plesseria}, J.-Y. and {Pylyser}, E. and {Rabecki}, F. and {Ravet-Krill}, M.-F. and {Rebellato}, J. and {Renotte}, E. and {Rodriguez}, L. and {Roose}, S. and {Rosin}, J. and {Rossi}, L. and {Roth}, P. and {Rouesnel}, F. and {Roulliay}, M. and {Rousseau}, A. and {Ruane}, K. and {Scanlan}, J. and {Schlatter}, P. and {Seaton}, D.~B. and {Silliman}, K. and {Smit}, S. and {Smith}, P.~J. and {Solanki}, S.~K. and {Spescha}, M. and {Spencer}, A. and {Stegen}, K. and {Stockman}, Y. and {Szwec}, N. and {Tamiatto}, C. and {Tandy}, J. and {Teriaca}, L. and {Theobald}, C. and {Tychon}, I. and {van Driel-Gesztelyi}, L. and {Verbeeck}, C. and {Vial}, J.-C. and {Werner}, S. and {West}, M.~J. and {Westwood}, D. and {Wiegelmann}, T. and {Willis}, G. and {Winter}, B. and {Zerr}, A. and {Zhang}, X. and {Zhukov}, A.~N.},
        title = "{The Solar Orbiter EUI instrument: The Extreme Ultraviolet Imager}",
      journal = {\aap},
         year = 2020,
        month = oct,
       volume = {642},
          eid = {A8},
        pages = {A8},
          doi = {10.1051/0004-6361/201936663},
       adsurl = {https://ui.adsabs.harvard.edu/abs/2020A&A...642A...8R}
}

@ARTICLE{Russell13,
       author = {{Russell}, A.~J.~B. and {Fletcher}, L.},
        title = "{Propagation of Alfv{\'e}nic Waves from Corona to Chromosphere and Consequences for Solar Flares}",
      journal = {\apj},
         year = 2013,
        month = mar,
       volume = {765},
       number = {2},
          eid = {81},
        pages = {81},
          doi = {10.1088/0004-637X/765/2/81},
archivePrefix = {arXiv},
       eprint = {1302.2458},
 primaryClass = {astro-ph.SR},
       adsurl = {https://ui.adsabs.harvard.edu/abs/2013ApJ...765...81R}
}

@ARTICLE{Ryan25,
       author = {{Ryan}, Daniel F. and {Hayes}, Laura A. and {Collier}, Hannah and {Kerr}, Graham S. and {Inglis}, Andrew R. and {Williams}, David and {Walsh}, Andrew P. and {Janvier}, Miho and {M{\"u}ller}, Daniel and {Berghmans}, David and {Verbeeck}, Cis and {Kraaikamp}, Emil and {Young}, Peter R. and {Kucera}, Therese A. and {Krucker}, S{\"a}m and {Stiefel}, Muriel Z. and {Calchetti}, Daniele and {Reeves}, Katharine K. and {Savage}, Sabrina and {Polito}, Vanessa},
        title = "{Solar Orbiter's 2024 Major Flare Campaigns: An Overview}",
      journal = {\solphys},
         year = 2025,
        month = nov,
       volume = {300},
       number = {11},
          eid = {152},
        pages = {152},
          doi = {10.1007/s11207-025-02561-6},
archivePrefix = {arXiv},
       eprint = {2505.07472},
 primaryClass = {astro-ph.SR},
       adsurl = {https://ui.adsabs.harvard.edu/abs/2025SoPh..300..152R}
}

@ARTICLE{Sandlin86,
       author = {{Sandlin}, G.~D. and {Bartoe}, J. -D.~F. and {Brueckner}, G.~E. and {Tousey}, R. and {Vanhoosier}, M.~E.},
        title = "{The High-Resolution Solar Spectrum, 1175--1710 Angstrom}",
      journal = {\apjs},
         year = 1986,
        month = aug,
       volume = {61},
        pages = {801},
          doi = {10.1086/191131},
       adsurl = {https://ui.adsabs.harvard.edu/abs/1986ApJS...61..801S}
}

@ARTICLE{Scipy20,
  author  = {Virtanen, Pauli and Gommers, Ralf and Oliphant, Travis E. and
            Haberland, Matt and Reddy, Tyler and Cournapeau, David and
            Burovski, Evgeni and Peterson, Pearu and Weckesser, Warren and
            Bright, Jonathan and {van der Walt}, St{\'e}fan J. and
            Brett, Matthew and Wilson, Joshua and Millman, K. Jarrod and
            Mayorov, Nikolay and Nelson, Andrew R. J. and Jones, Eric and
            Kern, Robert and Larson, Eric and Carey, C J and
            Polat, {\.I}lhan and Feng, Yu and Moore, Eric W. and
            {VanderPlas}, Jake and Laxalde, Denis and Perktold, Josef and
            Cimrman, Robert and Henriksen, Ian and Quintero, E. A. and
            Harris, Charles R. and Archibald, Anne M. and
            Ribeiro, Ant{\^o}nio H. and Pedregosa, Fabian and
            {van Mulbregt}, Paul and {SciPy 1.0 Contributors}},
  title   = {{{SciPy} 1.0: Fundamental Algorithms for Scientific
            Computing in Python}},
  journal = {Nature Methods},
  year    = {2020},
  volume  = {17},
  pages   = {261--272},
  adsurl  = {https://rdcu.be/b08Wh},
  doi     = {10.1038/s41592-019-0686-2},
}

@misc{Shih20,
  author       = {Shih, Albert Y. and
                  Glesener, Lindsay and
                  Krucker, Säm and
                  Guidoni, Silvina and
                  Christe, Steven and
                  Reeves, Katharine and
                  Gburek, Szymon and
                  Caspi, Amir and
                  Alaoui, Meriem and
                  Allred, Joel and
                  Battaglia, Marina and
                  Baumgartner, Wayne and
                  Dennis, Brian and
                  Drake, James and
                  Goetz, Keith and
                  Golub, Leon and
                  Hannah, Iain and
                  Hayes, Laura and
                  Holman, Gordon and
                  Inglis, Andrew and
                  Ireland, Jack and
                  Kerr, Graham and
                  Klimchuk, James and
                  McKenzie, David and
                  Moore, Christopher and
                  Musset, Sophie and
                  Reep, Jeffrey and
                  Ryan, Daniel and
                  Saint-Hilaire, Pascal and
                  Savage, Sabrina and
                  Schwartz, Richard and
                  Seaton, Daniel and
                  Stęślicki, Marek and
                  Woods, Thomas},
  title        = {Combined Next-Generation X-ray and EUV
                   Observations with the FIERCE Mission Concept
                  },
  month        = jan,
  year         = 2020,
  publisher    = {Zenodo},
  doi          = {10.5281/zenodo.3674079},
  url          = {https://doi.org/10.5281/zenodo.3674079},
}

@ARTICLE{Scherrer12,
   author = {{Scherrer}, P.~H. and {Schou}, J. and {Bush}, R.~I. and {Kosovichev}, A.~G. and 
	{Bogart}, R.~S. and {Hoeksema}, J.~T. and {Liu}, Y. and {Duvall}, T.~L. and 
	{Zhao}, J. and {Title}, A.~M. and {Schrijver}, C.~J. and {Tarbell}, T.~D. and 
	{Tomczyk}, S.},
    title = "{The Helioseismic and Magnetic Imager (HMI) Investigation for the Solar Dynamics Observatory (SDO)}",
  journal = {Solar Physics},
     year = 2012,
    month = jan,
   volume = 275,
    pages = {207-227},
      doi = {10.1007/s11207-011-9834-2},
   adsurl = {http://adsabs.harvard.edu/abs/2012SoPh..275..207S}
}

@ARTICLE{Sturrock66,
   author = {{Sturrock}, P.~A.},
    title = "{Model of the High-Energy Phase of Solar Flares}",
  journal = {Nature},
     year = 1966,
    month = aug,
   volume = 211,
    pages = {695-697},
      doi = {10.1038/211695a0},
   adsurl = {http://adsabs.harvard.edu/abs/1966Natur.211..695S}
}

@ARTICLE{Sunpy20,
       author = {{SunPy Community} and {Barnes}, Will T. and {Bobra}, Monica G. and {Christe}, Steven D. and {Freij}, Nabil and {Hayes}, Laura A. and {Ireland}, Jack and {Mumford}, Stuart and {Perez-Suarez}, David and {Ryan}, Daniel F. and {Shih}, Albert Y. and {Chanda}, Prateek and {Glogowski}, Kolja and {Hewett}, Russell and {Hughitt}, V. Keith and {Hill}, Andrew and {Hiware}, Kaustubh and {Inglis}, Andrew and {Kirk}, Michael S.~F. and {Konge}, Sudarshan and {Mason}, James Paul and {Maloney}, Shane Anthony and {Murray}, Sophie A. and {Panda}, Asish and {Park}, Jongyeob and {Pereira}, Tiago M.~D. and {Reardon}, Kevin and {Savage}, Sabrina and {Sip{\H{o}}cz}, Brigitta M. and {Stansby}, David and {Jain}, Yash and {Taylor}, Garrison and {Yadav}, Tannmay and {Rajul} and {Dang}, Trung Kien},
        title = "{The SunPy Project: Open Source Development and Status of the Version 1.0 Core Package}",
      journal = {\apj},
         year = 2020,
        month = feb,
       volume = {890},
       number = {1},
          eid = {68},
        pages = {68},
          doi = {10.3847/1538-4357/ab4f7a},
       adsurl = {https://ui.adsabs.harvard.edu/abs/2020ApJ...890...68S}
}

@ARTICLE{Szaforz25,
       author = {{Szaforz}, {\.Z}. and {Mrozek}, T. and {Tomczak}, M.},
        title = "{Statistical investigation of solar flares showing quasi-periodicity based on STIX quick-look light curves}",
      journal = {\aap},
         year = 2025,
        month = feb,
       volume = {694},
          eid = {A251},
        pages = {A251},
          doi = {10.1051/0004-6361/202452046},
       adsurl = {https://ui.adsabs.harvard.edu/abs/2025A&A...694A.251S}
}

@ARTICLE{Tandberg83,
       author = {{Tandberg-Hanssen}, E. and {Reichmann}, E. and {Woodgate}, B.},
        title = "{Behavior of Transition Region Lines during Impulsive Solar Flares}",
      journal = {\solphys},
         year = 1983,
        month = jul,
       volume = {86},
       number = {1-2},
        pages = {159-171},
          doi = {10.1007/BF00157184},
       adsurl = {https://ui.adsabs.harvard.edu/abs/1983SoPh...86..159T}
}

@ARTICLE{Tassev17,
       author = {{Tassev}, Svetlin and {Savcheva}, Antonia},
        title = "{QSL Squasher: A Fast Quasi-separatrix Layer Map Calculator}",
      journal = {\apj},
         year = 2017,
        month = may,
       volume = {840},
       number = {2},
          eid = {89},
        pages = {89},
          doi = {10.3847/1538-4357/aa6f06},
archivePrefix = {arXiv},
       eprint = {1609.00724},
 primaryClass = {astro-ph.SR},
       adsurl = {https://ui.adsabs.harvard.edu/abs/2017ApJ...840...89T}
}

@ARTICLE{Temmer07,
       author = {{Temmer}, M. and {Veronig}, A.~M. and {Vr{\v{s}}nak}, B. and {Miklenic}, C.},
        title = "{Energy Release Rates along H{\ensuremath{\alpha}} Flare Ribbons and the Location of Hard X-Ray Sources}",
      journal = {\apj},
         year = 2007,
        month = jan,
       volume = {654},
       number = {1},
        pages = {665-674},
          doi = {10.1086/509634},
       adsurl = {https://ui.adsabs.harvard.edu/abs/2007ApJ...654..665T}
}

@ARTICLE{Titov02,
       author = {{Titov}, Vyacheslav S. and {Hornig}, Gunnar and {D{\'e}moulin}, Pascal},
        title = "{Theory of magnetic connectivity in the solar corona}",
      journal = {Journal of Geophysical Research (Space Physics)},
         year = 2002,
        month = aug,
       volume = {107},
       number = {A8},
          eid = {1164},
        pages = {1164},
          doi = {10.1029/2001JA000278},
       adsurl = {https://ui.adsabs.harvard.edu/abs/2002JGRA..107.1164T}
}

@ARTICLE{VanDoorsselaere16,
       author = {{Van Doorsselaere}, Tom and {Kupriyanova}, Elena G. and {Yuan}, Ding},
        title = "{Quasi-periodic Pulsations in Solar and Stellar Flares: An Overview of Recent Results (Invited Review)}",
      journal = {\solphys},
         year = 2016,
        month = nov,
       volume = {291},
       number = {11},
        pages = {3143-3164},
          doi = {10.1007/s11207-016-0977-z},
archivePrefix = {arXiv},
       eprint = {1609.02689},
 primaryClass = {astro-ph.SR},
       adsurl = {https://ui.adsabs.harvard.edu/abs/2016SoPh..291.3143V}
}

@ARTICLE{Warren16,
       author = {{Warren}, Harry P. and {Reep}, Jeffrey W. and {Crump}, Nicholas A. and {Sim{\~o}es}, Paulo J.~A.},
        title = "{Transition Region and Chromospheric Signatures of Impulsive Heating Events. I. Observations}",
      journal = {\apj},
         year = 2016,
        month = sep,
       volume = {829},
       number = {1},
          eid = {35},
        pages = {35},
          doi = {10.3847/0004-637X/829/1/35},
archivePrefix = {arXiv},
       eprint = {1606.09045},
 primaryClass = {astro-ph.SR},
       adsurl = {https://ui.adsabs.harvard.edu/abs/2016ApJ...829...35W}
}

@ARTICLE{Wyper21,
       author = {{Wyper}, P.~F. and {Pontin}, D.~I.},
        title = "{Is Flare Ribbon Fine Structure Related to Tearing in the Flare Current Sheet?}",
      journal = {\apj},
         year = 2021,
        month = oct,
       volume = {920},
       number = {2},
          eid = {102},
        pages = {102},
          doi = {10.3847/1538-4357/ac1943},
archivePrefix = {arXiv},
       eprint = {2108.10966},
 primaryClass = {astro-ph.SR},
       adsurl = {https://ui.adsabs.harvard.edu/abs/2021ApJ...920..102W}
}

@ARTICLE{Young18,
       author = {{Young}, P.~R. and {Keenan}, F.~P. and {Milligan}, R.~O. and {Peter}, H.},
        title = "{A Si IV/O IV Electron Density Diagnostic for the Analysis of IRIS Solar Spectra}",
      journal = {\apj},
         year = 2018,
        month = apr,
       volume = {857},
       number = {1},
          eid = {5},
        pages = {5},
          doi = {10.3847/1538-4357/aab556},
archivePrefix = {arXiv},
       eprint = {1803.01721},
 primaryClass = {astro-ph.SR},
       adsurl = {https://ui.adsabs.harvard.edu/abs/2018ApJ...857....5Y}
}

@ARTICLE{ZhangP22,
       author = {{Zhang}, PeiJin and {Chen}, Jun and {Liu}, Rui and {Wang}, ChuanBing},
        title = "{FastQSL: A Fast Computation Method for Quasi-separatrix Layers}",
      journal = {\apj},
         year = 2022,
        month = sep,
       volume = {937},
       number = {1},
          eid = {26},
        pages = {26},
          doi = {10.3847/1538-4357/ac8d61},
archivePrefix = {arXiv},
       eprint = {2208.12569},
 primaryClass = {astro-ph.SR},
       adsurl = {https://ui.adsabs.harvard.edu/abs/2022ApJ...937...26Z}
}

@ARTICLE{ZhangY25,
       author = {{Zhang}, Yining and {Li}, Ting and {Hou}, Yijun and {Duan}, Xuchun and {Sun}, Zheng and {Zhou}, Guiping},
        title = "{Quasiperiodic Super-Alfv{\'e}nic Slippage along Flare Ribbons Observed by the Interface Region Imaging Spectrograph}",
      journal = {\apjl},
         year = 2025,
        month = mar,
       volume = {982},
       number = {1},
          eid = {L9},
        pages = {L9},
          doi = {10.3847/2041-8213/adbb6e},
archivePrefix = {arXiv},
       eprint = {2502.16579},
 primaryClass = {astro-ph.SR},
       adsurl = {https://ui.adsabs.harvard.edu/abs/2025ApJ...982L...9Z}
}

@ARTICLE{Zhao16,
       author = {{Zhao}, Jie and {Gilchrist}, Stuart A. and {Aulanier}, Guillaume and {Schmieder}, Brigitte and {Pariat}, Etienne and {Li}, Hui},
        title = "{Hooked Flare Ribbons and Flux-rope-related QSL Footprints}",
      journal = {\apj},
         year = 2016,
        month = may,
       volume = {823},
       number = {1},
          eid = {62},
        pages = {62},
          doi = {10.3847/0004-637X/823/1/62},
archivePrefix = {arXiv},
       eprint = {1603.07563},
 primaryClass = {astro-ph.SR},
       adsurl = {https://ui.adsabs.harvard.edu/abs/2016ApJ...823...62Z}
}

@ARTICLE{Zimovets21,
       author = {{Zimovets}, I.~V. and {McLaughlin}, J.~A. and {Srivastava}, A.~K. and {Kolotkov}, D.~Y. and {Kuznetsov}, A.~A. and {Kupriyanova}, E.~G. and {Cho}, I. -H. and {Inglis}, A.~R. and {Reale}, F. and {Pascoe}, D.~J. and {Tian}, H. and {Yuan}, D. and {Li}, D. and {Zhang}, Q.~M.},
        title = "{Quasi-Periodic Pulsations in Solar and Stellar Flares: A Review of Underpinning Physical Mechanisms and Their Predicted Observational Signatures}",
      journal = {\ssr},
         year = 2021,
        month = aug,
       volume = {217},
       number = {5},
          eid = {66},
        pages = {66},
          doi = {10.1007/s11214-021-00840-9},
       adsurl = {https://ui.adsabs.harvard.edu/abs/2021SSRv..217...66Z}
}

@ARTICLE{Polito18,
       author = {{Polito}, V. and {Testa}, P. and {Allred}, J. and {De Pontieu}, B. and {Carlsson}, M. and {Pereira}, T.~M.~D. and {Go{\v{s}}i{\'c}}, Milan and {Reale}, Fabio},
        title = "{Investigating the Response of Loop Plasma to Nanoflare Heating Using RADYN Simulations}",
      journal = {\apj},
         year = 2018,
        month = apr,
       volume = {856},
       number = {2},
          eid = {178},
        pages = {178},
          doi = {10.3847/1538-4357/aab49e},
archivePrefix = {arXiv},
       eprint = {1804.05970},
 primaryClass = {astro-ph.SR},
       adsurl = {https://ui.adsabs.harvard.edu/abs/2018ApJ...856..178P}
}

\end{document}